\documentclass[fleqn,usenatbib]{mnras}

\usepackage{newtxtext,newtxmath}

\usepackage[T1]{fontenc}

\DeclareRobustCommand{\VAN}[3]{#2}
\let\VANthebibliography\thebibliography
\def\thebibliography{\DeclareRobustCommand{\VAN}[3]{##3}\VANthebibliography}

\usepackage{graphicx}	
\usepackage{amsmath}	

\title[Production of hot Jupiter candidates from high-eccentricity mechanisms]{Production of hot Jupiter candidates from high-eccentricity mechanisms for different initial planetary mass configurations}
\author[H. Garz\'on et al.]{
H. Garz\'on,$^{1}$
Adri\'an Rodr\'iguez,$^{1}$\thanks{E-mail: adrian@ov.ufrj.br}
and G. C. de El\'ia$^{2,3}$
\\
$^{1}$Observat\'orio do Valongo, Universidade Federal do Rio de Janeiro, Ladeira do Pedro Ant\^onio 43, 20080-090, Rio de Janeiro, Brazil\\
$^{2}$Instituto de Astrof\'isica de La Plata, CCT La Plata-CONICET-UNLP, Paseo del Bosque S/N (1900), La Plata, Argentina\\
$^{3}$Facultad de Ciencias Astron\'omicas y Geof\'isicas, Universidad Nacional de La Plata, Paseo del Bosque S/N (1900), La Plata, Argentina
}

\date{Accepted XXX. Received YYY; in original form ZZZ}

\pubyear{2022}

\begin{document}
\label{firstpage}
\pagerange{\pageref{firstpage}--\pageref{lastpage}}
\maketitle

\begin{abstract}
Hot Jupiters (HJs) are giant planets with orbital periods of the order of a few days with semimajor axis within $\sim$0.1 au.  Several theories have been invoked in order to explain the origin of this type of planets, one of them being the high-eccentricity migration. This migration can occur through different high-eccentricity mechanisms. Our investigation focused on six diﬀerent kinds of high-eccentricity mechanisms, namely, direct dispersion, coplanar, Kozai-Lidov, secular chaos, E1 and E2 mechanisms. We investigated the eﬃciency of these mechanisms for the production of HJ candidates in multi-planet systems initially tightly-packed in the semimajor axis, considering a large set of numerical simulations of the exact equations of motion in the context of the N-body problem. In particular, we analyzed the sensitivity of our results to the initial number of planets, the initial semimajor axis of the innermost planetary orbit, the initial conﬁguration of planetary masses, and to the inclusion of general relativity eﬀects. We found that the E1 mechanism is the most eﬃcient in producing HJ candidates both in simulations with and without the contribution of general relativity, followed by the Kozai-Lidov and E2 mechanisms. Our results also revealed that, except for the initial equal planetary mass conﬁguration, the E1 mechanism was notably eﬃcient in the other initial planetary mass conﬁgurations considered in this work. Finally, we investigated the production of HJ candidates with prograde, retrograde, and alternating orbits.  According to our statistical analysis, the Kozai-Lidov mechanism has the highest probability of significantly exciting the orbital inclinations of the HJ candidates. 
\end{abstract}

\begin{keywords}
Planetary Systems -- Planets and Satellites: Dynamical Evolution and Stability -- Planets and Satellites:  Gaseous Planets
\end{keywords}


\section{Introduction}

Representing about 10\% of confirmed exoplanets (statistics from the \textit{NASA Exoplanet Archive} catalog), Hot Jupiters \footnote{In this paper, planets with mass (or minimum mass) in the range [0.3-13.0] $M_J$, with orbital periods or semimajor axis less than or equal to 10 days or 0.1 au, respectively.} (HJs) are part of the key evidence of the migratory behavior of giant planets, even though they have an occurrence rate in the range 0.4-1.5$\%$ \citep[][]{Marcy2005,Cumming2008,Mayor2011,Wright2012,Howard2012,Mortier2012,Fressin2013,Petigura2018,Zhou2019}. This estimates differ depending on detection method used \citep[][]{Wang2015}.  In some cases, transmission and occultation spectroscopies have provided information about their atmospheres \citep[see][and references therein]{Fortney2021}.   Furthermore, Doppler and photometric monitoring allowed the discovery of some of them with mean densities that exceed the mean density of rocky planets in the Solar System \citep[e.g.][]{Hebrard2013,Rodriguez2019}, and others with eccentric orbits ($e>0.3$) \citep[see, e.g.][]{Queloz2010,Deleuil2012,Hellier2019}, being this orbital parameter one of the evidences considered in studies on the origins of HJs \citep[see Table 1 in][]{Dawson2018}. 

Different theories have been proposed to explain the origins of HJs.  One of them is in-situ formation.  In this theory, HJ originates without a significant process of orbital migration, but likely through a core accretion mechanism of planetary formation.  Given the conditions so close to the star, in-situ formation needs appropriate conditions for its effectiveness \citep[see, e.g.][]{Batygin2016}. Therefore, the most common theories for the origins of HJs involve extensive planetary migration processes and it can occur through gravitational and dissipative interactions between the planet and the protoplanetary disk, as well as through gravitational planet-planet interactions (or planet-star companion) and dissipative interactions with the host star, in the post-disk phase.  The first case it is known as disk migration and allows the planet to attain semimajor axis smaller than  0.1 au \citep[e.g.][]{Lin1996,Heller2019}.  Planet-disk interactions also generate giant planets on low eccentricity orbits ($e \lesssim 0.1$) \citep[][]{Duffell2015}. The second case, called high-eccentricity migration, involves two phases: the eccentricity excitation phase, where the planet gets a highly eccentric orbit (usually $e \gtrsim 0.9$) due to gravitational interaction with neighboring planets or stellar companion, and the decay phase of the semimajor axis, in which the planet dissipates its orbital energy during the periastron passage due to the tidal interaction with the host star \citep[][]{Ivanov2004}.  There are several dynamical mechanisms that have the ability to excite the planet’s eccentricity and, therefore, reduce its orbital angular momentum.  These include planet-planet scattering, planet-planet and planet-star companion Kozai-Lidov, secular chaos, coplanar, E1 and E2 mechanisms \citep[see][]{Rasio1996,Marzari2002,Wu2003,Fabrycky2007,Chatterjee2008,Nagasawa2008,Wu2011,Naoz2011,Naoz2012,Beauge2012,Li2014,Petrovich2015a,Petrovich2015b,Xue2017,Wang2017,Carrera2019,Teyssandier2019,Vick2019}.  Each of these high-eccentricity mechanisms has its specific characteristics and can be classified as secular and non-secular mechanisms.

In non-secular mechanisms, planet-planet scattering is the mechanism commonly invoked to explain such eccentricity excitation from a series of multiple close encounters between planets that can lead to collisions, impacts with the host star or ejection of some of them \citep[e.g.][]{Weidenschilling1996,Rasio1996,Marzari2002,Ford2008,Chatterjee2008,Juric2008,Nagasawa2008,Raymond2009,Nagasawa2011,Beauge2012,Petrovich2014,Zanardi2017,Carrera2019,Marzari2019,Anderson2020}.  Planet-planet scattering has the ability to modify the semimajor axes of planets but with a high probability of not reaching the characteristic value of  HJs ($a \lesssim 0.1$ au) without first entering the stellar tidal dissipation phase.  In this dynamic mechanism, the typical timescales for the planet to decouple from its perturber are on the order of magnitude of thousands of years \citep[][]{Beauge2012}.  On the other hand, when the dynamics is secular, the exchange of angular momentum between the widely separated planets is slow and therefore decoupling occurs on typical time scales of the order of magnitude of millions of years \citep[][]{Naoz2011,Fabrycky2007,Petrovich2015b,Wu2011}.  This exchange can be periodic or chaotic.  In the first case, we have the planet-planet and planet-star companion Kozai-Lidov \citep[e.g.][]{Naoz2011,Naoz2012,Vick2019}, coplanar \citep[][]{Petrovich2015b,Xue2017}, E1 and E2 mechanisms \citep[][]{Wang2017}.  The second case includes the secular chaos mechanism \citep[][]{Wu2011, Teyssandier2019}.  In secular mechanisms the orbital energies of the planets remain constant as long as dissipative interactions are not taken into account.  Within these high-eccentricity mechanisms, the least known in the literature are the E1 and E2 mechanisms. Then, it is important to mention that the E1 and E2 mechanisms were reported by \citet{Wang2017} as a result of studying the efficiencies of the other high-eccentricity mechanisms in HJs production.  In E1 mechanism, the phase diagram $e$ vs $\Delta \varpi$, where $\Delta \varpi$ is given by the difference between the longitudes of periastron $\varpi_{HJ} - \varpi_{perturber}$, presents the typical characteristics of the coplanar mechanism while the argument of periastron $\omega$ of the HJ circulates, but its inclination can be higher than $30^{\circ}$ and even higher than $90^{\circ}$ (retrograde orbits).  According to \citet{Petrovich2015b}, the inclinations of HJs produced through the coplanar mechanism are mostly $\leq 30^{\circ}$. In the case of the E2 mechanism, the phase diagrams $e$ vs $\Delta \varpi$ and $e$ vs $\omega$ of the HJ do not show the characteristics of the coplanar and Kozai-Lidov mechanisms, respectively.  Furthermore, the orbits of the HJs formed through the E2 mechanism can also be retrograde.

\citet{Wang2017}, motivated by the fact that previous studies on high-eccentricity migration focused only on the total efficiency of HJs formed, analyzed the efficiency of each high-eccentricity mechanism, trying to understand which mechanism is the dominant one in HJ formation.  They considered multi-planetary systems containing from 2 to 5 planets with equal masses ($1 M_J$), initially in circular and near-coplanar orbits with a host star of $1 M_\odot$ and $1 R_\odot$.  Morever, various initial semimajor axis conditions were considered, depending on the initial mutual separation between the planets. Their numerical simulations were performed using the classical version of the Mercury code \citep[][]{Chambers1999}, which does not include general relativity (GR) effects and tidal interaction with the central star.  They studied how the initial number of planets, the spatial separation between them, and the location of the inner planet influence the efficiencies of high-eccentricity mechanisms. As a result, they found that the Kozai-Lidov mechanism plays the most important role in HJ production. The restriction made in the previous study, regarding not including the contribution of GR and realistic initial mass configurations (unequal masses), makes us wonder about the implications of these contributions on the efficiency of each high-eccentricity mechanism in the activation of the HJ production process.  The main effect of GR is to cause the apsidal precession of planetary orbit,  where the rate of precession is faster for planets close to the star and with eccentric orbits due to the term ${a}^{5/2}$($1-e^2$) in the denominator, being $a$ and $e$ the planet's semimajor axis and eccentricity, respectively \citep[][]{Einstein1916,Misner1973}.  Studies related to the influence of GR on the dynamic evolution of systems with more than one massive planet have been carried out, but only considering some particular systems \citep[e.g.][]{Adams2006,Migaszewski2009,Veras2010,Zhang2013,Marzari2019,Marzari2020}.  Therefore, the aim of our study is to obtain a broader and more detailed description of the efficiency of each high- eccentricity mechanism in the activation of the HJ production process, including the contribution of GR and different initial planetary mass configurations. The analysis is made by solving the numerical simulation of the exact equations of motion, in the context of general N-body problem. Several initial conditions are considered, changing the initial mass and number of planets, the semimajor axis of the inner planet and the location of the other planets in the system.

This paper is organized as follows.  Section~\ref{sec:methodology} presents the description of the methodology including numerical model and initial conditions, selection criteria as HJ candidate and procedure to identify the high-eccentricity mechanism.  Section~\ref{sec:results} contains our statistical results about efficiencies of high-eccentricity mechanisms in the production of HJ candidates, its dependence with respect the initial planetary mass configurations, and the production of HJ candidates with prograde, retrograde and alternating orbits.  The discussions and conclusions can be found in Section~\ref{sec:conclusions_discussions}.

\section{Methodology}
\label{sec:methodology}

\subsection{Numerical model and initial conditions}
\label{sec:conditions} 

Our numerical model included multi-planet systems composed initially of Cold Jupiters (CJs) with densities equal  to the Jupiter's mean density ($1.33$\,g\, cm$^{-3}$), orbiting a star of $1 M_\odot$ and $1 R_\odot$.  Between three and five CJs were initially located in tightly-packed configurations in order to induce dynamic instability events during their evolutions, which can lead to planet-planet collisions, impacts with the host star, or ejections from the system. To do this, we carried out an extensive set of N-body simulations performed with the Bulirsch-Stoer (BS) algorithm \citep[][]{Press1992} that has high accuracy using a modified version of the Mercury code, which allows to select the inclusion of GR effects.  Those effects were modeled by a perturbation (relativistic perturbation) to the Newtonian acceleration.  Specifically, the expression for the relativistic perturbation incorporated in the Mercury code is given by:
\begin{equation} 
   \Delta \ddot{\vec{r}}\ = \frac{\mu}{r^{3}c^{2}} \left[ \left( \frac{4\mu}{r}-{\vec{v}}^{2}\right)\vec{r}+4\left(\vec{v}\cdot\vec{r} \right)\vec{v} \right]  \ ,
   \label{eq:perturbation}
\end{equation}
where $\mu = \textit{G}M_{*}$ with $M_{*}=1 M_\odot$, $c$ is the speed of light in  vacuum,  and $\vec{r}$, $\vec{v}$ are  the astrocentric position and velocity vectors, respectively.  This expression is valid when considering the relativistic effects generated only by a spherically symmetric and non rotating star \citep[][]{Anderson1975,Shahid1994}.  The Mercury code evolves the orbits of planets allowing collisions between them. Such collisions were treated as perfect mergers, conserving the mass of the interacting bodies. In our N-body simulations, one planet was assumed to be ejected from the system if it reached a distance from the central star greater than 1000 au.  In the present study, we carried out 96000 N-body simulations. Of this total, 48000 runs corresponded to numerical experiments with GR effects.  For comparison purposes, the numerical experiments of the above set were also carried out without GR effects.  This corresponds to the other 48000 runs. These two groups of numerical experiments will be referred to as GR and No-GR simulations in the present manuscript. Finally, each numerical simulation was integrated for 50 Myr.

As initial conditions, we considered multi-planet systems with three, four and five CJs, where the innermost CJ semimajor axis was 1.0 au and 5.0 au.  The initial semimajor axis of the others CJs was defined in units of mutual Hill radius ($R_{H,mut}$),  so that
\begin{equation}
    a_{i+1} - a_i = K R_{H,mut} \ , 
    \label{eq:separation}
\end{equation} 
where 
\begin{equation}
    R_{H,mut} = \left(\frac{m_i + m_{i+1}}{3M_*}\right)^{1/3} \frac{(a_i+a_{i+1})}{2} \ , \label{eq:hill}
\end{equation}
being $a_i$ and $m_i$ the semimajor axis and the mass of the \textit{i}th CJ, respectively,  $M_*$ the mass of the host star, and \textit{K} a dimensionless number.  Combining ~(\ref{eq:separation}) and ~(\ref{eq:hill}):
\begin{equation}
    a_{i+1} = a_i \left[\frac{1 + \frac{K}{2} \left(\frac{m_i + m_{i+1}}{3M_*}\right)^{1/3}}{1 - \frac{K}{2}\left(\frac{m_i + m_{i+1}}{3M_*}\right)^{1/3}}\right] \ .
    \label{eq:recurrence}
\end{equation}
Equation~(\ref{eq:recurrence}) is a recurrence relation that allowed us the calculation of the initial semimajor axis of the  others CJs in the system.  This relation is commonly used in studies concerning the early dynamic stability of multi-planetary systems \citep[see, e.g.][]{Chambers1996,Zhou2007,Pu2015} with the stability limit being measured in the number \textit{K}.  Even though the stability limit depends on many factors, such as the number of planets, the mass of each one, their eccentricities and mutual inclinations \citep[][]{Pu2015}, we maintain the separation between adjacent CJs in the range established by \citet{Wang2017}, that is, $2 \leq{\textit{K}} \leq{6}$, but with an interval $ \Delta \textit{K} = 0.002$ and not $ \Delta \textit{K} = 0.001$. Such \textit{K} values were selected in order that the simulated systems become unstable quickly due to computational reasons. In fact, we have considered initial configurations of CJs with different masses in systems that dynamically evolved over 50 Myr. We remark that the range of values of \textit{K} parameter adopted in the present investigation is consistent with that used in different previous works such as \citet{Chatterjee2008}, \citet{Raymond2008}, \citet{Raymond2009}, \citet{Raymond2013}, \citet{Gong2013}, \citet{Carrera2016}, \citet{Wang2017}, \citet{Carrera2019}, among others. We would like to mention that \citet{Marzari2014a} suggested that a two dimensional approach in terms of \textit{K} parameter would be more appropriated to find the transition between stable and unstable systems that may lead to dynamical instabilities. In fact, studies such as those developed by \citet{Marzari2014b}, \citet{Zanardi2017}, among others, have used two different values of \textit{K} parameter to define the initial semimajor axes of the planets of each system. The sensitivity of the results to a two dimensional approach in terms of \textit{K} parameter will be studied in a forthcoming paper.
The choice of these initial conditions produces tightly-packed systems, which is widely consistent with an initial set of massive planets that remain stable throughout the gas stage, and later undergoes dynamical instabilities once the gas disk dissipates.

In the present study, we aimed to focus on realistic mass distributions for the simulated CJs.  Thus, based on the statistics of the \textit{NASA Exoplanet Archive} catalog \footnote{1167 exoplanets with mass (or minimum mass) in the range [0.3-13.0] $M_J$.  Approximately 83\% of them have a mass (or minimum mass) between 0.3-5.0 $M_J$.  Review date 26/August/2021}, we selected four different groups of initial planetary mass configurations, namely: 1- equal mass, 2- random mass, 3- increasing mass, and 4- decreasing mass. The range of planetary masses associated with each of such configurations is described in Table~\ref{tab:configurations}. The mass assigned to each planet in each configuration was randomly selected from an uniform distribution in such a range.

\begin{table*}
	\centering
	\caption{Distribution of initial masses of the simulated CJs according to the type of initial planetary mass configuration.  The initial planetary configuration of random mass does not include situations of increasing, decreasing and equal mass.}
	\label{tab:configurations}
	\begin{tabular}{cccc} 
		\hline
		 Init. planet. & $m_i (M_J)$ & $m_i (M_J)$ & $m_i (M_J)$\\
		 mass config. & ($i$ to 3) & ($i$ to 4) & ($i$ to 5)\\
		\hline
		Equal & $m_1$ = $m_2$ = $m_3$ & $m_1$ = ... = $m_4$ & $m_1$ = ... = $m_5$\\ & $1.0$ & $1.0$ & $1.0$\\ \\
		Random & $m_1$ , $m_2$ , $m_3$ & $m_1$,...,$m_4$ & $m_1$,...,$m_5$\\ & $[0.30-3.12)$ & $[0.30-4.06)$ & $[0.30-5.00)$\\ \\
		Increasing & $m_1 = [0.30-1.24)$ & $m_1 = [0.30-1.24)$ & $m_1 = [0.30-1.24)$\\ & $m_2 = [1.24-2.18)$ & $m_2 = [1.24-2.18)$ & $m_2 = [1.24-2.18)$ \\  & $m_3 = [2.18-3.12)$ & $m_3 = [2.18-3.12)$ & $m_3 = [2.18-3.12)$ \\ & & $m_4 = [3.12-4.06)$ & $m_4 = [3.12-4.06)$ \\ & & & $m_5 = [4.06-5.00)$ \\ \\
		Decreasing & $m_1 = [2.18-3.12)$ & $m_1 = [3.12-4.06)$ & $m_1 = [4.06-5.00)$\\ & $m_2 = [1.24-2.18) $ & $m_2 = [2.18-3.12)$ & $m_2 = [3.12-4.06)$ \\ & $m_3 = [0.30-1.24)$ & $m_3 = [1.24-2.18)$ & $m_3 = [2.18-3.12)$ \\ & & $m_4 = [0.30-1.24)$ & $m_4 = [1.24-2.18)$ \\ & & & $m_5 = [0.30-1.24)$  \\
		\hline
	\end{tabular}
\end{table*}

For each of the four groups of planetary mass configurations, the initial orbits of CJs were assumed to be circular with inclinations $i$ randomly selected in the range $(0^{\circ},2^{\circ}]$ following an uniform distribution. The initial values of the argument of periastron $\omega$, the longitude of the ascending node $\Omega$, and the mean anomaly $M$ of CJs were randomly chosen between $0^{\circ}$ and $360^{\circ}$, also following an uniform distribution. We remark that the 50 Myr of dynamic evolution considered in the integrations allowed us to clearly distinguish which was the high-eccentricity mechanism that led to the formation of each HJ candidate.  In many cases, the periodic behavior in the variation of the HJ candidate's periastron is not fully defined in the first 10 Myr of dynamic evolution, which was one of the integration time adopted by \citet{Wang2017}.  In addition, and based on several tests, a significant number of systems continued to experience dynamic instabilities after 10 Myr, but remained stable after 50 Myr of evolution.

\subsection{Selection criteria as hot Jupiter candidate}
\label{sec:criteria} 

In the present study, we focused our analysis on the efficiency of each high-eccentricity mechanism in the production of HJ candidates during gravitational planet-planet interactions. According to the idea proposed by \citet{Wang2017}, we did not include tidal effects in our N-body simulations. In fact, we considered that the inclusion of tides may cover up the characteristics of the dynamical mechanisms that the systems have experienced. According to this and following the criteria from \citet{Wang2017}, we just set a boundary of tidal effects in our study. Hence, a given giant planet will be considered a HJ candidate if, during its orbital evolution, it reaches pericentric distances ($q$) small enough to stellar tidal dissipation operate efficiently on a time scale ($\Delta t$) approximately equal to the orbital decay time scale of the dynamic tide ($ \tau_a$).  In quantitative terms, a  HJ candidate is produced when $q\lesssim0.05$ au \citep[][]{Rasio1996,Marzari2002} and $\Delta t(q \lesssim 0.05$ au) $\approx \tau_a$, where $\tau_a$ is given by:
\begin{equation} 
    \tau_a\ = \frac{GmM_{*}}{2a} \frac{P_{orb}}{\left(-\Delta E_{tide} \right)} \ ,
    \label{eq:tau}
\end{equation}
where $P_{orb}$ is the orbital period and $\Delta E_{tide}$ simplified to
\begin{equation}
    \Delta E_{tide}\  \sim - \frac{16\sqrt{2}}{15} {\widetilde{w_0}}^{3}{\widetilde{Q}}^{2} \xi \exp\left(- \frac{4\sqrt{2}}{3}\widetilde{w_0}\xi \right)\frac{Gm^{2}}{R}\ , 
    \label{eq:e_tide}
\end{equation}
with $\widetilde{w_0}\simeq 0.53(R/R_{J})+0.68$ and $\widetilde{Q}\simeq -0.12(R/R_{J})+0.68$ for Jovian mass planet, where $R_{J}$ is Jovian radius, and $\xi = (mq^{3})^{1/2}(M_{*}R^{3})^{-1/2}$.  The value $\widetilde{w_0}$ is a dimensionless frequency of fundamental mode, $\widetilde{Q}$ is a dimensionless overlap integral that depends on the planetary interior model, $m$ and $R$ are the mass and radius of the planet, respectively \citep[][]{Nagasawa2008}.

\subsection{Procedure to identify the high-eccentricity mechanism}
\label{sec:identify_mechanism} 

Based on the initial conditions established in this work, it is expected that most multi-planet systems simulated will have an initial phase of orbital instabilities dominated by close encounters between the planets, where one or more of them will be ejected, suffer collisions or impact the host star.  Therefore, the initial configuration of most systems will be dramatically altered and the surviving planets will generally undergo secular interactions, possibly obtaining high eccentricities and inclinations.  We remark that the inclinations of the planets of the resulting systems were defined with respect to the invariant plane of the system.  Some of the surviving planets of our simulations met the selection criteria as HJ candidate described in the Section~\ref{sec:criteria} in the first 50 Myr of evolution due to the activation of a high-eccentricity mechanism.  According to that mentioned in previous works, we identify different high-eccentricity mechanisms, which are defined by the following criteria:

\begin{enumerate}
    \item If only one planet survives in the planetary system and reaches a pericentric distance $q\lesssim0.05$ au, then the mechanism is identified as direct dispersion.
    \\
    \item For planetary systems with more than one surviving planet, if  the high-eccentricity of the HJ candidate undergoes very low amplitude oscillations such that the evolution of its pericentric distance will always be in the interval $q\lesssim0.05$ au after the dynamic instability event occurs, the mechanism is also identified as direct dispersion.
    \\
    \item If during the dynamic evolution of the planets there are multiple close encounters, leading some of them to reach a pericentric distance $q\lesssim0.05$ au over a time scale $\Delta t \approx \tau_a$, the direct dispersion is the responsible mechanism.
    \\
    \item  In situations with more than one surviving planet, if the phase diagram of $e$ vs $\Delta \varpi$ of the HJ candidate has the characteristics of libration of $\Delta \varpi$ around $0^{\circ}$ or $180^{\circ}$ (case 1) as well as situations of circulation of $\Delta \varpi$ with a maximum (or minimum) value of $e$ when $\Delta \varpi$ is around $0^{\circ}$ or $180^{\circ}$ (case 2), but having the mutual inclination  $i_{tot} \lesssim 30^{\circ}$ for a time $\Delta t(q \lesssim 0.05$ au) $\approx \tau_a$, the mechanism is identified as coplanar.
    \\
    \item If the planet meets the selection criteria as HJ candidate in scenarios where there is a libration of its argument of periastron $\omega$, the Kozai-Lidov is the responsible mechanism.
    \\
    \item If the semimajor axis of the surviving planets remain constant, the evolution of their eccentricities and inclinations are non-periodic and the phase diagrams $e$ vs $\omega$ and $e$ vs $\Delta \varpi$ of the planet that meets the criteria $\Delta t(q \lesssim 0.05$ au) $\approx \tau_a$ does not have the characteristics of Kozai-Lidov and coplanar, respectively, the mechanism is identified as secular chaos.
    \\
    \item For planetary systems with more than one surviving planet, if the phase diagram of $e$ vs $\Delta \varpi$ of the HJ candidate  has typical characteristics of the coplanar mechanism (cases 1 and 2), its argument of periastron $\omega$ circulates and the evolution of its inclination includes values higher than $30^{\circ}$ reaching the maximum every time its periastron distance reaches the minimum value, the mechanism is identified as E1.
    \\
    \item The other scenarios with HJ candidates not classified within the situations mentioned above are identified as E2 mechanism.
\end{enumerate}

Fig.~\ref{fig:scattering} to ~\ref{fig:E2_1} show examples of HJ candidates produced by each of the high-eccentricity mechanisms described above. It should be noted that our procedure has some similarity with the one proposed by \citet{Wang2017} but with the difference that we do not include HJ candidates selected from the check of their orbital parameters obtained at the end of the integration.  In the next section, we show the results.

\section{Results}
\label{sec:results}

\subsection{Efficiencies of high-eccentricity mechanisms in the production of hot Jupiter candidates}
\label{sec:efficiencies} 

As we have mentioned in previous section, a total of 96000 N-body simulations were carried out, of which half correspond to experiments that included GR effects, while the other half only took into account Newtonian gravitational effects.  On the one hand, 34356 multi-planet systems resulting from the No-GR simulations had at least some dynamic instability during the first 50 Myr of dynamic evolution, which corresponds to 71.58\% of the total simulated systems.  On the other hand, 34372 multi-planet systems resulting from the GR simulations experienced at least some dynamic instability event during the first 50 Myr of evolution, which  corresponds to 71.61\% of the total systems.  In both groups of systems with dynamic instability events, the planets that met the selection criteria as HJ candidate in Section~\ref{sec:criteria} were classified within the different high-eccentricity mechanisms using the procedure mentioned in Section~\ref{sec:identify_mechanism}.  Tables~\ref{tab:No_GR} and ~\ref{tab:GR} show the statistics obtained from this classification for No-GR and GR simulations, respectively. These tables show the statistics with all initial planetary mass configurations together. The statistics for each initial planetary mass configuration are shown in the Section~\ref{sec:mass_efficiencies}.

\begin{table*}
	\centering
	\caption{Number and fraction (\%) of HJ candidates produced (in increasing order) through different high-eccentricity mechanisms for No-GR simulations. The percentages were calculated with respect to the total number of HJ candidates produced in this No-GR scenario.}
	\label{tab:No_GR}
	\begin{tabular}{cccccc} 
	\hline
	Coplanar & Secular Chaos & Direct Dispersion & E2 & Kozai-Lidov & E1\\
	\hline
	45 & 73 & 125 & 212 & 248 & 438\\
	3.94\% & 6.40\% & 10.95\% & 18.58\% & 21.74\% & 38.39\%\\ 
	\hline
	\end{tabular}
\end{table*}

\begin{table*}
	\centering
	\caption{Number and fraction (\%) of HJ candidates produced (in increasing order) through different high-eccentricity mechanisms for GR simulations.  The percentages were calculated with respect to the total number of HJ candidates produced in this GR scenario.}
	\label{tab:GR}
	\begin{tabular}{cccccc} 
	\hline
	Coplanar & Secular Chaos & Direct Dispersion & Kozai-Lidov & E2  & E1\\
	\hline
	64 & 68 & 139 & 161 & 228 & 333\\
	6.44\% & 6.85\% & 14.0\% & 16.21\% & 22.96\% & 33.54\%\\ 
	\hline
	\end{tabular}
\end{table*}

\begin{table*}
	\centering
	\caption{Number and fraction (\%) of HJ candidates produced through coplanar and E1 mechanisms as a function of the characteristics of the phase diagram $e$ vs $\Delta \varpi$ for No-GR and GR scenarios.The percentages were calculated with respect to the total number of HJ candidates in each scenario and high-eccentricity mechanism.}
	\label{tab:coplanar_E1}
	\begin{tabular}{cccc} 
	\hline
	 High-eccentricity & Scenario  & Libration & No-Libration \\ mechanism & & of $\Delta \varpi$ & of $\Delta \varpi$ \\
	\hline
	 Coplanar & No-GR & 33 (73.33\%) & 12 (26.67\%)\\ 
	 & GR & 20 (31.25\%) & 44 (68.75\%) \\ 
	\hline
	 E1 & No-GR & 375 (85.62\%) &  63 (14.38\%) \\ 
	 & GR & 169 (50.75\%) & 164 (49.25\%) \\ 
	\hline
	\end{tabular}
\end{table*}

In total, 68728 systems suffered at least some dynamic instability event.  Of these, we obtained 2134 HJ candidates according to the implemented selection criteria. This means that in our study the efficiency in the production of HJs through high-eccentricity migration was 3.10\%.  This efficiency is within the order of magnitude of other theoretical studies already carried out.  In fact, \citet{Beauge2012} and \citet{Petrovich2015b} found efficiencies of approximately 3-5\% studying the planet-planet scattering and coplanar mechanism, respectively.  \citet{Munoz2016} derived an efficiency of HJ production of approximately 1-3\% through  the Kozai-Lidov mechanism due to a stellar companion.  \citet{Petrovich2016} determined an efficiency of approximately 5\% with the Kozai-Lidov mechanism due to a planetary companion.  In the case of the secular chaos mechanism, the efficiency of HJ production obtained by \citet{Teyssandier2019}  was of 4.5\%.  Finally, \citet{Wang2017} studied the high-eccentricity mechanisms previously mentioned without GR effects and found that 3.42\% of the simulated systems produced a HJ candidate.

In our study, 1141 HJ candidates were produced in No-GR scenarios and 993 in GR simulations.  In other words, the efficiencies of production of HJ candidates were 3.32\% and 2.89\% with respect to the total number of systems with dynamic instability events in No-GR and GR simulations, respectively.  This reduction in the number of HJ candidates in the GR scenario may be explained by the fact that GR contributes to reducing the periods and amplitudes of the eccentricity oscillations of the planets in secular evolution as \citet{Marzari2020} recently showed.

As shown in Table~\ref{tab:No_GR}, our results indicate that the number of HJ candidates produced in the No-GR scenario through the direct dispersion, coplanar, Kozai-Lidov, secular chaos, E1 and E2 mechanisms were 125, 45, 248, 73, 438 and 212, respectively.  Thus, on the one hand, the E1 mechanism was the most efficient  with 38.39\% of the total cases in this scenario, followed by the Kozai-Lidov and E2 mechanisms with 21.74\% and 18.58\%, respectively.  On the other hand, the least efficient mechanisms were the secular chaos with 6.40\% and coplanar with 3.94\%.  The inclusion of GR showed similarities and differences respect to that obtained in No-GR simulations.  Such as Table~\ref{tab:GR} indicates, the E1 mechanism remained the most efficient in the GR simulations with 33.54\%.  However, this percentage was less than that derived in the No-GR scenario.  Moreover, coplanar and secular chaos continued to be the least efficient mechanisms in the HJ candidate production in GR simulations, showing 6.44\% and 6.85\%, respectively.  In particular, GR increased the HJ candidate production efficiency for the coplanar mechanism (6.44\%) in comparison with that obtained without GR (3.94\%). In addition, E2 and Kozai-Lidov mechanisms maintained high percentages associated with the production of HJ candidates with the inclusion of GR, with efficiencies of 22.96\% and 16.21\%, respectively.  If we compare the previous efficiency of the Kozai-Lidov mechanism with the result in Table~\ref{tab:No_GR}, we can see that GR has the effect of inhibiting the libration of the argument of periastron $\omega$ of the HJ candidate.  Thus, the efficiency of production of HJ candidates for the Kozai-Lidov mechanism in GR simulations decreased respect to that obtained in No-GR scenario, while coplanar and E2 mechanisms increased their percentages associated with the HJ candidate production.  For the direct dispersion mechanism, the efficiency was 14.0\% with GR, that is, an increase of approximately 3\% compared to the efficiency obtained without GR.  The inclusion of GR also had the effect of decreasing (increasing) the number of HJ candidates with librations (circulations) of $\Delta \varpi$ for the coplanar and E1 mechanisms, which is shown in Table~\ref{tab:coplanar_E1}.  According to this, 49.25\% of HJ candidates produced through the E1 mechanism in GR simulations had circulations of $\Delta \varpi$.  This represents a very significant increase in comparison with that obtained in the No-GR scenario, which was 14.38\%.  For the case of the coplanar mechanism, 68.75\% and 26.67\% of its HJ candidates showed circulations of $\Delta \varpi$ for GR and No-GR scenarios, respectively.

\begin{figure}
	\includegraphics[width=\columnwidth]{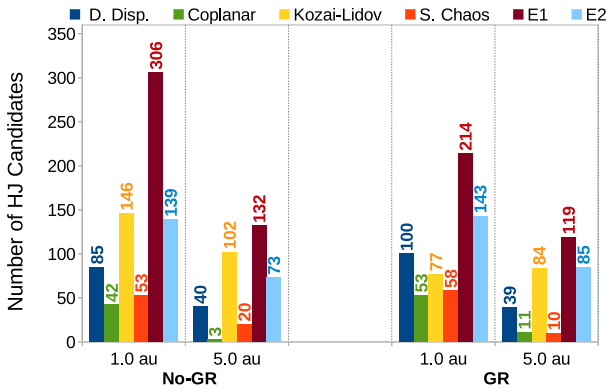}
    \caption{Statistical distribution of the number of HJ candidates produced through different high-eccentricity mechanisms as a function of the initial semimajor axis of the innermost planet for No-GR and GR scenarios.}
    \label{fig:total_inner}
\end{figure}

We explore how the efficiencies of high-eccentricity mechanisms depend on the initial semimajor axis of the innermost planet ($a_1$) and initial planet number (N) for GR and No-GR scenarios.  Fig.~\ref{fig:total_inner} shows the statistical distribution of the number of HJ candidates produced through different high-eccentricity mechanisms as a function of the initial semimajor axis of the innermost planet with and without GR effects.  In both scenarios, the results show that between 65\% and 68\%  of HJ candidates were produced with $a_1$ = 1.0 au.  According to this, the number of HJ candidates decreases with an increase in the initial value of $a_1$.  For both sets of simulations, E1 remains as the most efficient mechanism in the production of HJ candidates with an increase of $a_1$.  For No-GR simulations, the greater the semimajor axis of the innermost planet $a_1$, the smaller the efficiency of production of HJ candidates for any of the six high-eccentricity mechanisms.  For GR simulations, we observed the same negative correlation between the efficiencies of formation of HJ candidates and $a_1$, except for the Kozai-Lidov mechanism.  In fact, the number of HJ candidates produced from the Kozai-Lidov mechanism was slightly greater with an increase of $a_1$ in the GR simulations.  This result can be explained by the fact that GR inhibits the libration of $\omega$, which is a stronger effect for planets initially located closer to the host star.  

\begin{figure}
	\includegraphics[width=\columnwidth]{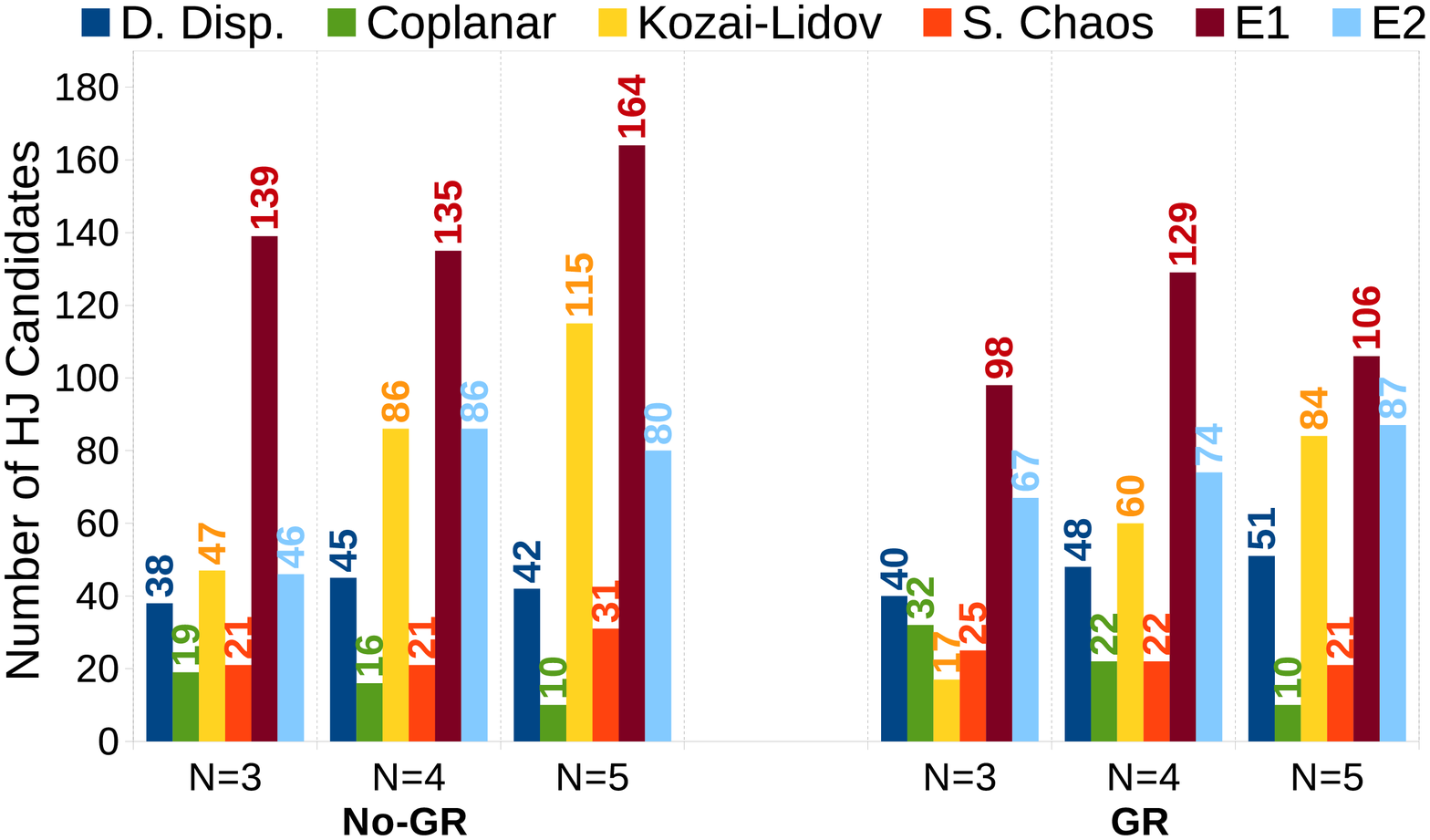}
    \caption{Statistical distribution of the number of HJ candidates produced through different high-eccentricity mechanisms as a function of the initial planet number (N) for No-GR and GR scenarios.}
    \label{fig:total_number}
\end{figure}

On the other hand, Fig.~\ref{fig:total_number} shows the statistical distribution of the number of HJ candidates produced through different high-eccentricity mechanisms as a function of the initial planet number N for each scenario.  Our results indicate that the number of HJ candidates increases with an increase in the initial number of planets both in simulations with and without GR effects.  In No-GR simulations,  the number of HJ candidates was 309, 389 and 442 for N=3, N=4 and N=5, respectively, while in the GR scenario, the number of HJ candidates was 279 for N=3, 355 for N=4 and 359 for N=5.  As the reader can see, the positive correlation between the HJ candidate production and the initial number of planets in the GR scenario when N changed from 4 to 5 was weaker than that derived for the No-GR simulations.  However, such a positive correlation was strong when N changed from 3 to 4 in both scenarios.  This behavior can be explained if we take into account the dependence of the initial space separations between adjacent planets with the initial planetary mass configurations.  A more detailed analysis of this dependency is carried out in Section~\ref{sec:mass_efficiencies}.  The statistical distribution in Fig.~\ref{fig:total_number} also shows that the number of HJ candidates produced by each high-eccentricity mechanism do not necessarily increases with an increase of N.  For example, the E1 mechanism produces more HJ candidates when N=4 in GR and for the coplanar mechanism a negative correlation has occurred in each scenario.  Only for the Kozai-Lidov mechanism does its efficiency increases with increasing N in both scenarios. 

\begin{figure}
	\begin{tabular}{cc}
	\includegraphics[width=\columnwidth]{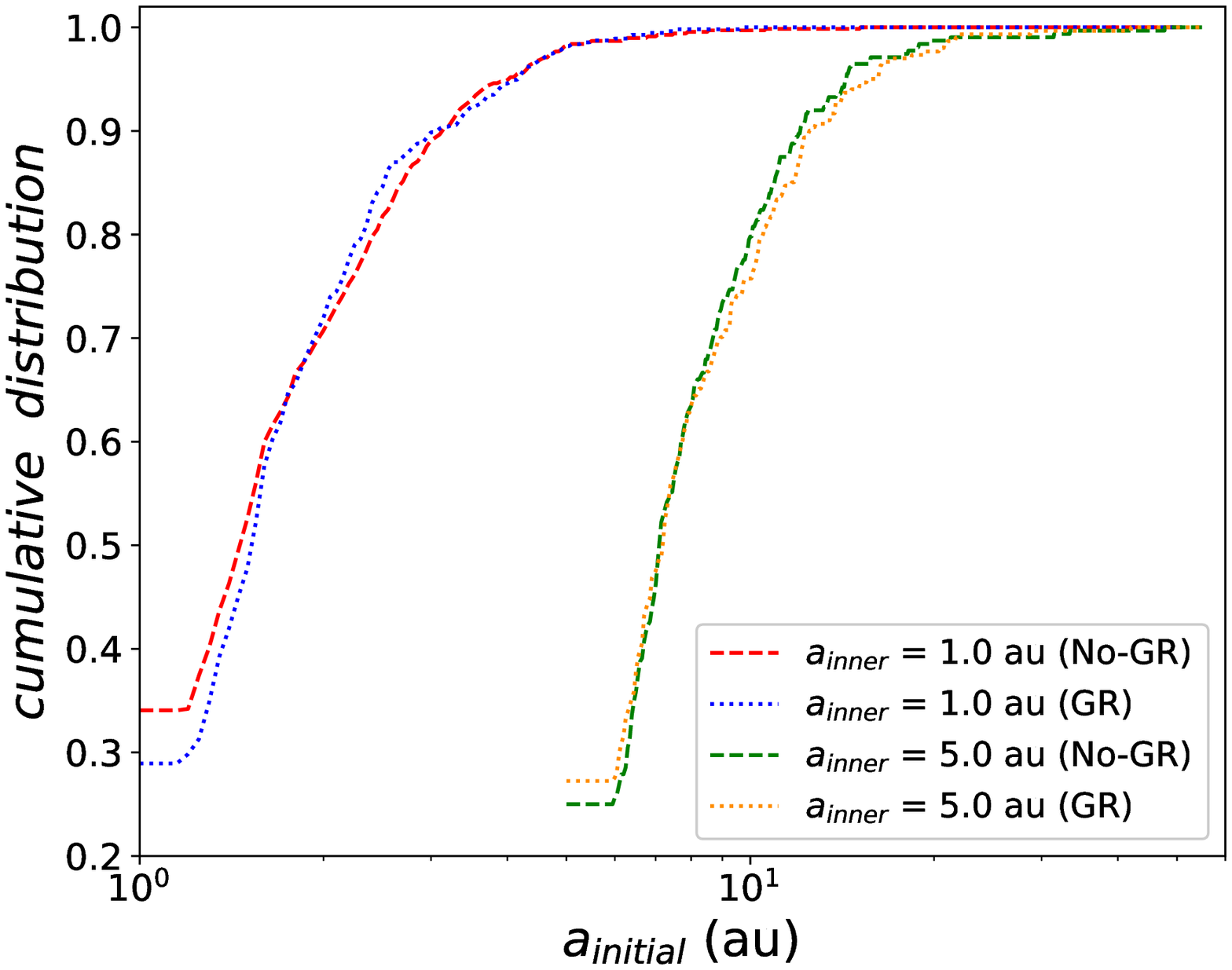} 
	\end{tabular}
    \caption{Cumulative distribution of initial semimajor axis for the HJ candidates. Different colours refer to the initial semimajor axis of the innermost planet for No-GR and GR scenarios.}
    \label{fig:a_initial}
\end{figure}

Finally, we describe the population of HJ candidates as a function of their initial semimajor axis $a_{\textrm{initial}}$. We carry out such investigation for the two initial values of semimajor axis of the innermost planet $a_{\textrm{inner}}$ assumed in the present study in both No-GR and GR scenarios. From this, Fig.~\ref{fig:a_initial} illustrates the cumulative distribution of HJ candidates as a function of their initial semimajor axis $a_{\textrm{initial}}$. Our results show that the fraction of HJ candidates produced from the innermost planet of the work systems for $a_{\textrm{inner}}$ = 1 au is 0.35 and 0.29 in No-GR and GR scenarios, respectively. These fractions slightly decrease for $a_{\textrm{inner}}$ = 5 au. Moreover, for any value of $a_{\textrm{inner}}$, the cumulative distributions do not show significant differences in No-GR and GR scenarios. It is interesting to remark that a low fraction ($\sim 0.1$) of HJ candidates come from regions with $a_{\textrm{initial}} > 3.0$ au and $a_{\textrm{initial}} > 12.0$ au for $a_{\textrm{inner}} = 1.0$ au and $a_{\textrm{inner}} = 5.0$ au, respectively.

\subsection{Efficiencies of high-eccentricity mechanisms in the production of hot Jupiter candidates as a function of the initial planetary mass configuration}
\label{sec:mass_efficiencies} 

We now examine how the efficiencies of high-eccentricity mechanisms in the production of HJ candidates depend on the initial planetary mass configuration.  Table~\ref{tab:mass_efficiencies} summarizes the results derived from our investigation.

\begin{table*}
	\centering
	\caption{Number and fraction (\%) of HJ candidates produced through different high-eccentricity mechanisms in No-GR and GR scenarios for different initial planetary mass configurations.  The percentages were calculated with respect to the total number of HJ candidates produced in each scenario and initial planetary mass configuration.}
	\label{tab:mass_efficiencies}
	\begin{tabular}{ccccccccc} 
	\hline
	 Init. planet. & Scenario & Coplanar & Secular Chaos & Direct Dispersion & Kozai-Lidov & E2 & E1 & Total HJ \\
	 mass config. & & & & & & & & candidates \\
	\hline
	 Equal & No-GR & 11 & 33 & 65 & 137 & 104 & 136 & 486 \\  &  & (2.26\%) & (6.79\%) & (13.37\%) & (28.19\%) & (21.40\%) & (27.99\%) &  \\  & GR & 18 & 25 & 84 & 81 & 127 & 85 & 420 \\  &  & (4.29\%) & (5.95\%) & (20.0\%) & (19.29\%) & (30.24\%) & (20.23\%) &  \\ \\
	 Random & No-GR & 11 & 14 & 15 & 45 & 46 & 103 & 234 \\  &  & (4.70\%) & (5.98\%) & (6.41\%) & (19.23\%) & (19.66\%) & (44.02\%) &  \\  & GR & 18 & 15 & 15 & 34 & 44 & 82 & 208 \\  &  & (8.65\%) & (7.21\%) & (7.21\%) & (16.35\%) & (21.15\%) & (39.43\%) &  \\ \\ Increasing & No-GR &  8 & 17 & 28 & 42 & 40 & 110 & 245 \\  &  & (3.26\%) & (6.94\%) & (11.43\%) & (17.14\%) & (16.33\%) & (44.90\%) &  \\  & GR & 15 & 13 & 20 & 26 & 38 & 85 & 197 \\  &  & (7.61\%) & (6.60\%) & (10.15\%) & (13.20\%) & (19.29\%) & (43.15\%) &  \\ \\ 
	 Decreasing & No-GR &  15 & 9 & 17 & 24 & 22 & 89 & 176 \\  &  & (8.52\%) & (5.11\%) & (9.66\%) & (13.64\%) & (12.50\%) & (50.57\%) &  \\  & GR & 13 & 15 & 20 & 20 & 19 & 81 & 168 \\  &  & (7.74\%) & (8.93\%) & (11.90\%) & (11.90\%) & (11.31\%) & (48.22\%) &  \\ 
	\hline
	\end{tabular}
\end{table*}

\begin{table*}
	\centering
	\caption{Number of HJ candidates produced through coplanar and E1 mechanisms as a function of the characteristics of the phase diagram $e$ vs $\Delta \varpi$ for No-GR and GR (in brackets) scenarios for different initial planetary mass configurations.}
	\label{tab:coplanar_E1_mass_config}
	\begin{tabular}{cccc} 
	\hline
	High-eccentricity & Init. planet.  & Libration & No-Libration \\ mechanism & mass config. & of $\Delta \varpi$ & of $\Delta \varpi$ \\
	\hline
	 & Equal & 8 (6) & 3 (12)\\ 
	 Coplanar & Random & 7 (5) & 4 (13) \\ 
	 & Increasing & 8 (6) & 0 (9)\\ 
	 & Decreasing & 10 (3) & 5 (10)\\ 
	\hline
	 & Equal & 122 (38) & 14 (47)\\ 
	 E1 & Random & 90 (37) & 13 (45) \\ 
	 & Increasing & 96 (50) & 14 (35)\\ 
	 & Decreasing & 67 (44) & 22 (37)\\ 
	\hline
	\end{tabular}
\end{table*}

Table~\ref{tab:mass_efficiencies} reveals that 42.46\% of the total HJ candidates produced in No-GR and GR simulations were derived from the initial equal mass configuration, which was approximately twice of the individual efficiency obtained with the initial random and increasing mass configurations.  By contrast, the initial decreasing mass configuration had the lowest efficiency with 16.12\% of the total HJ candidates.  Based on the recurrence relation established in Section~\ref{sec:conditions}, the high efficiency obtained with the initial equal mass configuration can be accounted to the fact that it generated initially more tightly-packed multi-planet systems, which significantly decreased the number of systems without dynamic instability events. Table~\ref{tab:mass_efficiencies} also shows that the E1 mechanism was notably efficient in the HJ candidate production in the initial random, increasing, and decreasing mass configurations in both scenarios No-GR and GR.  In fact, on the one hand, the efficiency of formation of HJ candidates by E1 mechanism derived from such initial planetary mass configurations ranged from 39\% and 51\%, where the highest percentages are associated with the initial decreasing mass configuration. On the other hand, the efficiencies of the other high-eccentricity mechanisms resulting from the initial random, increasing, and decreasing mass configurations did not reach 22\%.  In the initial equal mass configuration, the efficiency of HJ candidate production from the E1 mechanism (27.99\%) was slightly surpassed by the Kozai-Lidov mechanism (28.19\%) in the No-GR scenario, while, in the GR simulations, the efficiency of E2 mechanism (30.24\%) was more significant than that derived from the E1 mechanism (20.23\%).

Table~\ref{tab:coplanar_E1_mass_config} shows the number of HJ candidates obtained from the coplanar and E1 mechanisms that experienced librations and circulations of $\Delta \varpi$ in the different initial planetary mass configurations in the No-GR and GR simulations. On the one hand, in the No-GR (GR) scenario, the number of HJ candidates derived from the coplanar mechanism with librations of $\Delta \varpi$ was greater (less) than that associated with circulations of such an angle for any initial planetary mass configuration. On the other hand, in No-GR simulations, the number of HJ candidates resulting from the E1 mechanism with librations of $\Delta \varpi$ was significantly greater than the number of HJ candidates with circulations of such an angle regardless of the initial planetary mass configuration. The situation is something different when the E1 mechanism is analyzed considering GR effects. In fact, the efficiency of HJ candidate production in the GR scenario from the E1 mechanism with librations of $\Delta \varpi$ is less (greater) than that related to circulations of $\Delta \varpi$ in the equal and random (increasing and decreasing) initial planetary mass configurations. 

\begin{figure*}
	\begin{tabular}{cc}
	\includegraphics[width=\columnwidth]{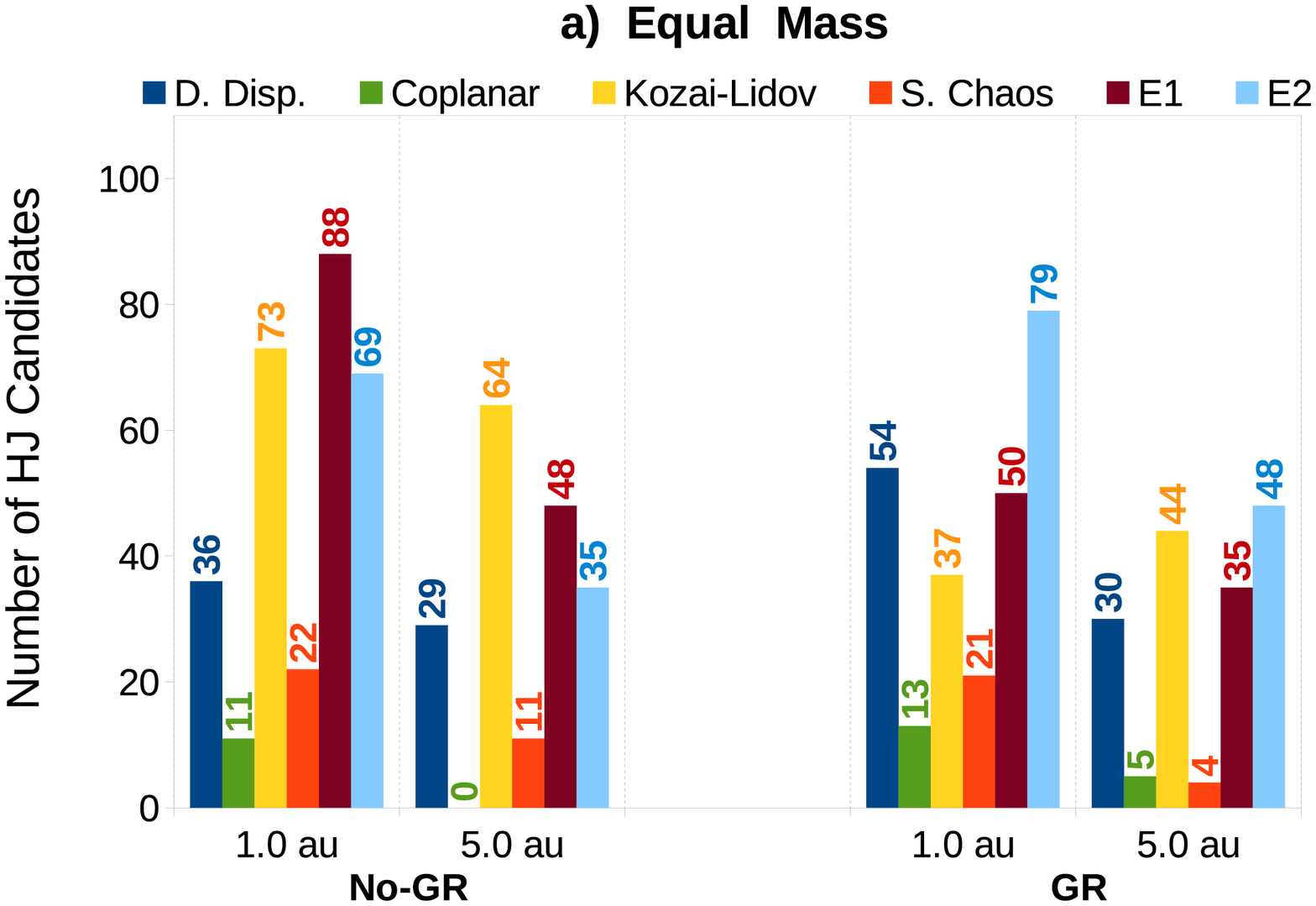} &
	\includegraphics[width=\columnwidth]{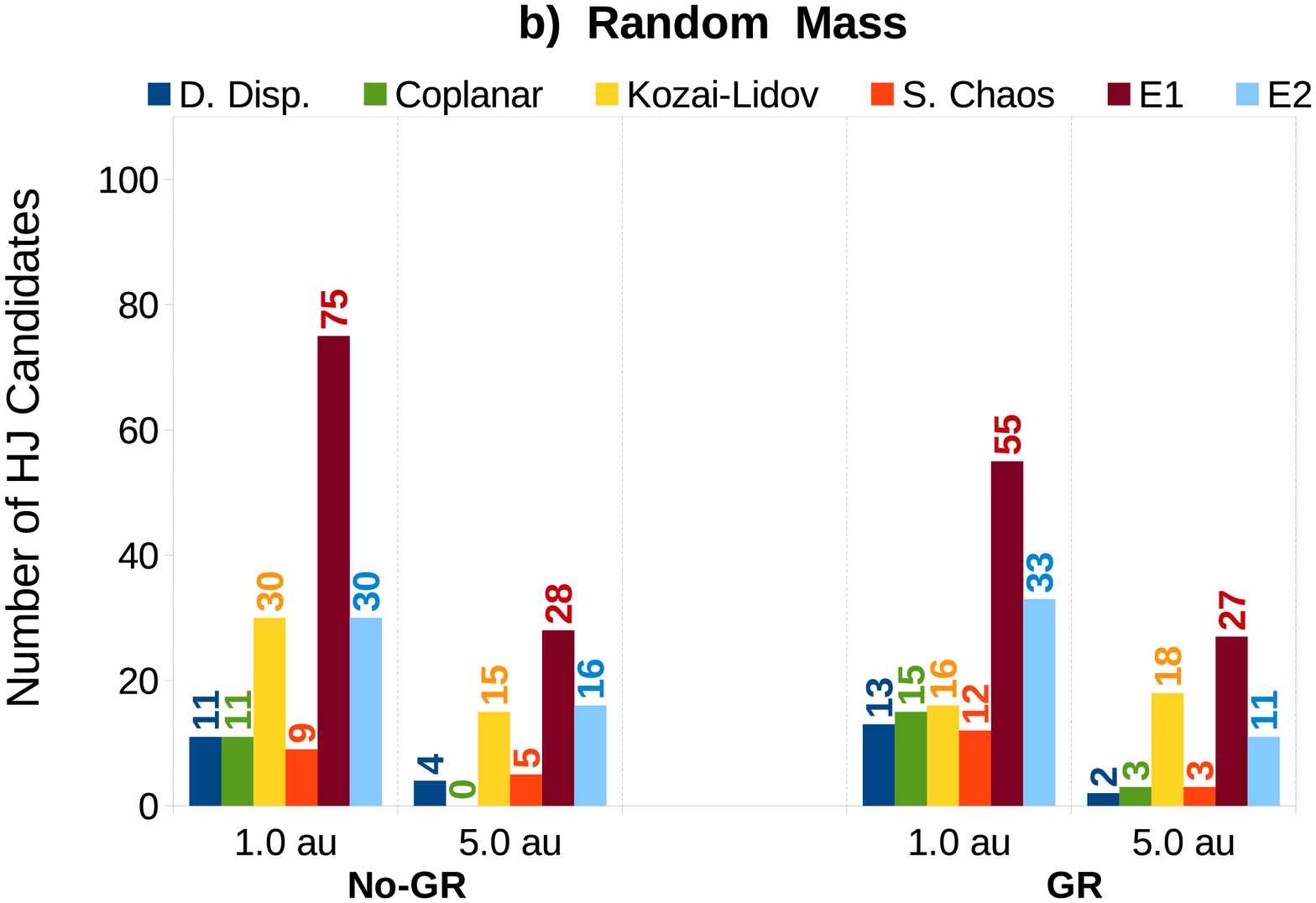}\\
	\includegraphics[width=\columnwidth]{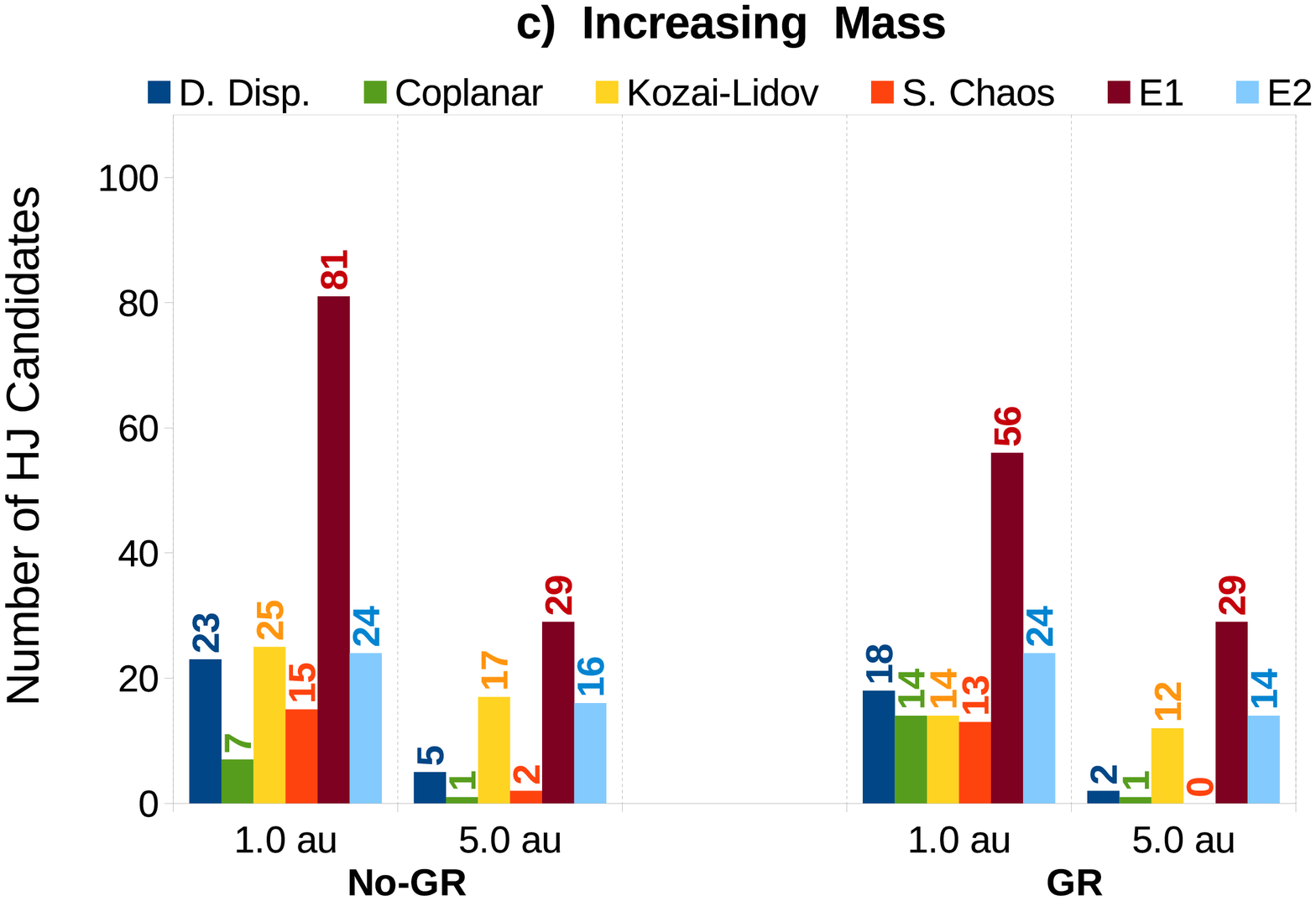}&
	\includegraphics[width=\columnwidth]{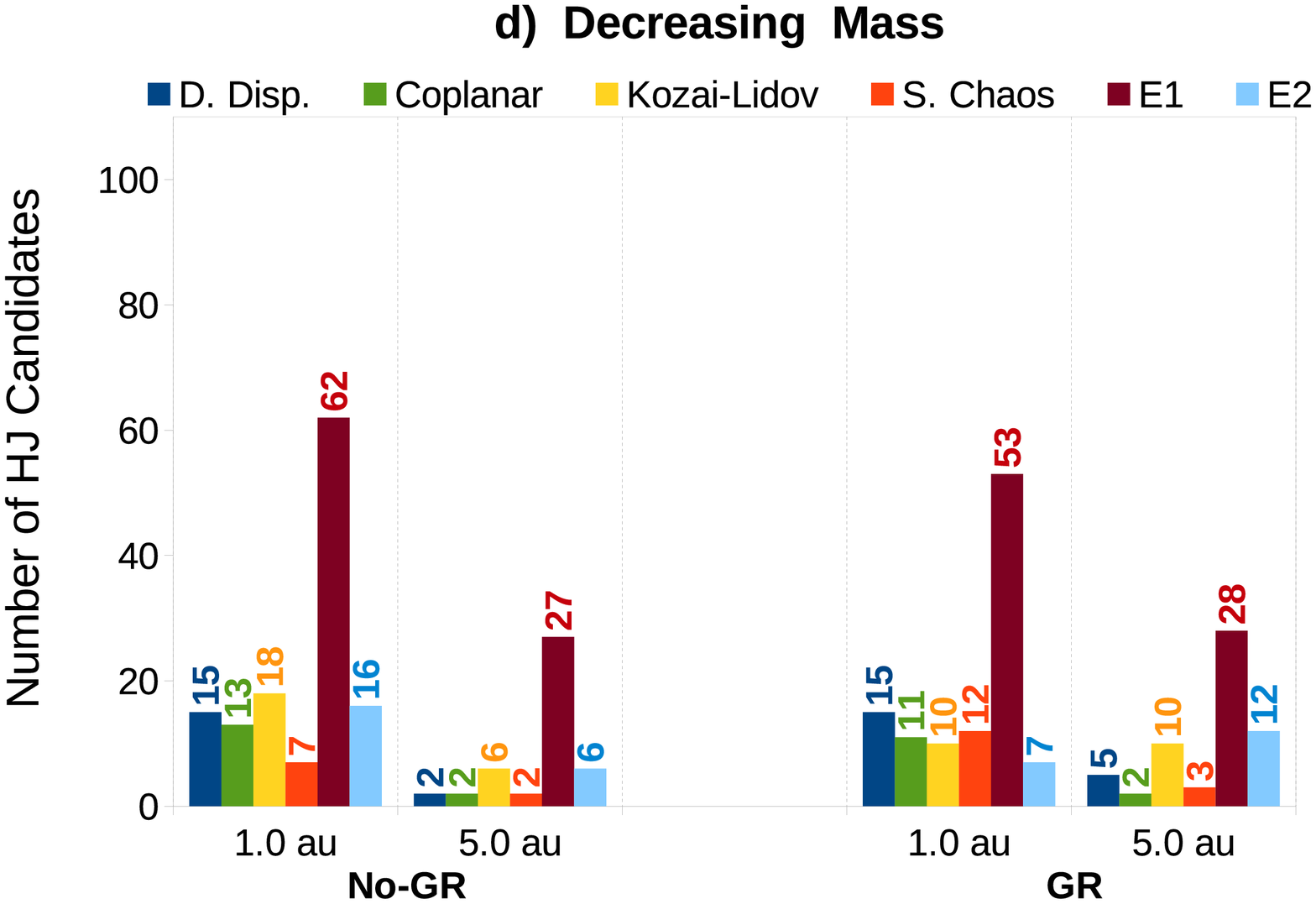}\\
	\end{tabular}
    \caption{Statistical distributions of the number of HJ candidates produced through different high-eccentricity mechanisms as a function of the initial semimajor axis of the innermost planet with and without GR effects for different initial planetary mass configurations.}
    \label{fig:mass_inner}
\end{figure*}

Fig.~\ref{fig:mass_inner} illustrates the number of HJ candidates as a function of the initial semimajor axis of the innermost planet $a_1$ in the No-GR and GR scenarios for each of the four initial planetary mass configurations. Our results shows that, except for the initial equal mass configuration, E1 was always the most efficient mechanism regardless of the value of $a_1$. For the initial equal mass configuration, the highest efficiency in the HJ candidate production was obtained from the E1 (Kozai-Lidov) mechanism for $a_1$ = 1.0 au (5.0 au) in the No-GR scenario, while E2 was the most efficient mechanism regardless of the $a_1$ value in GR simulations.

\begin{figure*}
	\begin{tabular}{cc}
	\includegraphics[width=\columnwidth]{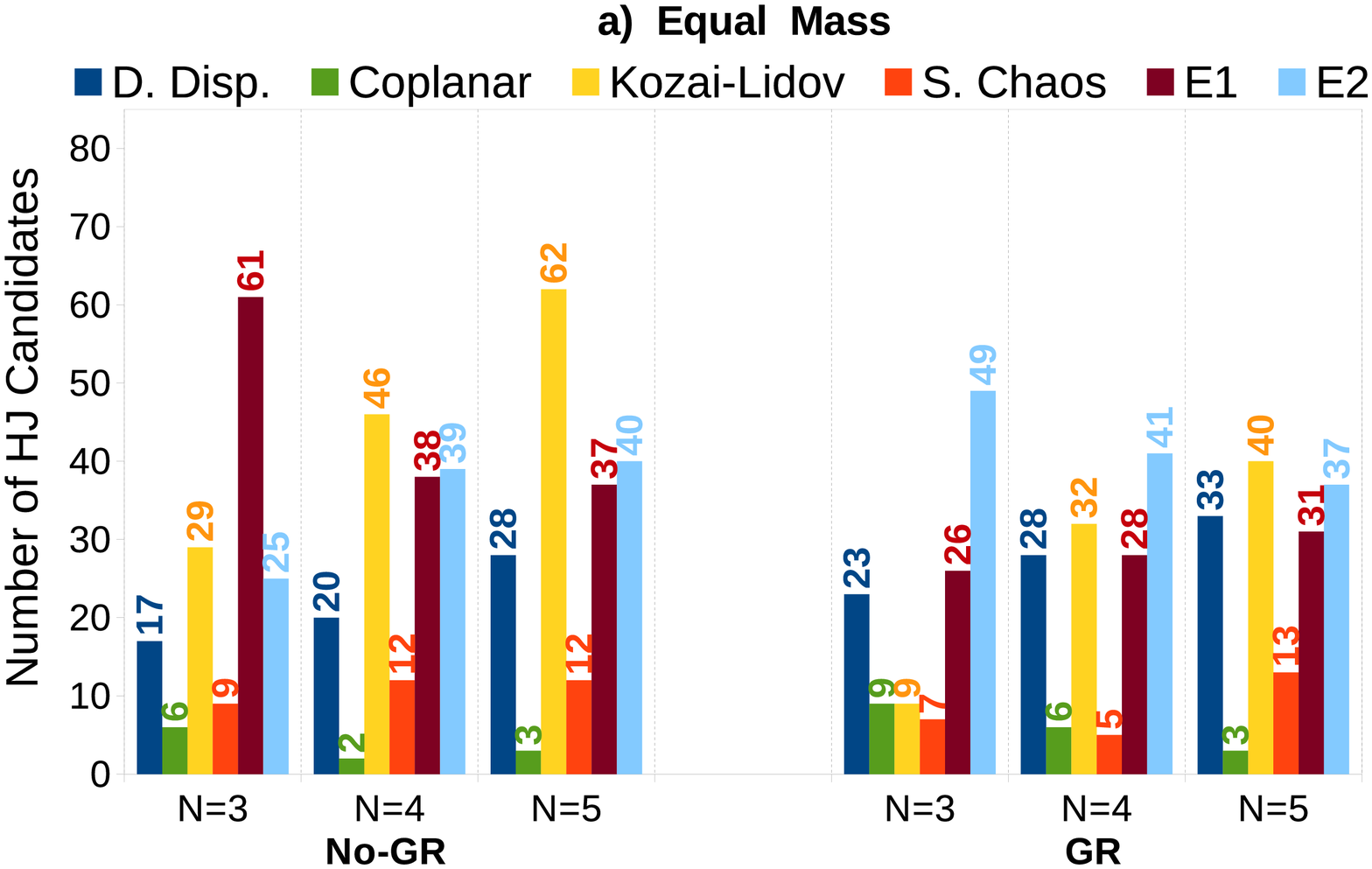} &
	\includegraphics[width=\columnwidth]{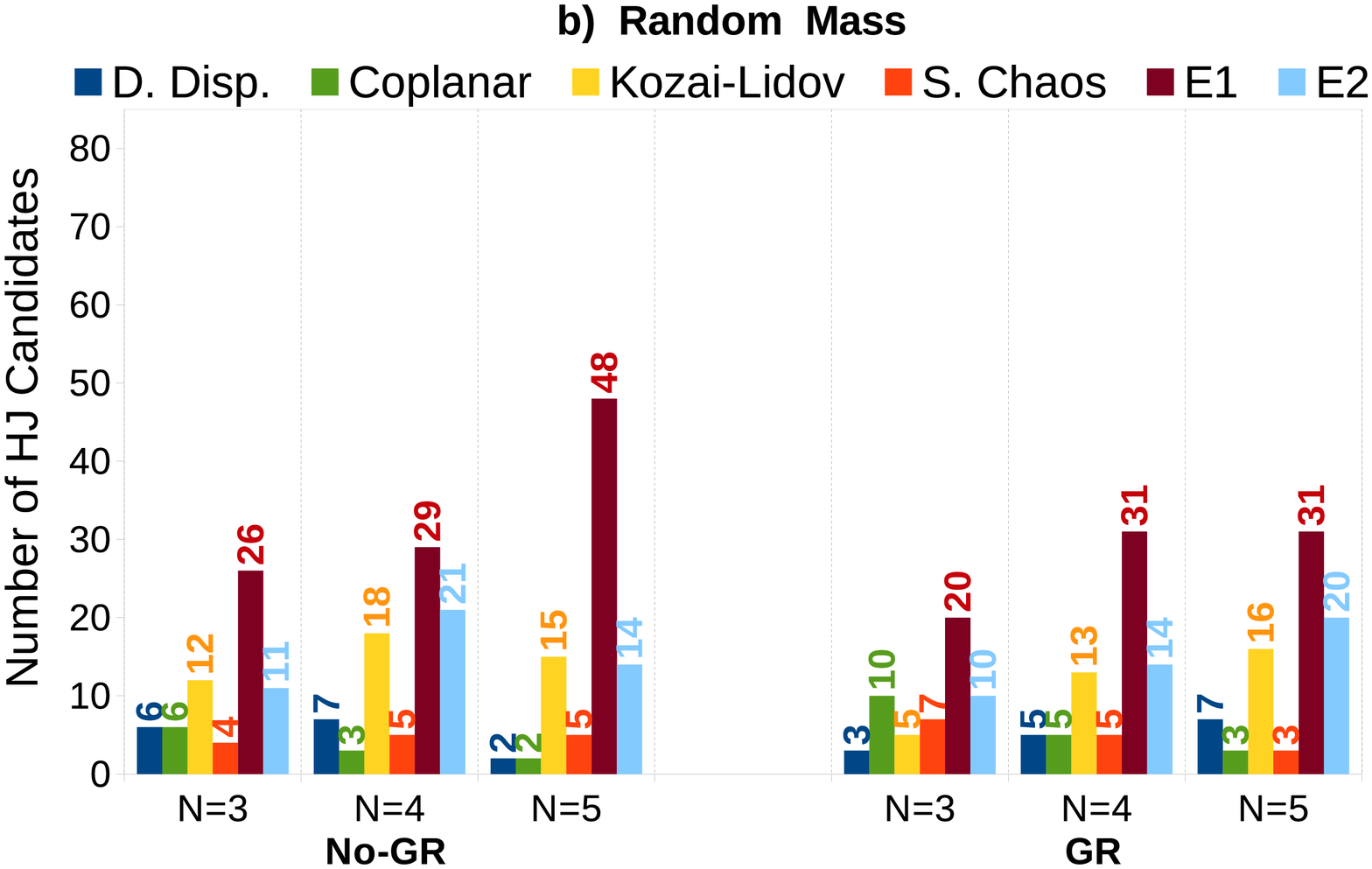}\\
	\includegraphics[width=\columnwidth]{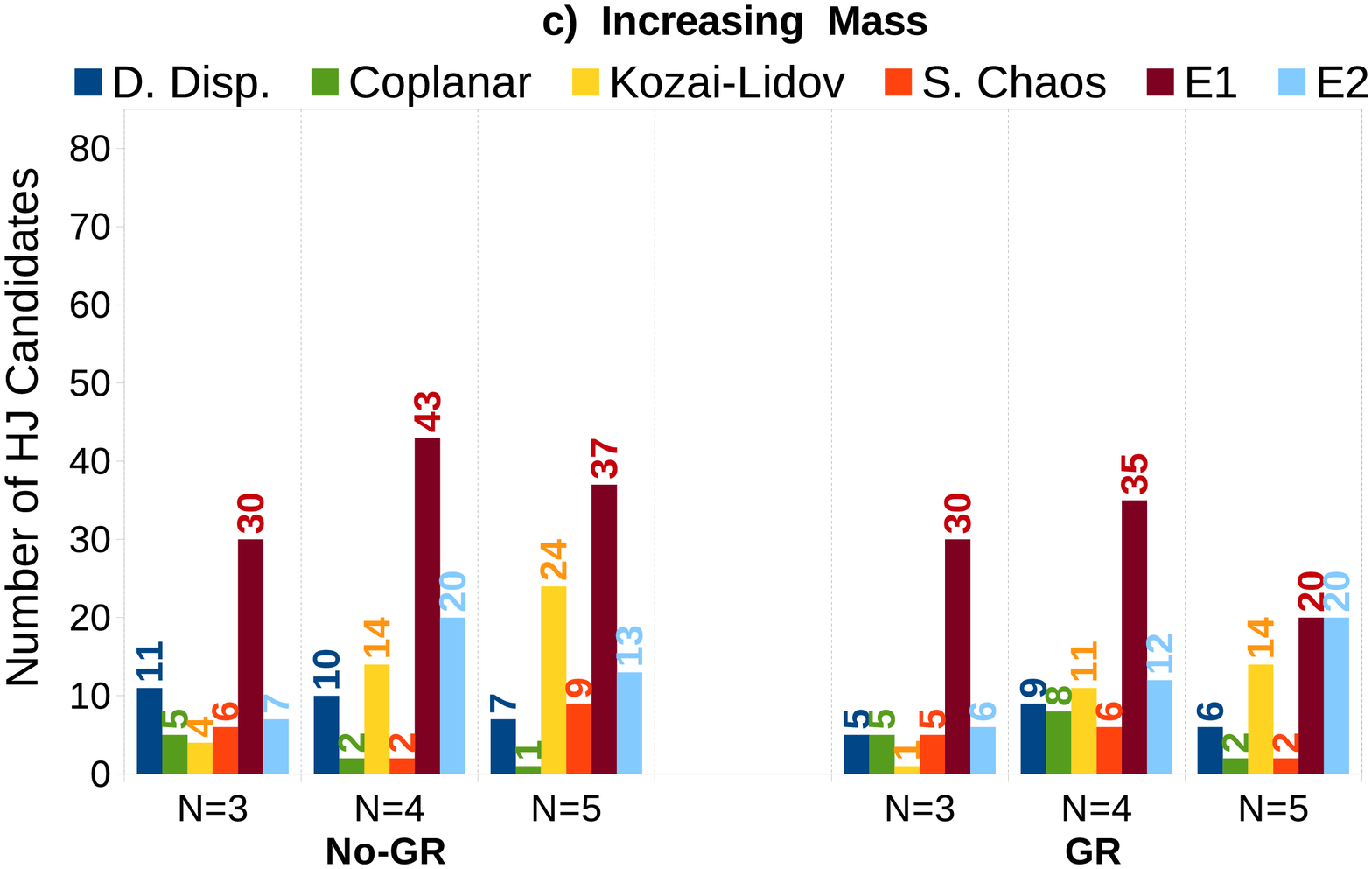}&
	\includegraphics[width=\columnwidth]{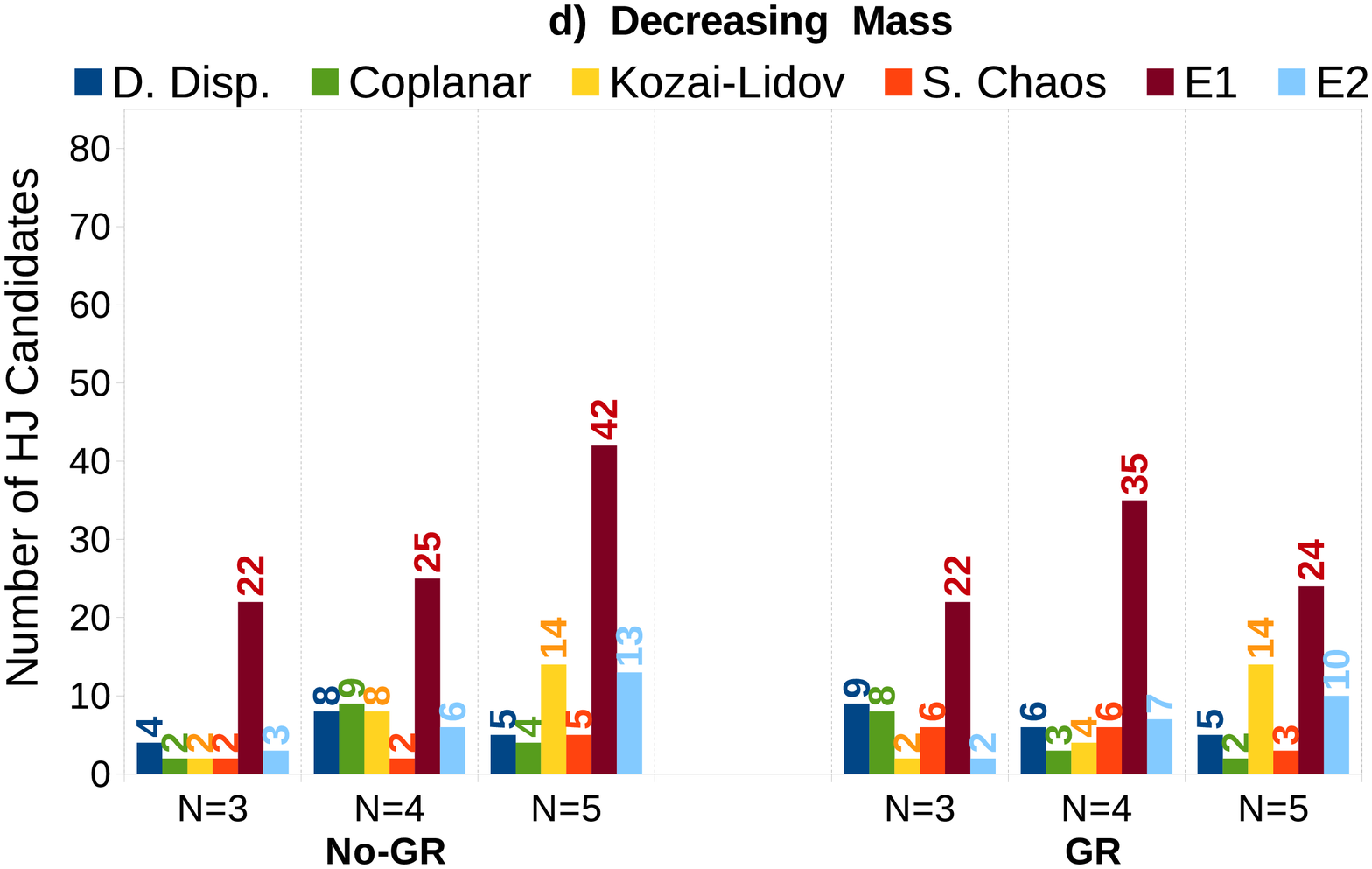}\\
	\end{tabular}
    \caption{Statistical distributions of the number of HJ candidates produced through different high-eccentricity mechanisms as a function of the initial planet number (N) with and without GR effects for different initial planetary mass configurations.}
    \label{fig:mass_number}
\end{figure*}

Fig.~\ref{fig:mass_number} shows the statistical distributions of the number of HJ candidates produced through different high-eccentricity mechanisms  as a function of the initial planet number N for each initial planetary mass configuration in No-GR and GR scenarios.  For the random, increasing, and decreasing initial planetary mass configurations, E1 was the most efficient mechanism in the HJ candidate production for each value of N in the No-GR and GR scenarios without necessarily having a direct relationship with this parameter.  Of the other high-eccentricity mechanisms, the efficiency of the Kozai-Lidov mechanism was highly favored with an increase of N in each initial planetary mass configuration.  For the initial equal mass configuration, there is a strong dependence between the dominant high-eccentricity mechanism able to produce HJ candidates and the value of N. In fact, on the one hand, E1 and Kozai-Lidov were the most efficient mechanisms when N = 3 and N = 4 in the No-GR scenario, while E2 was the dominant mechanism for such values of N in the GR simulations. On the other hand, Kozai-Lidov was the most efficient mechanism for N = 5 in No-GR and GR scenarios.

For comparative purposes, the most efficient high-eccentricity mechanisms obtained from the study developed by \citet{Wang2017} were Kozai-Lidov (36.6\%), scattering (34.9\%) and E1 (16.2\%), while coplanar (7.6\%), E2 (2.8\%) and secular chaos (1.9\%) were the least efficient in their simulations.   We remark that the investigation carried out by \citet{Wang2017} did not include GR effects and it only considered the initial equal mass configuration with $ \Delta \textit{K} = 0.001$ for the development of their simulations.  For the initial equal mass configuration and the No-GR scenario, we found as the most efficient mechanisms able to produce HJ candidates were Kozai-Lidov (28.19\%), E1 (27.99\%) and E2 (21.40\%), while the least efficient ones were direct dispersion (13.37\%), secular chaos (6.79\%) and coplanar (2.26\%).  We see that there are substantial differences between our results and the study carried out by \citet{Wang2017}.  This can be explained if we take into account that \citet{Wang2017} considered initial inclinations that obeyed Rayleigh distributions, a smaller $ \Delta \textit{K} $ interval for the generation of initial conditions, HJ candidates selected from the check of their orbital parameters obtained at the end of the integration, and a different procedure to classify HJ candidates as E2 mechanism.  If we consider only the $e$ vs $\Delta \varpi$ and $e$ vs $\omega$ characteristics shown in Fig. 3 by \citet{Wang2017}, then the E2 mechanism does not exceed 1\% in efficiency in our simulations.

\subsection{Production of hot Jupiter candidates with prograde, retrograde and alternating orbits}
\label{sec:retrograde_orbits} 

Table~\ref{tab:orbits} shows the number of HJ candidates with prograde, retrograde, and alternating orbits produced from each high-eccentricity mechanism in No-GR and GR scenarios for the different initial planetary mass configurations studied in the present paper.  In relation to total number of HJ candidates, we have found that 1737 (81.40\%) remained with prograde orbits ($i<90^{\circ}$), 112 (5.24\%) with retrograde orbits ($i\ge90^{\circ}$) and 285 (13.36\%) alternating between prograde and retrograde orbits in No-GR and GR simulations. As the reader can see, with the exception of the coplanar mechanism, all other high-eccentricity mechanisms produced HJ candidates with retrograde or alternating orbits, being the level of efficiency strongly dependent on the initial planetary mass configuration.  More specifically, assuming No-GR and GR scenarios, the secular chaos, direct dispersion, Kozai-Lidov and E2 mechanisms generated 2 ($\sim 0.1$\%), 27 (1.26\%), 38 (1.78\%) and 45 (2.11\%) HJ candidates with retrograde orbits, respectively, while the most efficient mechanisms in the production of HJ candidates with alternating orbits were Kozai-Lidov with 117 (5.48\%), E2 with 83 (3.89\%) and direct dispersion with 46 (2.16\%).  These results indicate that the Kozai-Lidov mechanism has the highest probability of significantly exciting the orbital inclinations of the HJ candidates.  We must take into consideration that, due to the fact of not including tidal effects in this study, the final inclinations of these HJ candidates after orbital circularization are not known.

\begin{table*}
	\centering
	\caption{Number of HJ candidates with prograde, retrograde and alternating orbits in No-GR and GR (in brackets) scenarios for different initial planetary mass configurations and each high-eccentricity mechanisms. The total percentages were calculated with respect to the total number of HJ candidates produced in each scenario and initial planetary mass configuration (see last column of Table 4).}
	\label{tab:orbits}
	\begin{tabular}{cccccccccc} 
	\hline
	 Orbit & Init. planet.  & Direct Dispersion & Coplanar & Kozai-Lidov & S. Chaos & E1 & E2 & Total \\ & mass config. & &  &  &  &  &  & (\%) \\
	\hline
	 & Equal & 34 (53) & 11 (18) & 79 (40) & 26 (23) & 130 (81) & 74 (81) & 72.84 (70.48) \\ 
	 Prograde & Random & 13 (14) & 11 (18) & 34 (21) & 12 (13) & 102 (80) & 32 (37) & 87.18 (87.98) \\ 
	 & Increasing & 24 (18) & 8 (15) & 31 (16) & 14 (11) & 109 (82) & 23 (31) & 85.31 (87.82) \\ 
	 & Decreasing & 17 (18) & 15 (13) & 18 (15) & 7 (13) & 87 (81) & 18 (16) & 92.05 (92.85) \\ 
	\hline
     & Equal & 9 (10) & 0 (0) & 15 (11) & 1 (1) & 0 (0) & 7 (17) & 6.58 (9.29) \\ 
	 Retrograde & Random & 1 (1) & 0 (0) & 2 (4) & 0 (0) & 0 (0) & 2 (2) & 2.14 (3.36) \\ 
	 & Increasing & 3 (1) & 0 (0) & 3 (0) & 0 (0) & 0 (0) & 9 (3) & 6.12 (2.03) \\ 
	 & Decreasing & 0 (2) & 0 (0) & 2 (1) & 0 (0) & 0 (0) & 3 (2) & 2.84 (2.98) \\ 
	\hline
     & Equal & 22 (21) & 0 (0) & 43 (30) & 6 (1) & 6 (4) & 23 (29) & 20.58 (20.23) \\ 
	 Alternating & Random & 1 (0) & 0 (0) & 9 (9) & 2 (2) & 1 (2) & 12 (5) & 10.68 (8.65) \\ 
	 & Increasing & 1 (1) & 0 (0) & 8 (10) & 3 (2) & 1 (3) & 8 (4) & 8.57 (10.15) \\ 
	 & Decreasing & 0 (0) & 0 (0) & 4 (4) & 2 (2) & 2 (0) & 1 (1) & 5.11 (4.17) \\ 
	\hline
	\end{tabular}
\end{table*}

Depending on the initial planetary mass configuration, the inclusion of GR can increase or decrease the production of HJ candidates with retrograde orbits.   In fact, from Table~\ref{tab:orbits}, we see that the GR increased the efficiency of production of HJ candidates with retrograde orbits for the initial equal and random mass configurations.  Conversely, the number of HJ candidates with retrograde orbits decreased with the inclusion of GR for the initial increasing mass configuration. Finally, the GR did not have a significant effect in the HJ candidate production on retrograde orbits for the initial decreasing mass configuration.

\section{Discussions and Conclusions}
\label{sec:conclusions_discussions}

\subsection{Discussions}
\label{sec:discussions}

The present study is aimed to investigate the efficiencies of different high-eccentricity mechanisms for forming HJ candidates in planetary systems that underwent strong dynamical instabilities. To do this, we have analyzed the role of the general relativity and the dependence of the HJ candidate production on different initial planetary mass configurations. We are aware that tides will play a very important role in the formation of HJs. It is worth noting that the inclusion of tides should be done taking into account the different stages of dynamical evolution during the formation process of HJs. On the one hand, for low eccentricities, the tidal evolution is well described by the so called equilibrium tidal model \citep[see][]{Darwin1880,Mignard1979,Mignard1980,Hut1981,Ferraz2008}. In this description, it is assumed that the equilibrium figure of the distorted body deviates from the instantaneous equipotential shape due to internal viscosity, leading to angular momentum exchange between orbital and rotational motions. On the other hand, for very high orbital eccentricities, the tidal deformation only occurs close to the pericenter of the orbit. In this case, the tidally deformed body can not attain an equilibrium figure, undergoing forced oscillations. This regime of tidal evolution is described by the dynamical tide model \citep[see][]{Ivanov2004,Ivanov2007,Ivanov2011}. The dynamic tide is much more complex to model than the equilibrium tides. We note that the dynamical mechanism leading to the formation of HJ candidates strongly raises the eccentricities of the planets involved in the initial instabilities, approaching to the limit of parabolic orbits ($e \sim 1$), which is the limit case for the inclusion of dynamic tides. As the orbit decays and circularizes, tides would be included in the frame of the equilibrium tide model. Since both regimes operate in very different ways, it is not easy to construct an unified model for the whole tidal evolution of HJ candidates \citep[see][for a detailed discussion]{Beauge2012}. We will study in detail the influence of tides in the formation of HJs from a set of candidates in a forthcoming paper.

Considering the four initial planetary mass configurations proposed in our study, we obtained two different populations with 993 and 1141 HJ candidates from numerical experiments with and without GR effects, respectively. We consider that it would be very interesting to carry out a statistical analysis of our results in order to evaluate potential differences between both populations of HJ candidates. To do this, we constructed two incremental mass distributions from the samples of HJ candidates derived with and without GR and then, a chi-square test was applied. The result of the statistical test indicates that the two binned mass distributions of HJ candidates resulting with and with GR are consistent, at 95 percent of significance, with a single distribution function.

We consider that it is necessary to carry out a discussion concerning the integration timescale of our numerical experiments, which has always been of 50 Myr for all simulations of each scenario. We are aware that a certain number of new HJ candidates should be produced in our systems of work if the simulations were extended for longer timescales. However, we consider that to specify the number of new HJ candidates produced over timescales of 100 Myr or 1 Gyr is not a simple task in the framework of the present study, which is based on 96000 N-body simulations with and without GR, with different initial configurations related to the number of planets, semimajor axis of the innermost one, and planetary mass distribution. According to this, we will carry out an efficient study that allows us to quantify the number of new HJ candidates from the different high-eccentricity mechanisms over longer timescale in a forthcoming paper.  

In this line of analysis, the study of the convergence of the secular chaos mechanism is of particular interest. In fact, different previous studies, such as those developed by \citet{Wu2011} and \citet{Teyssandier2019}, considered timescales greater than 0.1 Gyr to analyze such a mechanism. In order to analyze the sensitivity of the efficiency of the secular chaos mechanism to the integration timescale adopted in the numerical experiments, we decided to select a sub-sample of our systems of work and then, to extend the simulations to 0.2 Gyr. To do this, we focused on the 12000 numerical experiments associated with the initial planetary equal mass configuration without GR since such an scenario showed the most efficiency of production of HJ candidates over an integration timescale of 50 Myr. Of those 12000 simulations, 8748 did not produce HJ candidates. From this, we adopted a sub-sample of systems of those 8748 in order to extend the simulations for 0.2 Gyr and then, to analyze the formation of HJ candidates on that extended timescale. To select the systems of work, we considered the criterion based on the angular momentum deficit \citep[AMD,][]{Laskar1997} proposed by \citet{Wu2011}. According to that discussed by these authors, the requirement for the HJ formation in a secularly interacting system is a sufficient amount of AMD, since it limits the maximum values of eccentricity and inclination that a given planet can attain. In fact, only when AMD is large enough can it be shared among the planets of the whole system, driving the inner planet to experience secular chaos. The AMD criterion is determined by Eq. 5 from \citet{Wu2011}. We implemented this criterion using the orbital elements obtained at 50 Myr for the planets of the 8748 systems of work. Of these systems, 1808 satisfy the AMD criterion, of which we randomly selected a sub-sample of 282 systems. Our results showed that 19.9\% of the total sub-sample systems underwent some dynamical instability event after 50 Myr of evolution. It is particularly interesting to analyze the results obtained as a function of the initial semimajor axis of the innermost planet $a_1$. On the one hand, 95 systems of the sub-sample had $a_1$ = 1 au, of which 7 systems (2.5\%) experienced some dynamical instability event after 50 Myr of evolution, producing only 4 HJ candidates between 50 Myr and 200 Myr. On the other hand, 187 systems of the sub-sample had $a_1$ = 5 au, of which 49 systems (17.4\%) underwent some dynamical instability after 50 Myr, leading to the production of only 7 HJ candidates between 50 Myr and 200 Myr. These results show that the greater the initial semimajor axis of the innermost planet, the greater the percentage of systems that experience some dynamical instability event after 50 Myr. A very important result of our study indicates that only 1 of 11 new HJ candidates produced from the 282 systems of the sub-sample between 50 Myr and 200 Myr was associated with the secular chaos mechanism. According to this, if the sub-sample of 282 systems that satisfy the AMD criterion is considered, the efficiency of the production of HJ candidates between 50 Myr and 200 Myr from secular chaos is $\sim$ 0.35\%. If this percentage is assumed for the total sample of 1808 systems that satisfy the AMD criterion, $\sim$ 6 new HJ candidates should be produced between 50 Myr and 200 Myr from secular chaos in our scenario associated with the initial planetary equal mass configuration without GR. This number is small in comparison with the 486 HJ candidates produced during the first 50 Myr in the scenario associated with the sub-sample, for which the total efficiency of generation of HJ candidates will not show significant changes over 200 Myr, at least due to the contribution of secular chaos.  

Finally, we want to remark that our study on the production of HJ candidates from high-eccentricity mechanisms was based on multi-planet systems initially located in the cold region of the system ($a\geq1.0$ au). Recently, \citet{Anderson2020} analyzed the in situ scattering of warm Jupiters from multi-planet systems, with the innermost planet placed between 0.1 au and 1 au. These authors found that the dynamical instabilities produce a comparable number of one-planet and two-planet systems, but one-planet systems show higher eccentricities. According to this, the efficiency of production of HJ candidates from the direct dispersion mechanism could be higher in multi-planet systems with the innermost planet in the warm region respect to that obtained in multi-planet systems initially located in the cold region. We think that it would be interesting to carry out a detailed study concerning the sensitivity of the formation efficiency of HJ candidates from high-eccentricity mechanisms to the initial location of the planets that compose the systems of work. However, this point is out of the scope of the present investigation.

\subsection{Conclusions}
\label{sec:conclusions}

In this paper, we have analyzed the formation efficiency of HJ candidates from high-eccentricity mechanisms in planetary systems that undergo strong dynamical instabilities between gaseous giants. In particular, our investigation focused on the efficiencies of six different kinds of high-eccentricity mechanisms, which are direct dispersion, coplanar, Kozai–Lidov, secular chaos, and E1 and E2 mechanisms. The present study is based on that developed by \citet{Wang2017}, though there are several significant differences in the initial conditions and in the methodology of both works. In particular, we analyzed the sensitivity of our results to the initial number of planets, the initial semimajor axis of the innermost planetary orbit, the initial configuration of planetary masses, and the general relativity (GR) effects. In fact, we have been able to construct a more detailed model to analyze the formation of HJ candidates from high-eccentricity mechanisms, which has allowed us to strengthen our understanding concerning the dynamical evolution of planetary systems that undergo early strong instabilities around solar-type stars.

In general terms, we have found that about 71.5\% of the systems resulting from our N-body simulations with or without GR undergo dynamical instability events within 50 Myr of evolution. From the study of such systems, our results indicated that the efficiency of formation of HJ candidates from high-eccentricity mechanisms is about 3.3\% (2.9\%) in numerical experiments without (with) GR effects. The slightly smaller number of HJ candidates in simulations with GR is consistent with studies developed by \citet{Marzari2020}, who showed the GR can significantly reduce the oscillations of eccentricity of close–in planets in their secular evolution. We remark that the percentages of HJ candidates produced in the present investigation are also in a good agreement with previous works that focused on different high-eccentricity mechanisms \citep[see][]{Beauge2012, Petrovich2015b,Munoz2016,Petrovich2016,Wang2017,Teyssandier2019}.

Of all mechanisms analyzed in our study, the E1 mechanism is the most efficient in producing HJ candidates both in simulations with GR and without GR, followed by Kozai-Lidov and E2 mechanisms, and direct dispersion in lesser order of importance. Since the GR has the effect of inhibiting the libration of the argument of pericenter of the HJ candidate, the Kozai-Lidov mechanism is less (more) relevant than E2 mechanism in simulations with (without) GR effects. The coplanar mechanism and secular chaos are the least efficient high-eccentricity mechanisms in forming HJ candidates in No-GR and GR simulations.

Our analysis regarding the sensitivity of the results to the initial semimajor axis of the innermost planetary orbit $a_1$ allowed us to derive some considerations of interest. On the one hand, the greater the value of $a_1$, the smaller the number of HJ candidates both in simulations with GR and without GR. On the other hand, the greater the value of $a_1$, the less efficient the high-eccentricity mechanism that produce HJ candidates in No-GR and GR scenarios. The only exception to this point was the Kozai-Lidov mechanism in GR simulations. In fact, the GR effect tends to inhibit the libration of the argument of pericenter of the planetary orbit, which is more efficient for planets closer to the central star.

We also study the dependence of our results on the initial number of giant planets that compose the simulated systems. We showed that the greater the initial number of planets, the greater the number of HJ candidates both in No-GR and GR scenarios. We did not observe a general dependence between the initial number of planets and the efficiency of HJ candidate production that is valid for all the high-eccentricity mechanisms.

One of the most important analysis of our investigation was that concerning the sensitivity of the formation efficiency of HJ candidates from high-eccentricity mechanisms resulting from different initial planetary mass configurations.  Our results showed that, on the one hand, the highest efficiency in the production of HJ candidates in No-GR and GR simulations was derived from the initial equal mass configuration. On the other hand, the initial random and increasing mass configurations had an individual efficiency for the HJ candidate production that was about half of that associated with the initial equal mass configuration. Finally, the initial decreasing mass configuration showed the lowest efficiency in the present study. We also found differences in the dominant mechanism that led to the HJ candidate production for the different initial planetary mass configurations. On the one hand, E1 mechanism was notably efficient in the generation of HJ candidates in the initial random, increasing, and decreasing mass configurations in No-GR and GR scenarios. On the other hand, in the initial equal mass configuration, the efficiency of HJ candidate production from the E1 mechanism was slightly surpassed by the Kozai-Lidov mechanism in the No-GR scenario, while the efficiency of E2 mechanism was more relevant than that obtained from the E1 mechanism in the GR simulations.

Finally, we studied the production of HJ candidates with prograde, retrograde, and alternating orbits. We found that the highest efficiencies of production correspond to HJ candidates on prograde orbits in No-GR and GR simulations. In both of such scenarios, the generation of HJ candidates on alternating orbits is also possible, while the lowest efficiencies of production correspond to HJ candidates on retrograde orbits. We also distinguished the most efficient mechanisms able to produce HJ candidates on retrograde and alternating orbits. In fact, in No-GR and GR scenarios, Kozai-Lidov is the dominant mechanism in the generation of HJ candidates on alternating orbits, while E2 and also Kozai-Lidov showed the highest efficiencies in the HJ candidate production on retrograde orbits. Finally, we analyzed the sensitivity of the efficiency of HJ candidate production with retrograde orbits to the inclusion of GR effects in the different initial planetary mass configurations. We found that the GR effects increased (decreased) the number of HJ candidates on retrograde orbits for the initial equal and random (increasing) mass configurations, while the results associated with the initial decreasing mass configuration concerning the production of retrograde HJ candidates were not sensitive to the GR.

We consider that the present investigation have allowed us to strengthen our knowledge about the dynamics of massive planets in systems that experience early strong instability events.

\section*{Acknowledgements}

We thanks to Julia Venturini and Tabar\'e Gallardo for providing the modified version of the Mercury code used in this work. We also want to thank Dr. Daniel D. Carpintero for his valuable comments about the statistic analysis implemented in our investigation. We acknowledge the use of the Lobo Carneiro supercomputer from NACAD - Coppe, Universidade Federal do Rio de Janeiro (UFRJ).  This study was financed in part by the Coordena\c{c}\~ao de Aperfei\c{c}oamento de Pessoal de N\'ivel Superior - Brasil (CAPES) - Finance Code 001.
GdE acknowledges the partial financial
support by Agencia Nacional de Promoción Científica y Tecnológica (ANPCyT), Argentina, through PICT 201-0505, and by Universidad Nacional de La Plata (UNLP), Argentina, through PID G172. Finally, we thank the anonymous reviewer for the helpful suggestions.

\section*{Data Availability}

The data underlying this article will be shared on reasonable request to the corresponding author.



\bibliographystyle{mnras}
\bibliography{references} 




\appendix

\section{Identification of high-eccentricity mechanisms}

\begin{figure*}
	\textbf{Direct dispersion mechanism}\par\medskip
	\begin{tabular}{cc}
	\includegraphics[width=6.35cm,height=4.4cm]{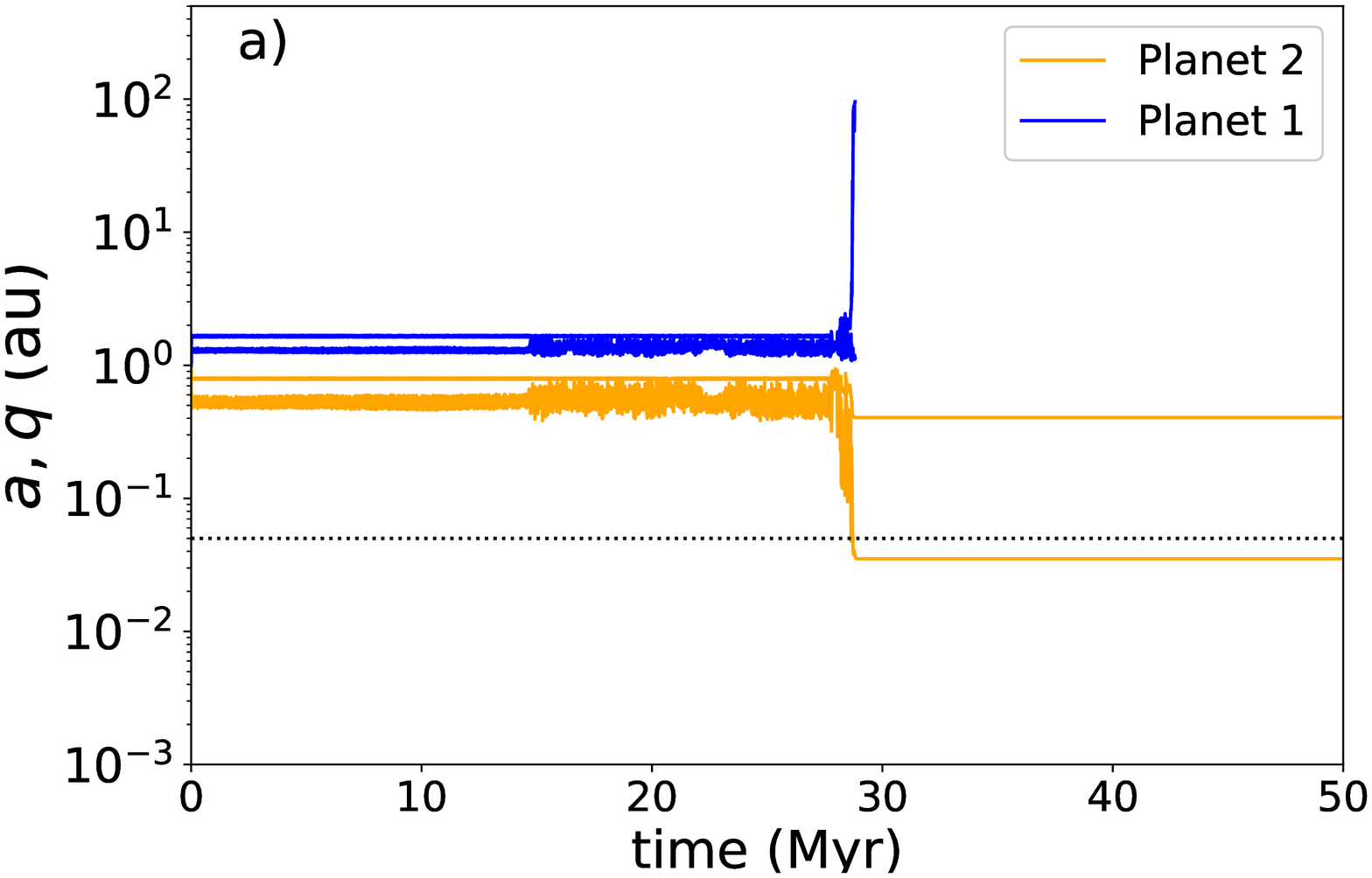} &
	\includegraphics[width=6.35cm,height=4.4cm]{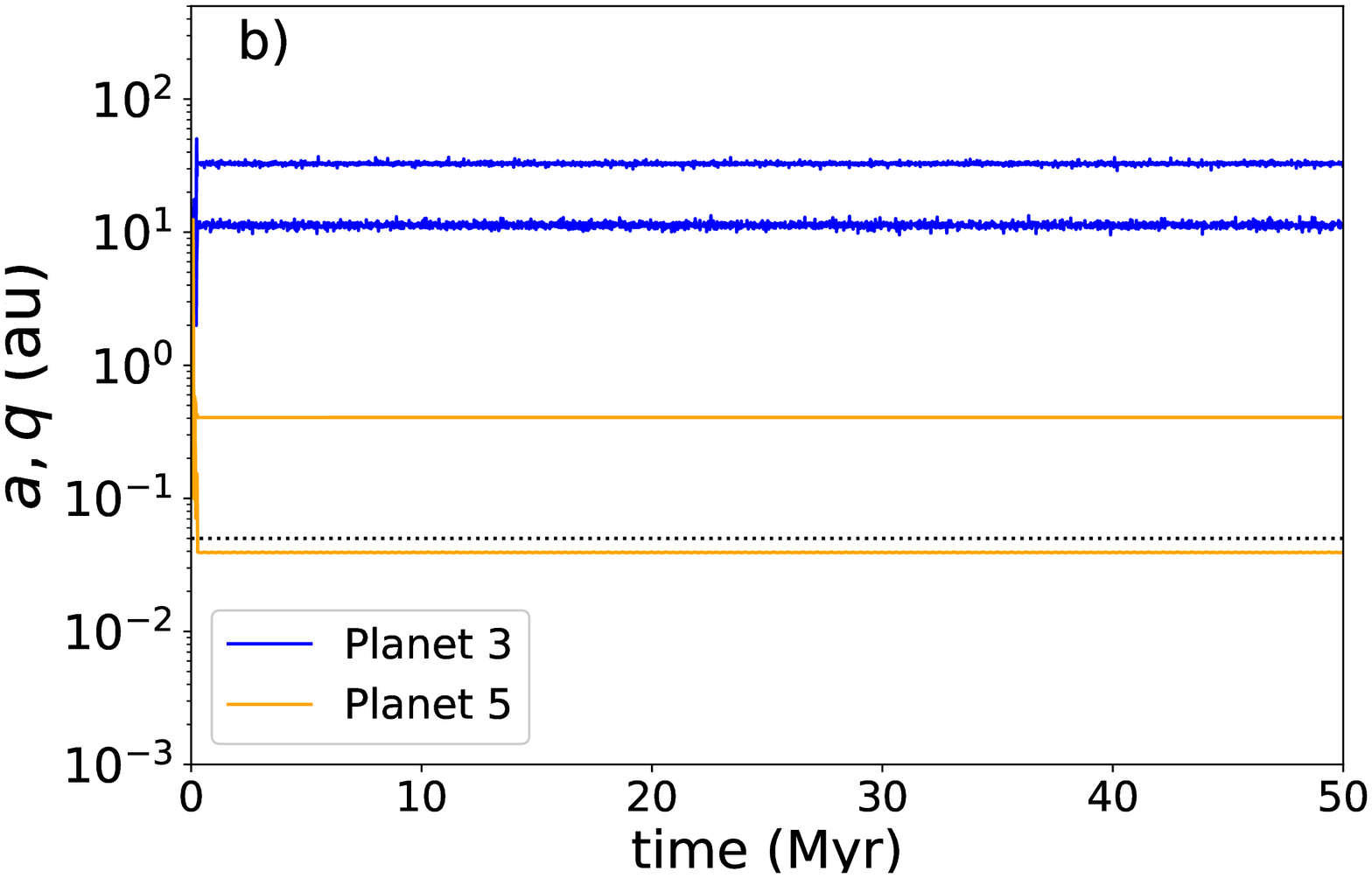}\\
	\includegraphics[width=6.35cm,height=4.4cm]{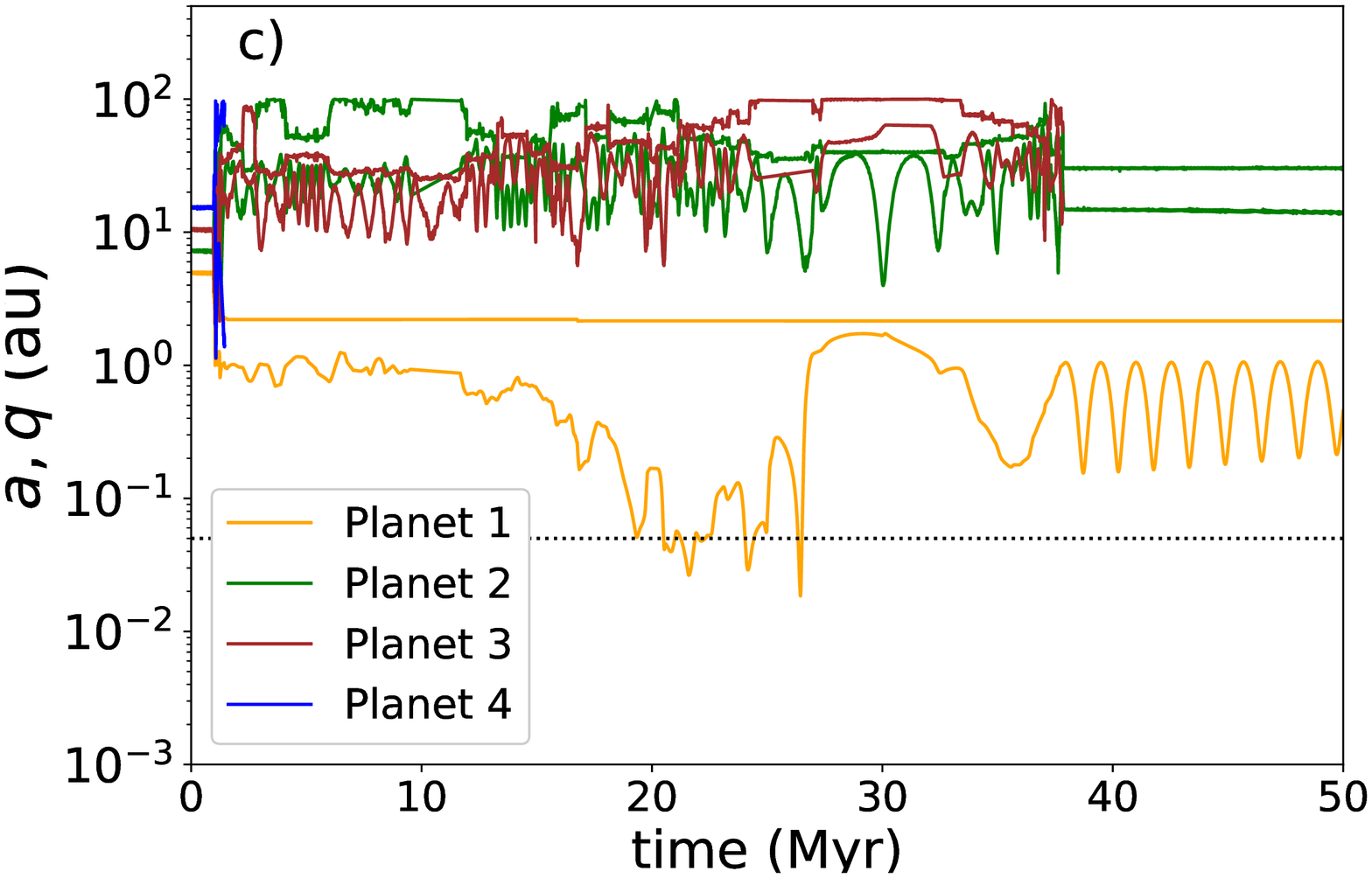}\\
	\end{tabular}
    \caption{Three different graphs of evolution of semimajor axis and pericentric distances in systems for which the direct dispersion mechanism was responsible for producing the HJ candidate based on the classification procedure described in Section 2.3. In panel a) planet 2 is left without companions in the system and with $q\lesssim0.05$ au after the ejection of planet 1 (criterion 1). Previously, planet 1 was hit by planet 3 at 147 years.  In panel b) planet 5 has a companion in the system and with $q\lesssim0.05$ au after the ejections of three planets (criterion 2). In panel c) planet 1 reaches a periastron $q\lesssim0.05$ au on a time scale $\Delta t \approx \tau_a$ due to the multiple close encounters that occurred in the system (criterion 3).}
    \label{fig:scattering}
\end{figure*}

\begin{figure*}
	\textbf{Coplanar mechanism - Case 1}\par\medskip
	\begin{tabular}{cc}
	\includegraphics[width=6.35cm,height=4.5cm]{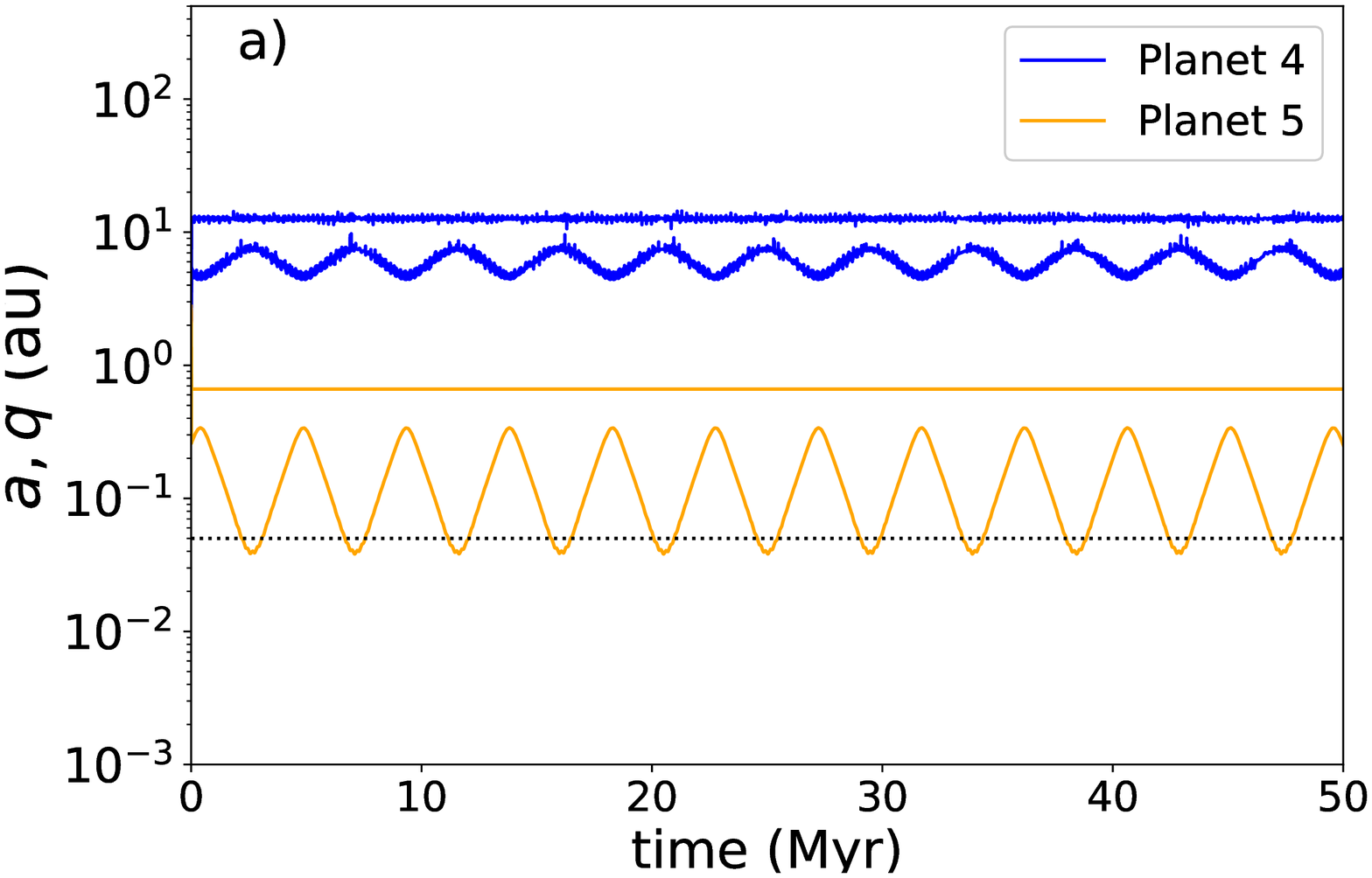} &
	\includegraphics[width=6.35cm,height=4.5cm]{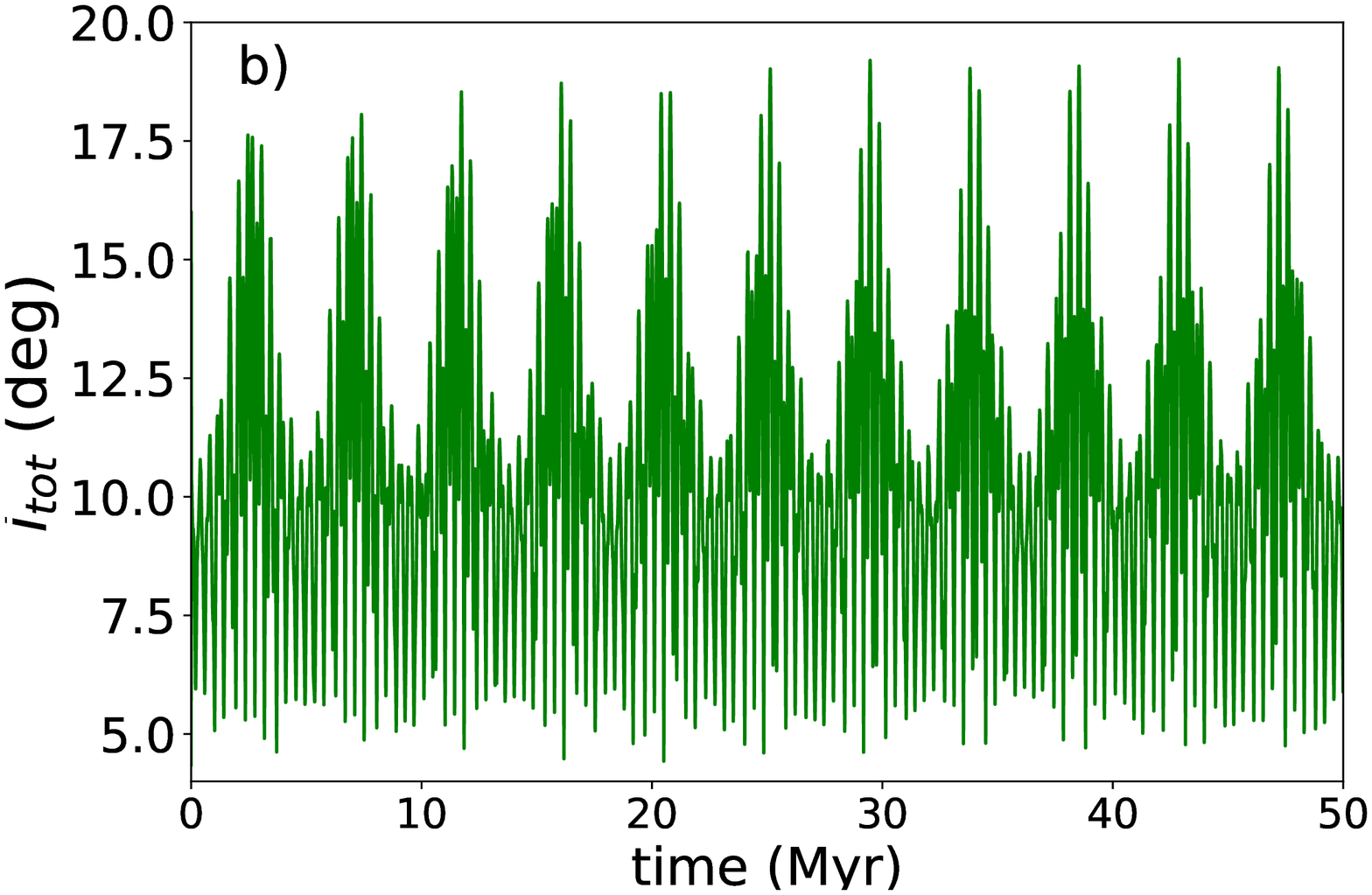}\\
	\includegraphics[width=6.35cm]{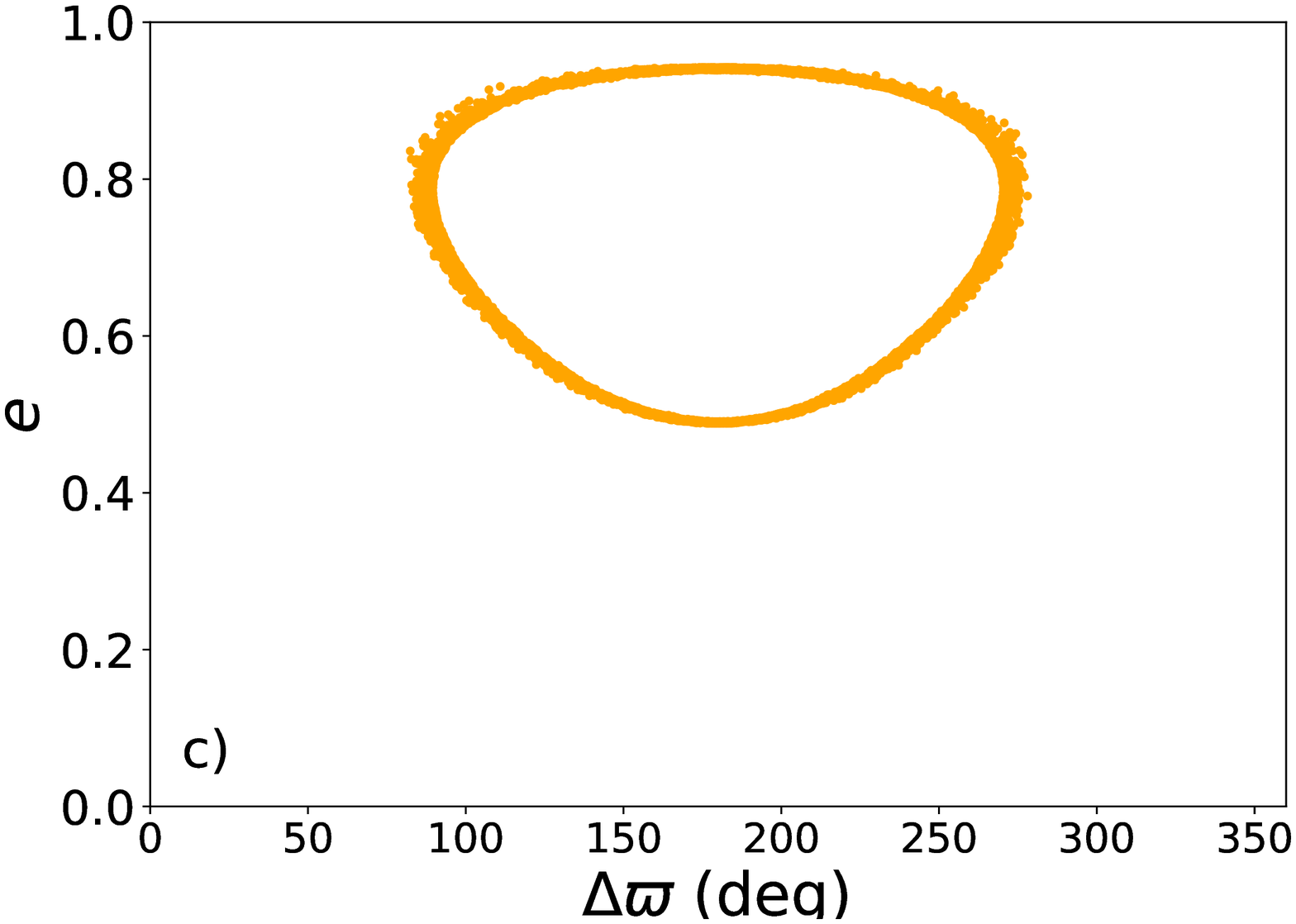} &
	\includegraphics[width=6.35cm]{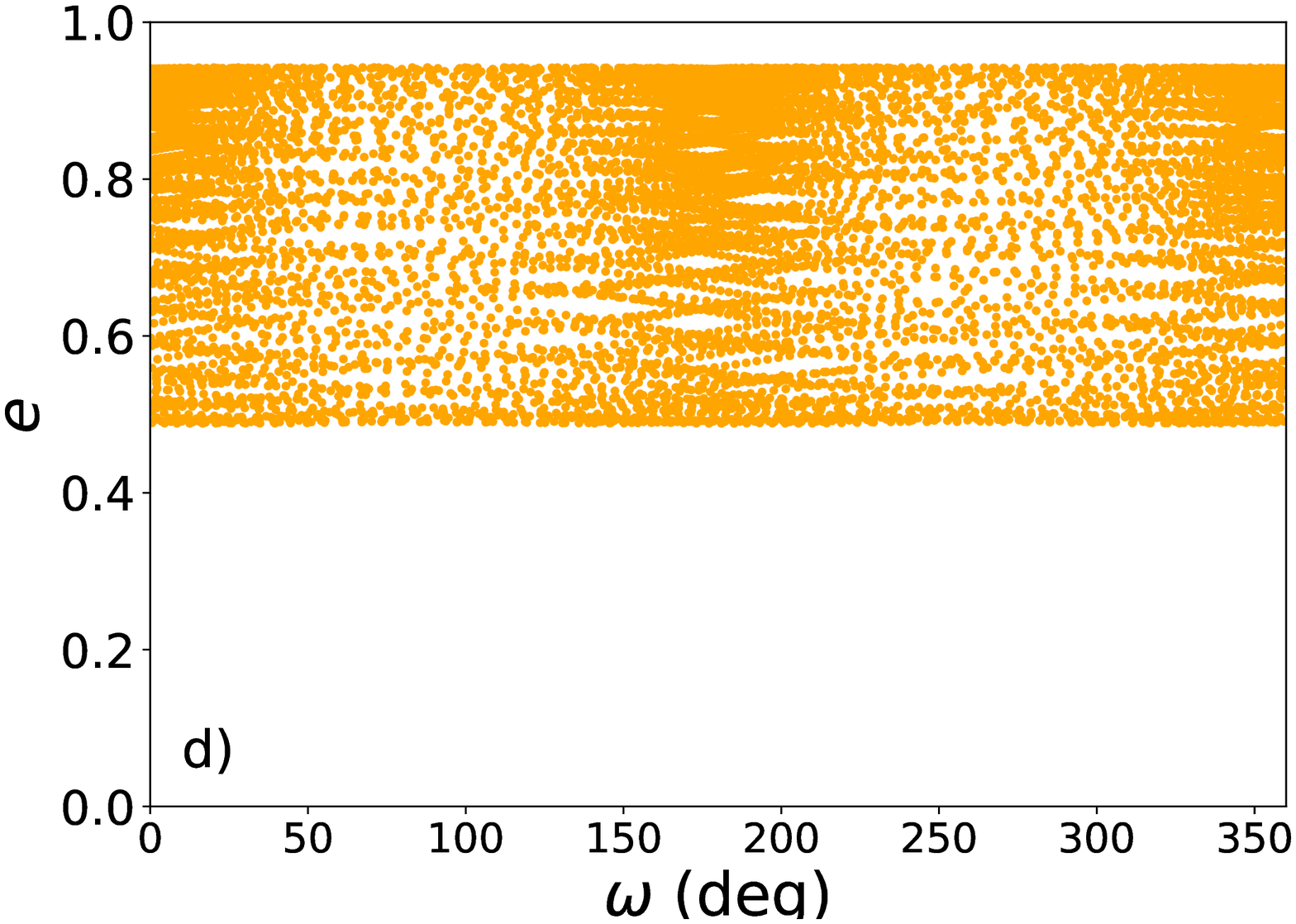}\\
	\end{tabular}
    \caption{Dynamic evolution of a system with five planets in an initial planetary configuration of random mass. In the dynamic instability phase, planets 1, 2 and 3 are ejected at 0.025, 0.013 and 0.012 Myr, respectively. In panel a) we show the evolution of the semimajor axis and pericentric distances for the surviving planets, being one of them the HJ candidate. The horizontal dashed line is located at 0.05 au. In panel b) the periodic evolution of the mutual inclination $i_{tot}=i_4+i_5$ involving the two surviving planets ($i_4$ and $i_5$ with respect to the invariant plane). Panels c) and d) show the phase diagrams $e$ vs $\Delta \varpi$ and $e$ vs $\omega$ of the HJ candidate after 0.025 Myr, respectively.}
    \label{fig:coplanar_1}
\end{figure*}

\begin{figure*}
	\textbf{Coplanar mechanism - Case 2}\par\medskip
	\begin{tabular}{cc}
	\includegraphics[width=6.35cm,height=4.5cm]{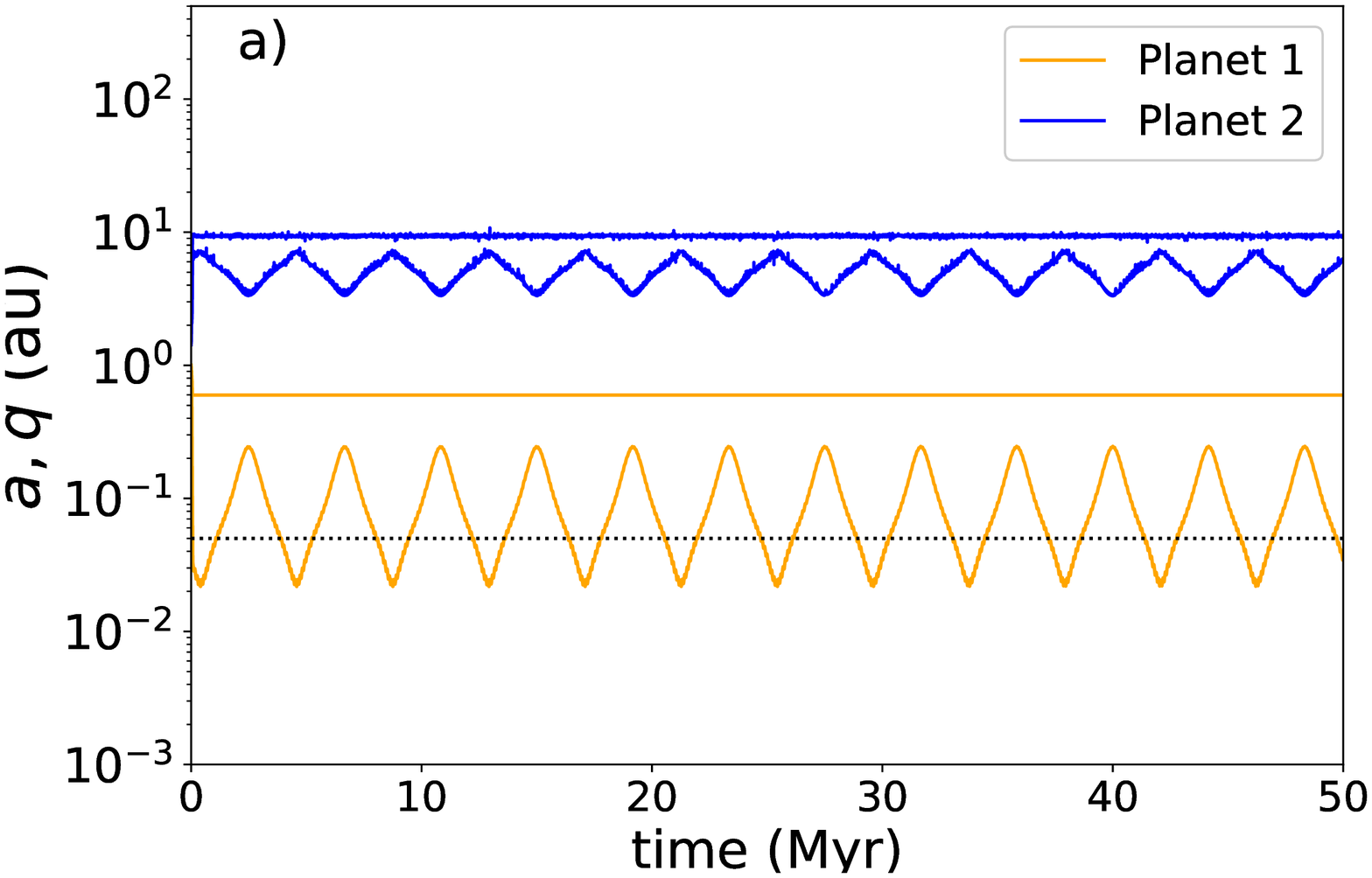} &
	\includegraphics[width=6.35cm,height=4.5cm]{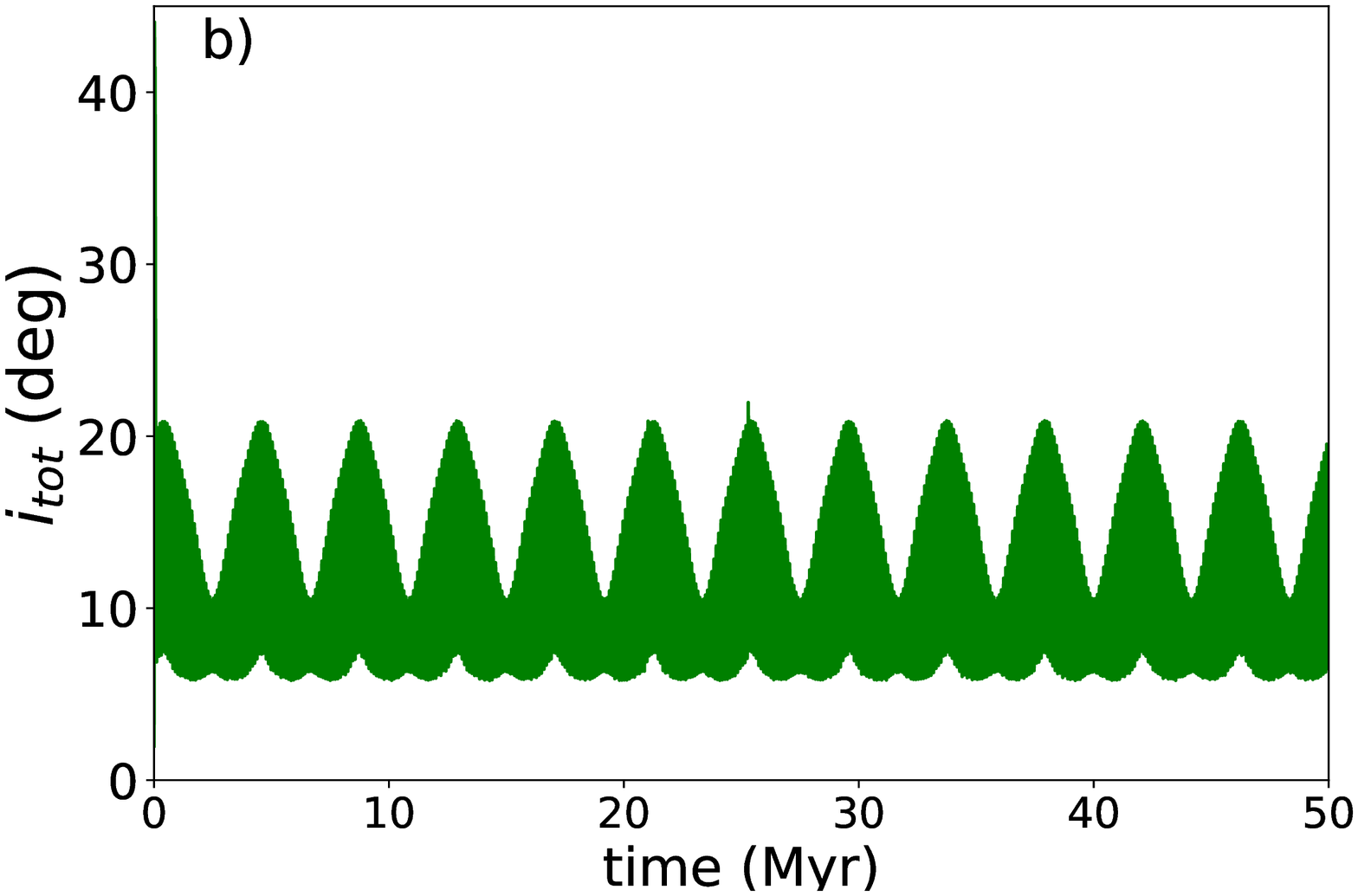}\\
	\includegraphics[width=6.35cm]{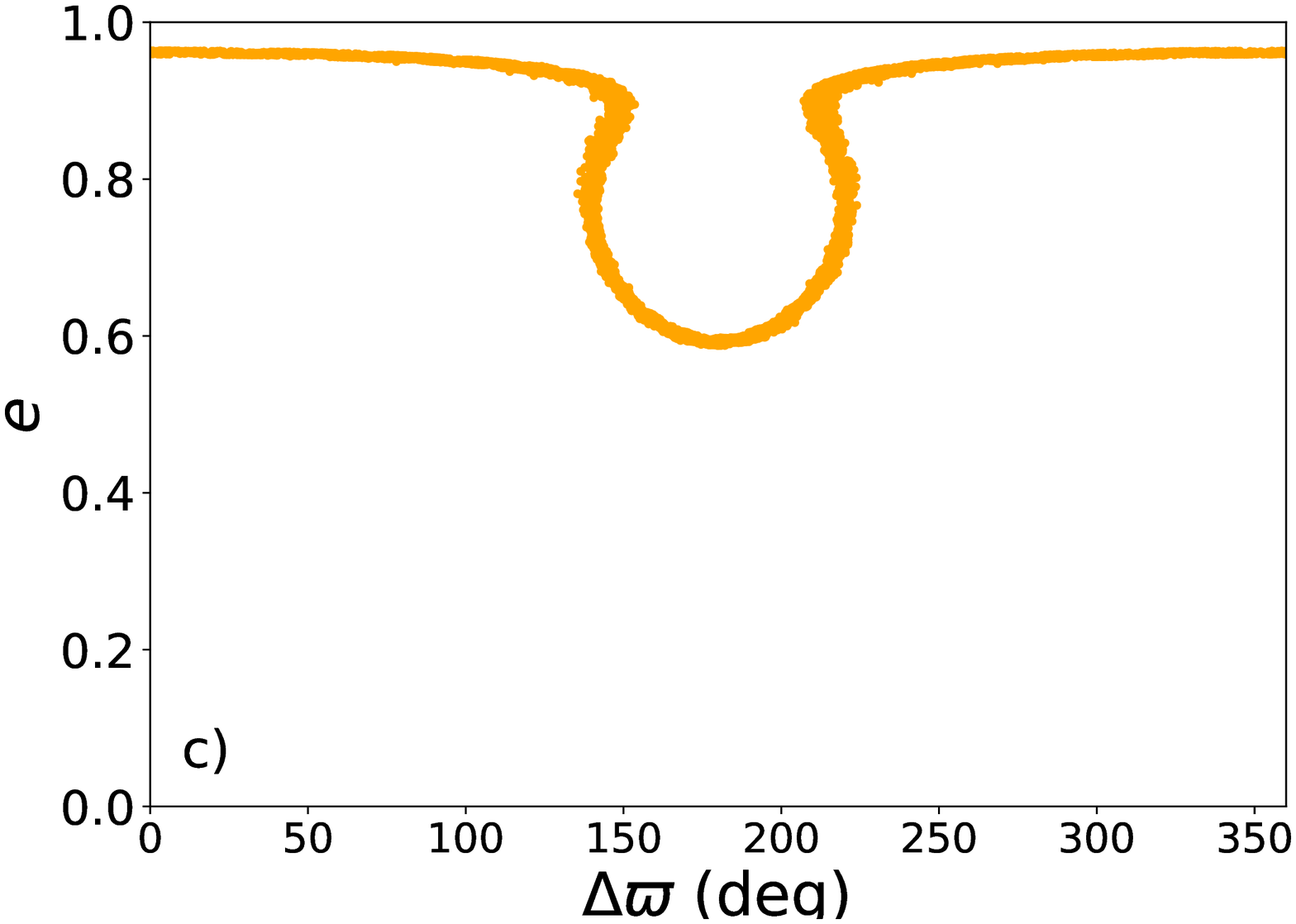} &
	\includegraphics[width=6.35cm]{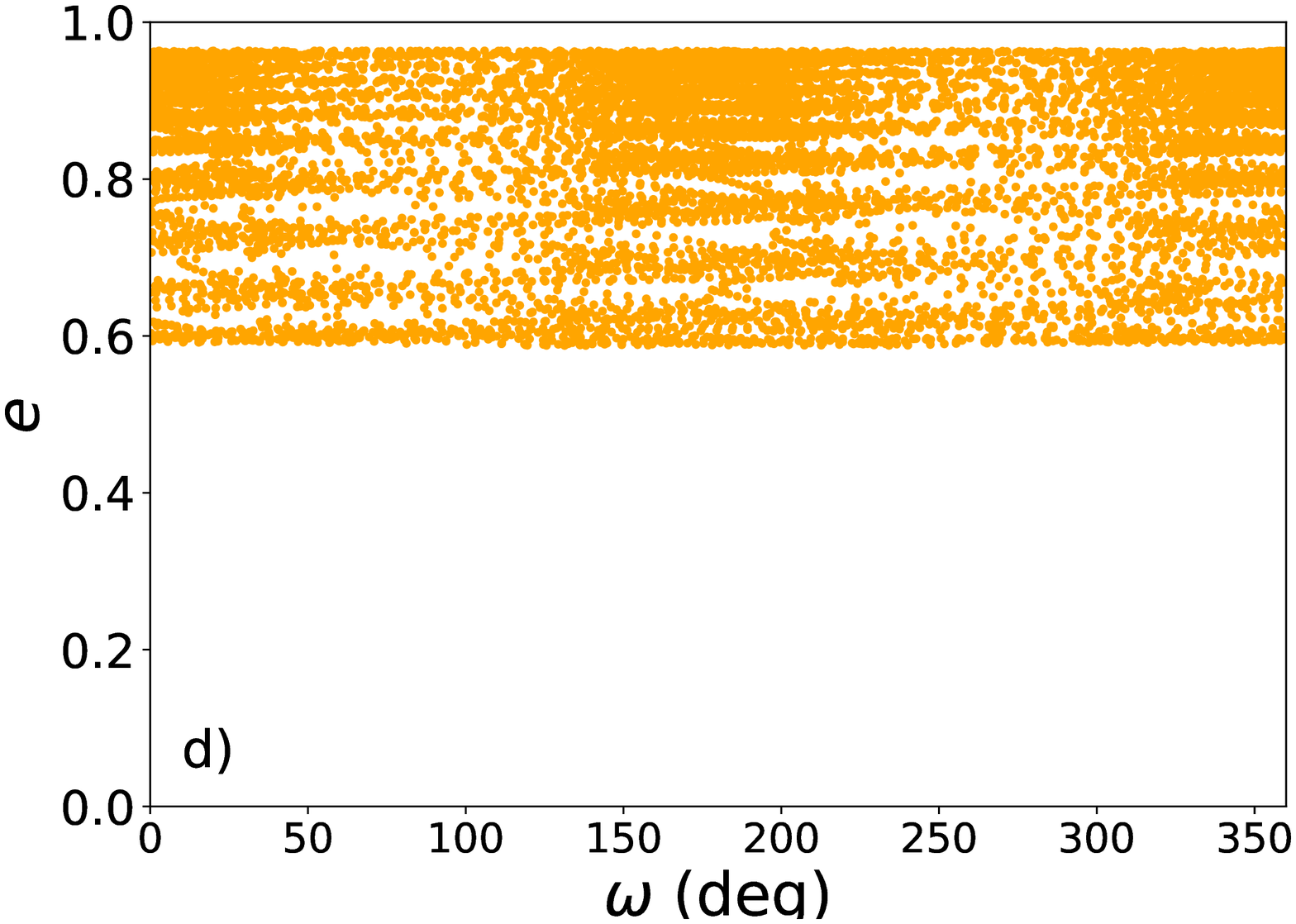}\\
	\end{tabular}
    \caption{Dynamic evolution of a system with three planets in an initial planetary configuration of decreasing mass. In the dynamic instability phase, planet 3 is ejected at approximately 0.06 Myr . In panel a) we show the evolution of the semimajor axis and pericentric distances for the surviving planets, being one of them the HJ candidate. The horizontal dashed line is located at 0.05 au. In panel b) the periodic evolution of the mutual inclination $i_{tot}=i_1+i_2$ involving the two surviving planets ($i_1$ and $i_2$ with respect to the invariant plane). Panels c) and d) show the phase diagrams $e$ vs $\Delta \varpi$ and $e$ vs $\omega$ of the HJ candidate after 0.06 Myr, respectively.}
    \label{fig:coplanar_2}
\end{figure*}

\begin{figure*}
	\textbf{Kozai-Lidov mechanism}\par\medskip
	\begin{tabular}{cc}
	\includegraphics[width=6.35cm,height=4.5cm]{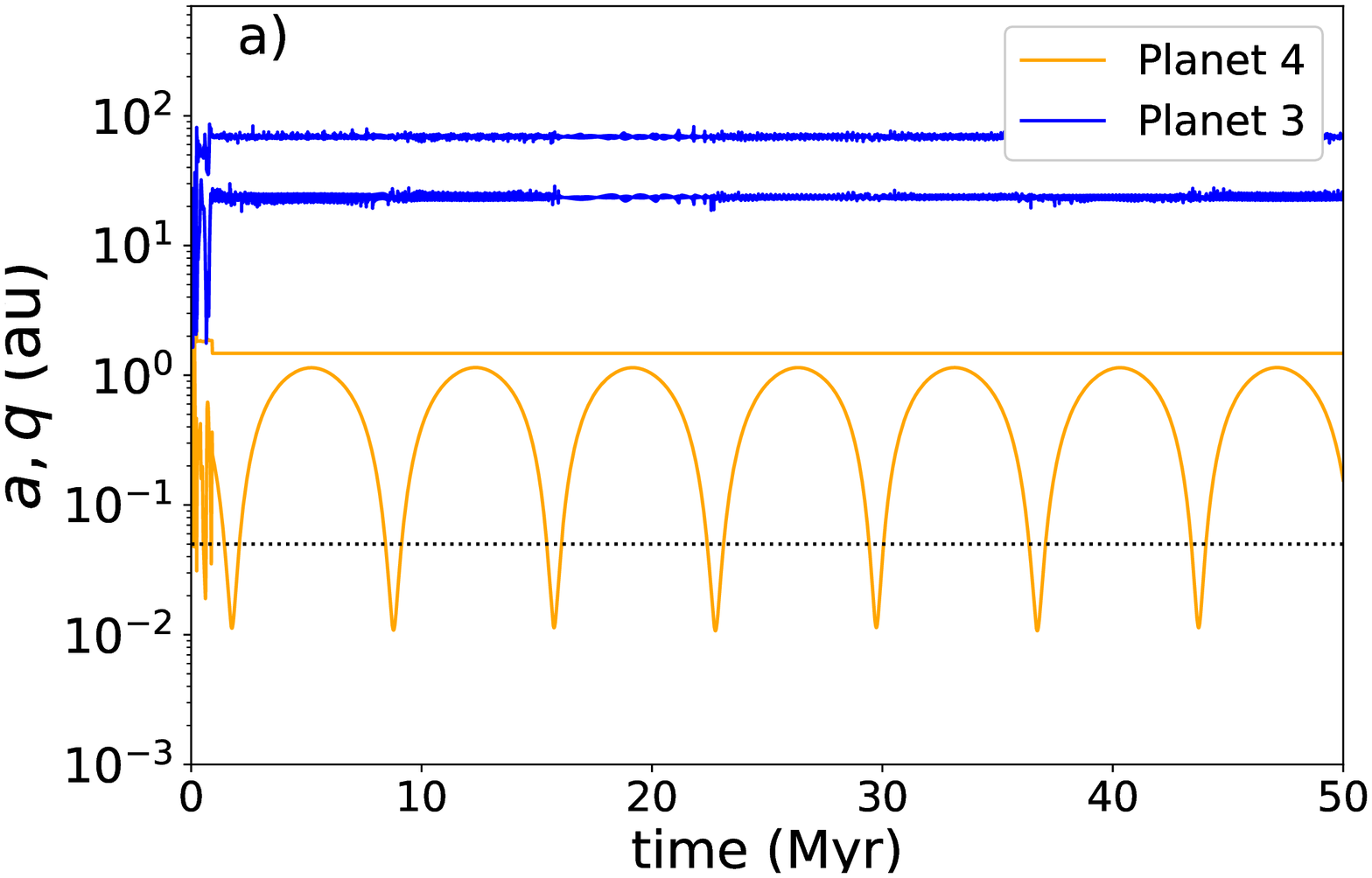} &
	\includegraphics[width=6.35cm,height=4.5cm]{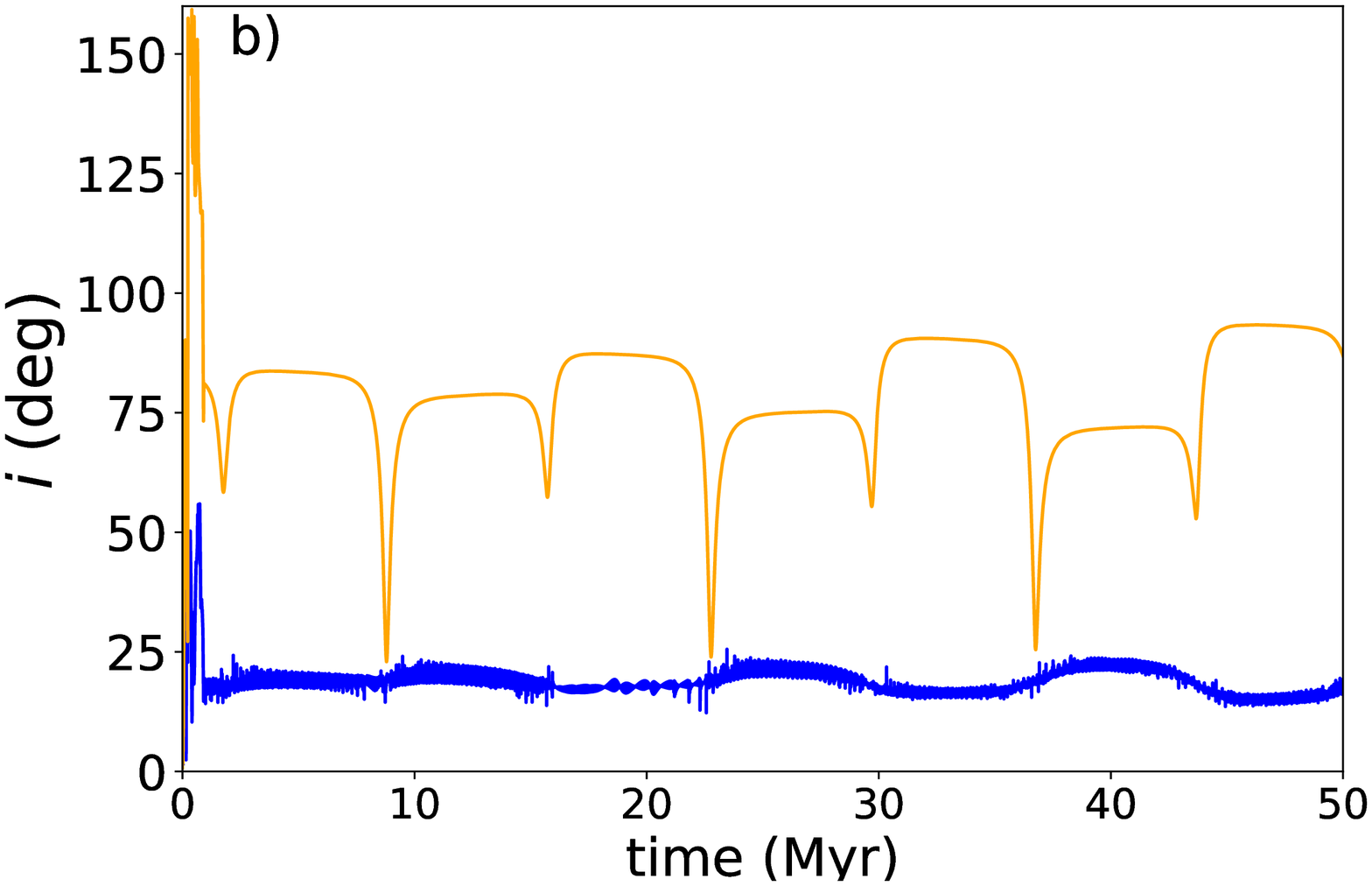}\\
	\includegraphics[width=6.35cm]{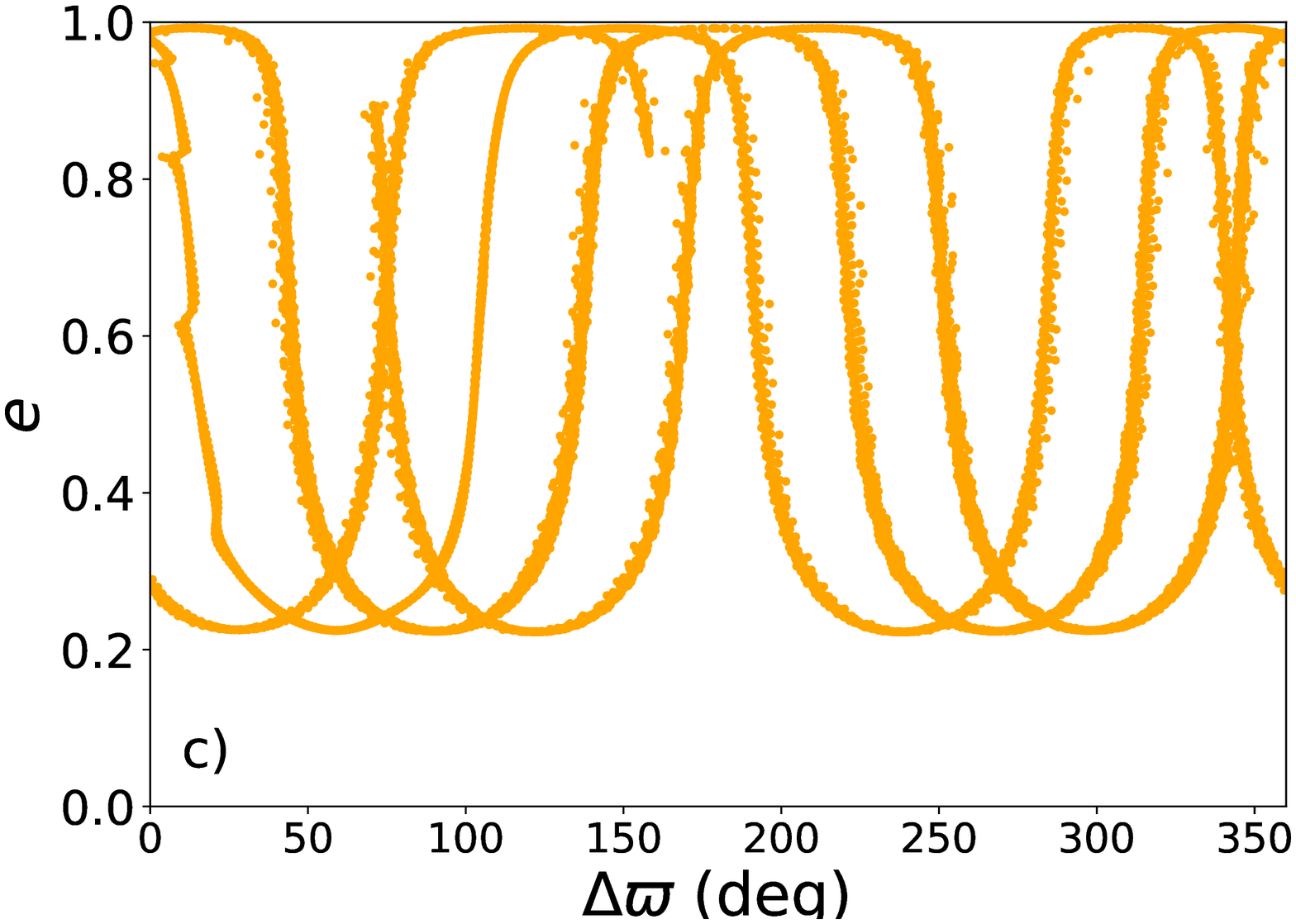} &
	\includegraphics[width=6.35cm]{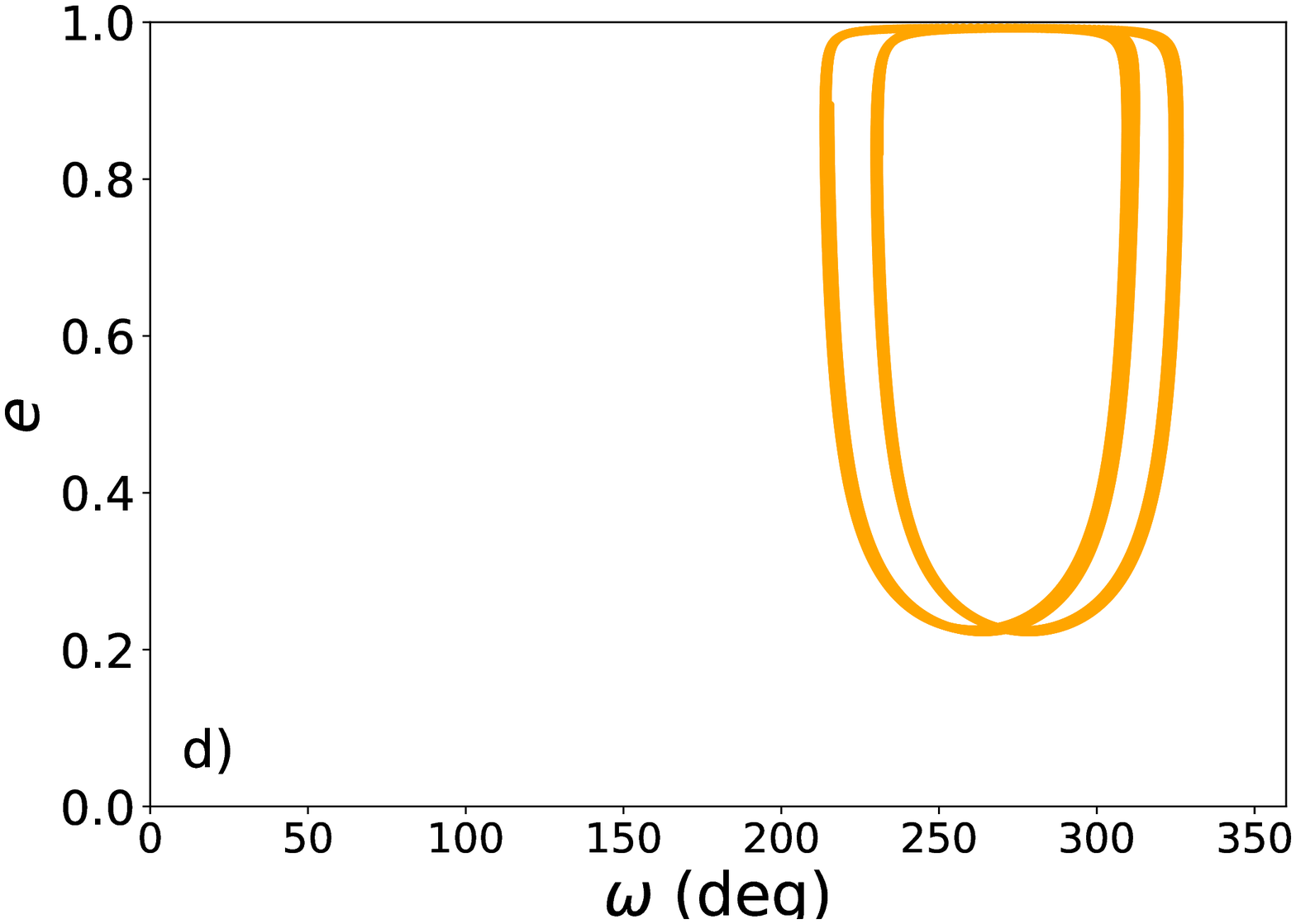}\\
	\end{tabular}
    \caption{Dynamic evolution of a system with five planets in an initial planetary configuration of random mass. In the dynamic instability phase, planets 1, 2 and 5 are ejected at 0.104, 0.915 and 0.058 Myr, respectively. In panel a) we show the evolution of the semimajor axis and pericentric distances for the surviving planets, being one of them the HJ candidate. The horizontal dashed line is located at 0.05 au. In panel b) the periodic evolution of its inclinations (with respect to the invariant plane). Panels c) and d) show the circulation of $\Delta \varpi$ and the libration of the argument of the periastron $\omega$ of the HJ candidate around $270^{\circ}$ after 0.915 Myr, respectively.}
    \label{fig:kozai}
\end{figure*}

\begin{figure*}
	\textbf{Secular chaos mechanism}\par\medskip
	\begin{tabular}{cc}
	\includegraphics[width=6.35cm,height=4.5cm]{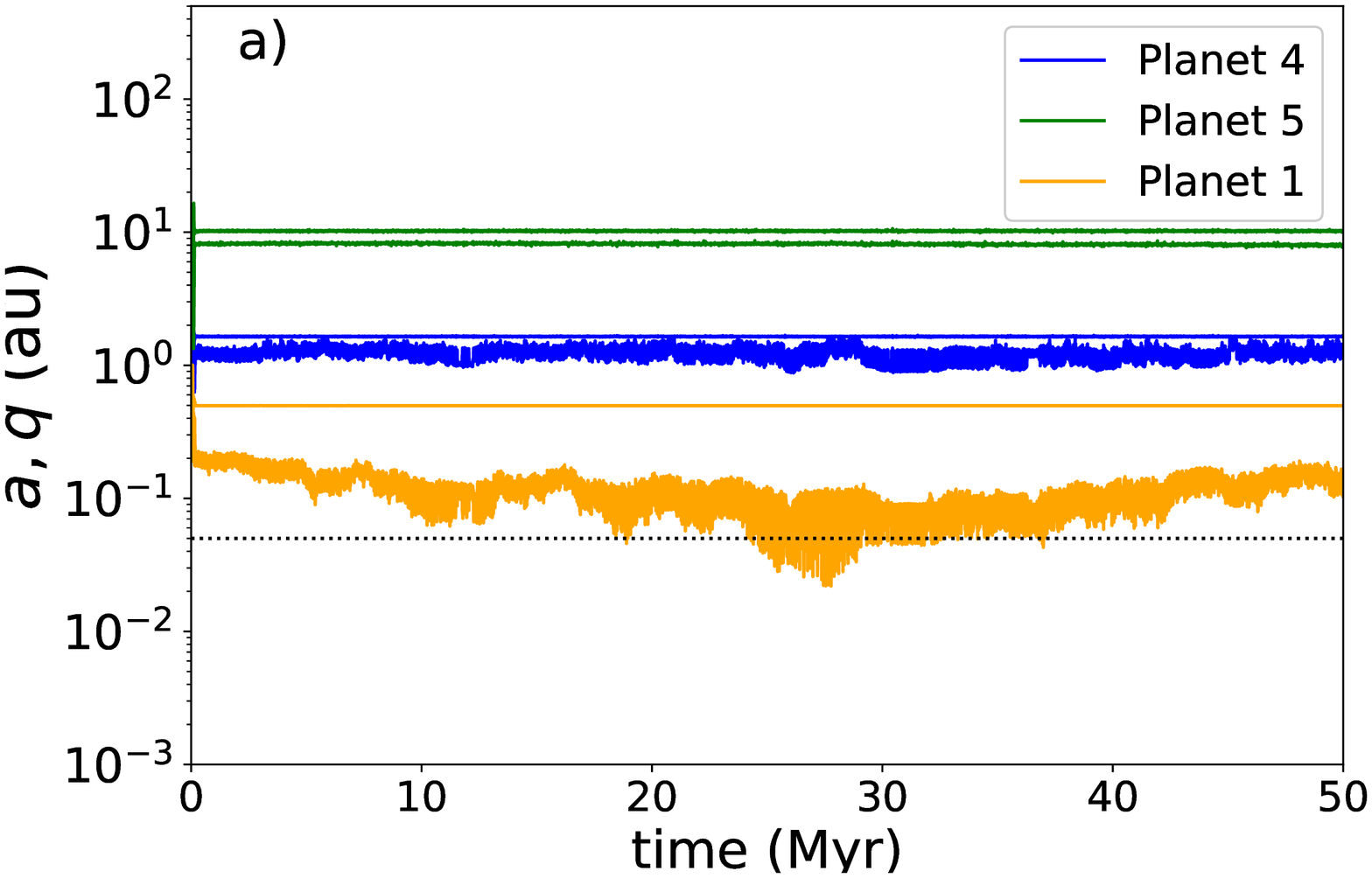} &
	\includegraphics[width=6.35cm,height=4.5cm]{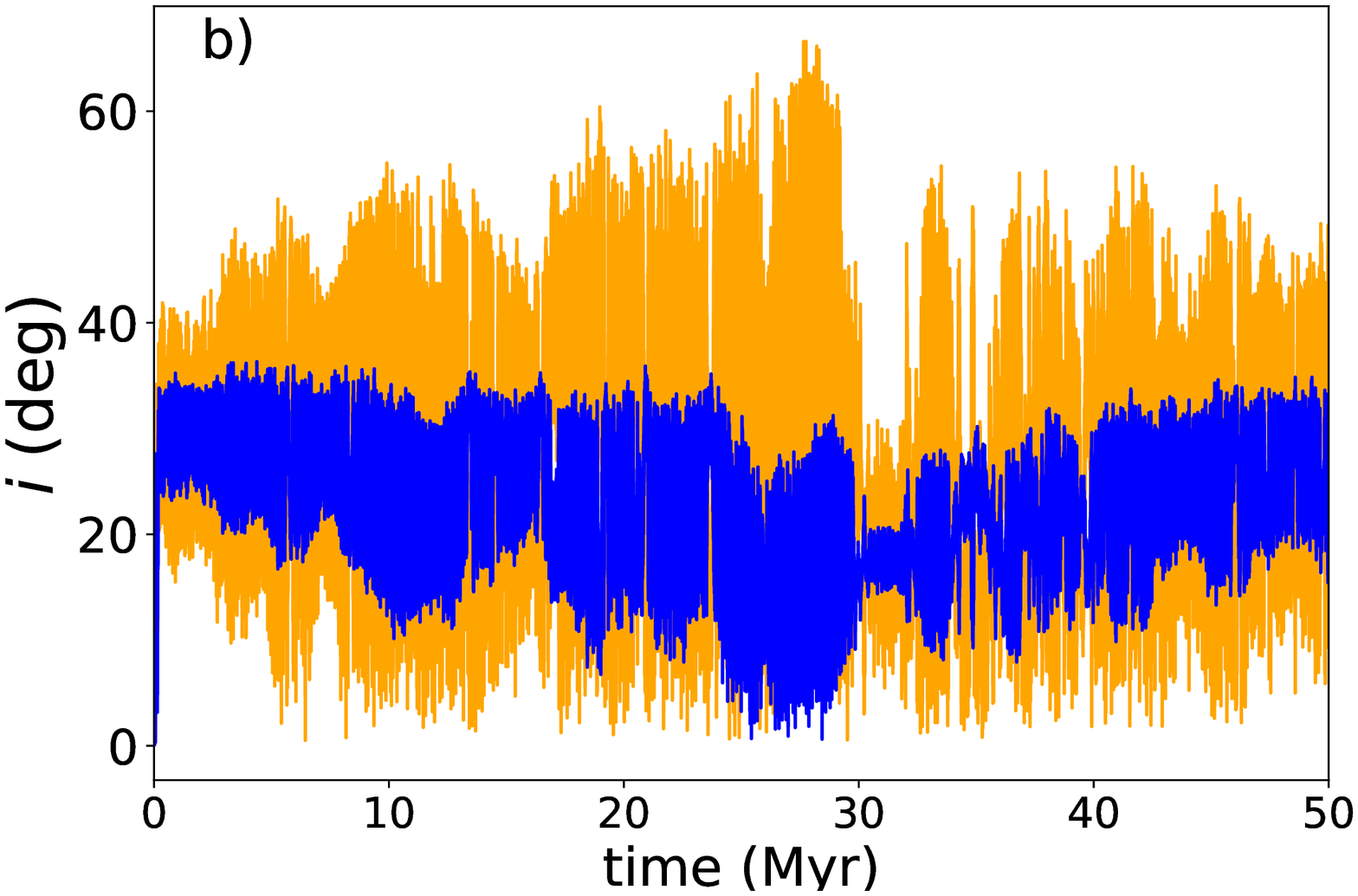}\\
	\includegraphics[width=6.35cm]{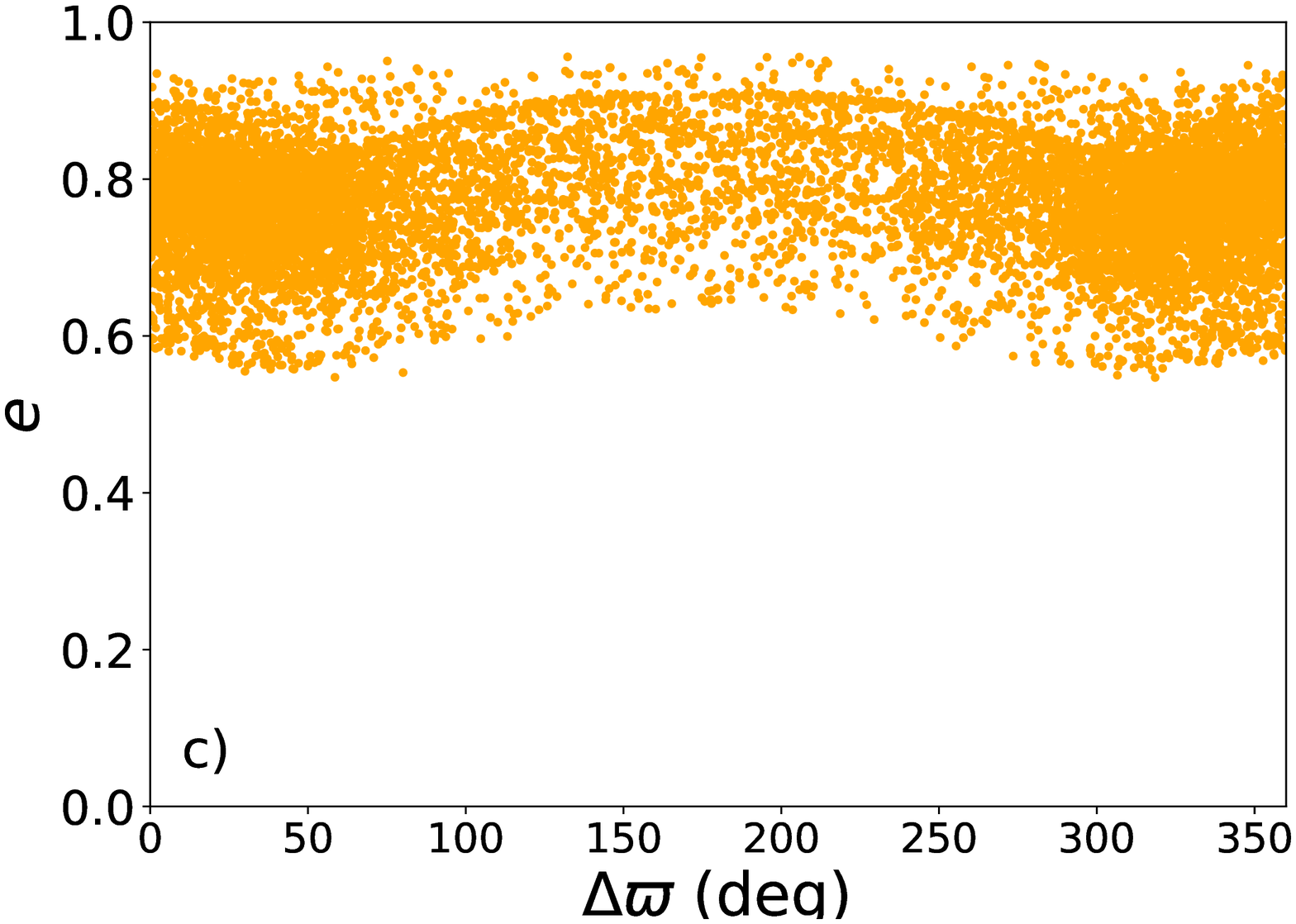} &
	\includegraphics[width=6.35cm]{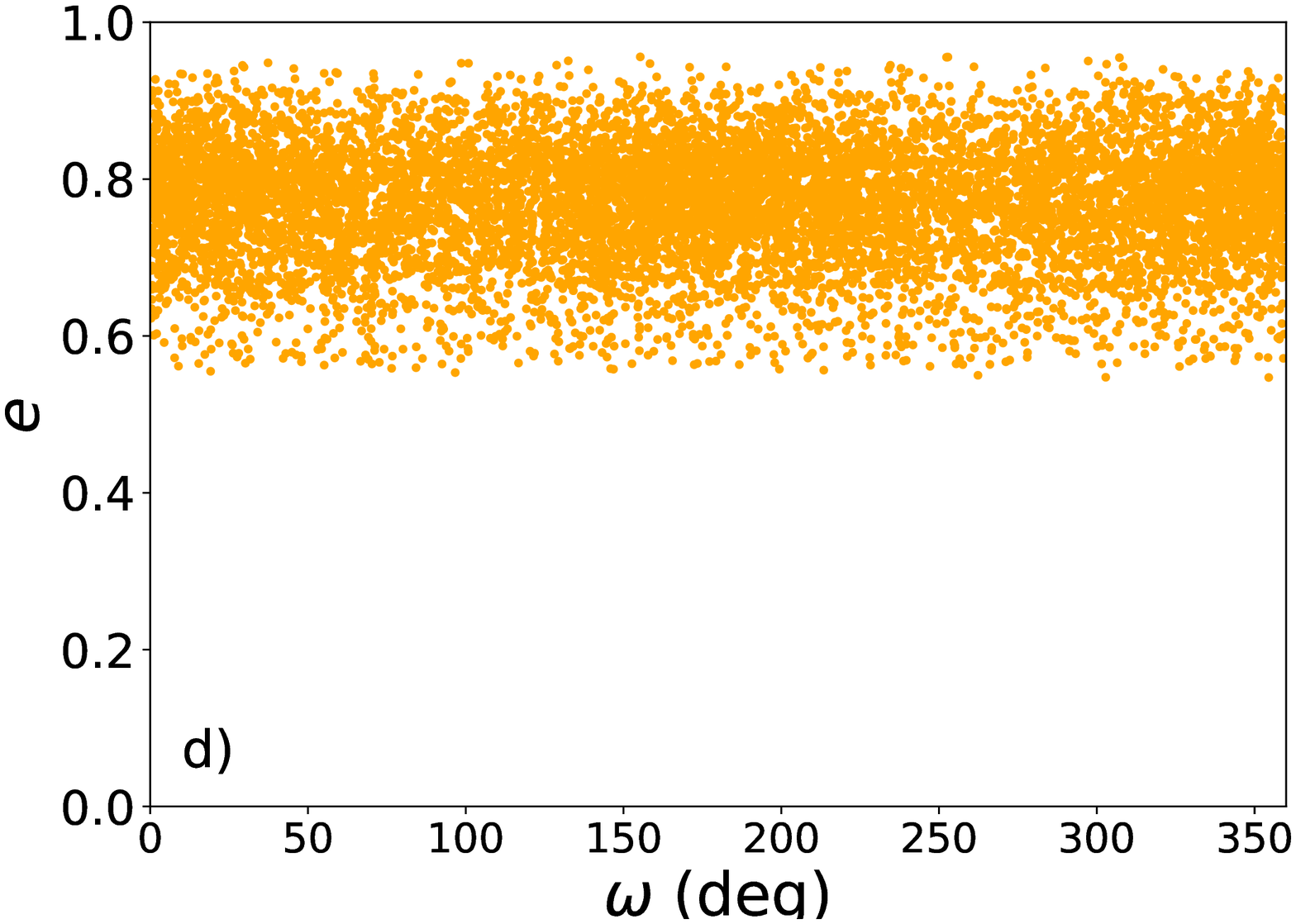}\\
	\end{tabular}
    \caption{Dynamic evolution of a system with five planets in an initial planetary configuration of equal mass. In the dynamic instability phase, planet 3 and 2 are ejected at 0.080 and 0.175 Myr, respectively. In panel a) we show the evolution of the semimajor axis and pericentric distances for the surviving planets, being one of them the HJ candidate. The horizontal dashed line is located at 0.05 au. In panel b) the non-periodic evolution of the inclinations (with respect to the invariant plane) of the HJ candidate and its closest companion. Panels c) and d) do not show the characteristics of coplanar and Kozai-Lidov mechanisms after 0.175 Myr, respectively.}
    \label{fig:secular_caos}
\end{figure*}

\begin{figure*}
	\textbf{E1 mechanism - case 1}\par\medskip
	\begin{tabular}{cc}
	\includegraphics[width=6.35cm,height=4.5cm]{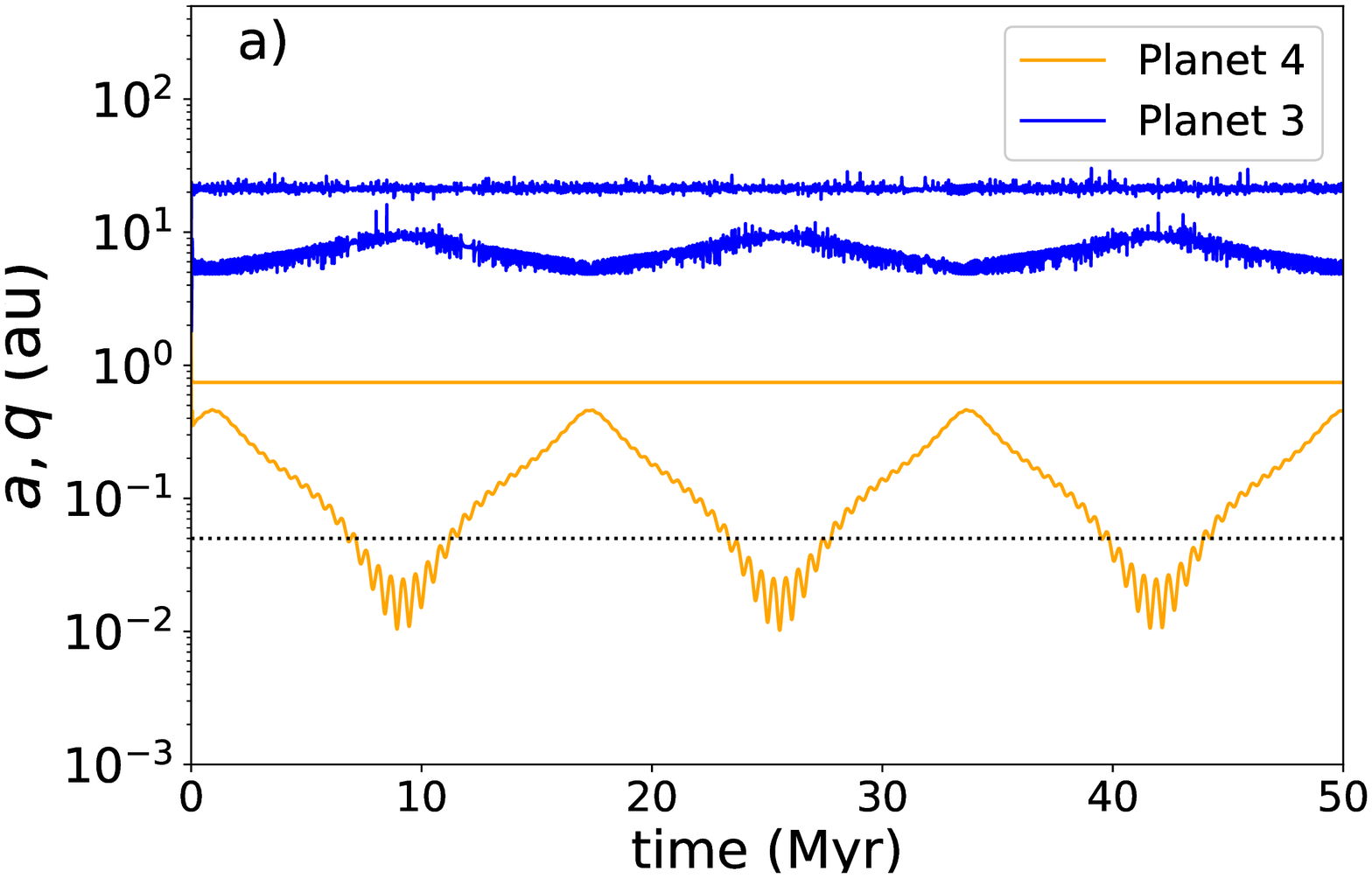} &
	\includegraphics[width=6.35cm,height=4.5cm]{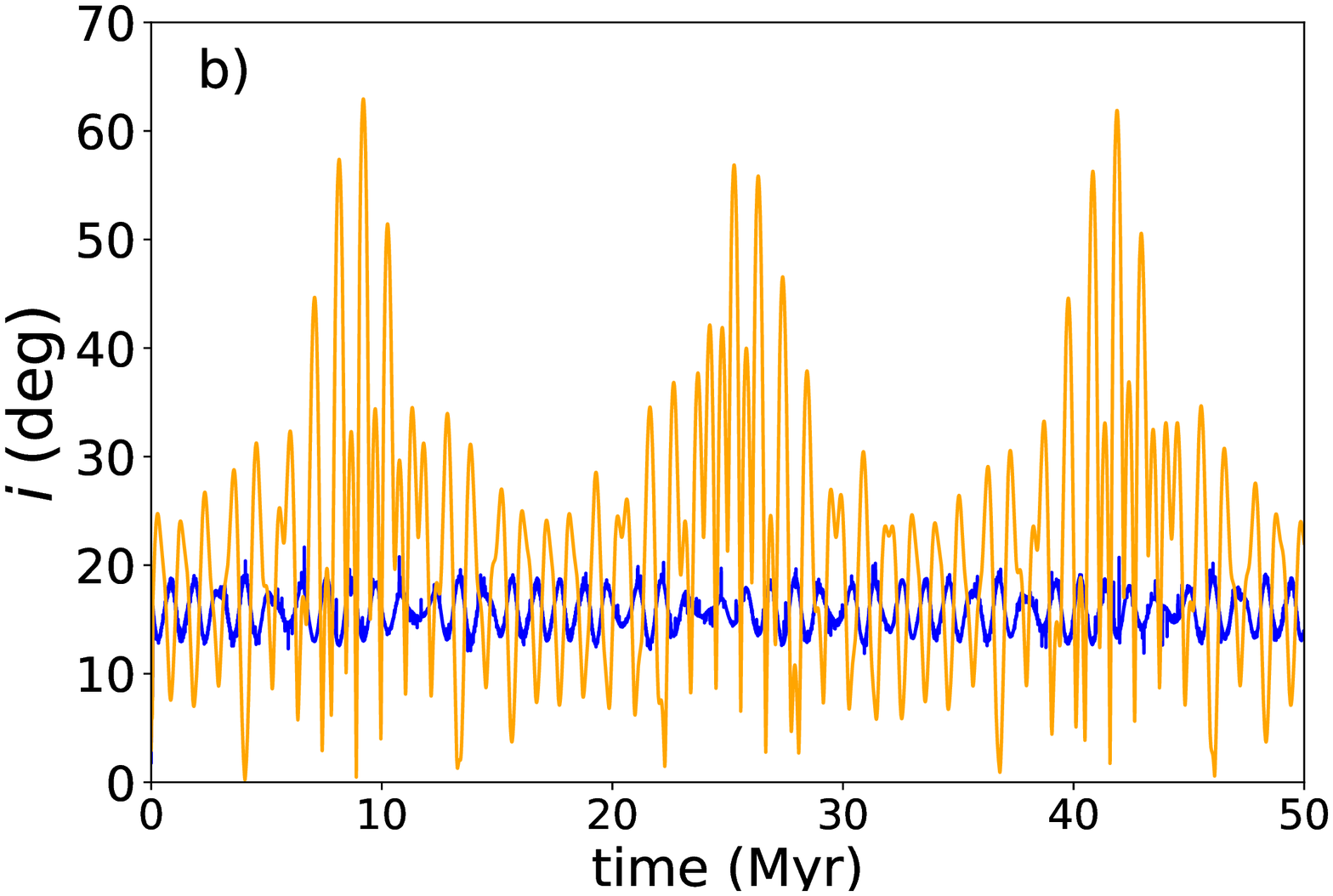}\\
	\includegraphics[width=6.35cm]{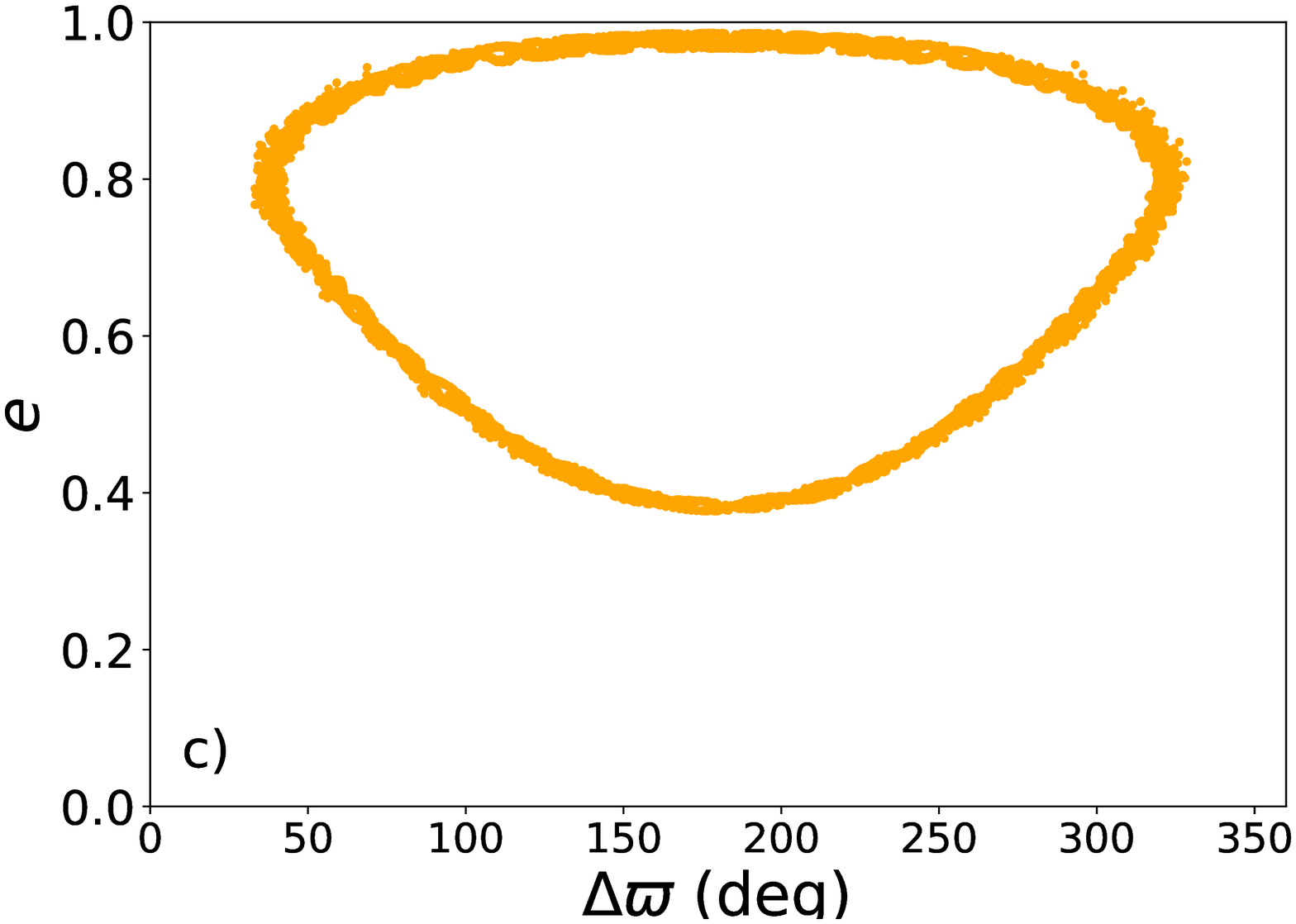} &
	\includegraphics[width=6.35cm]{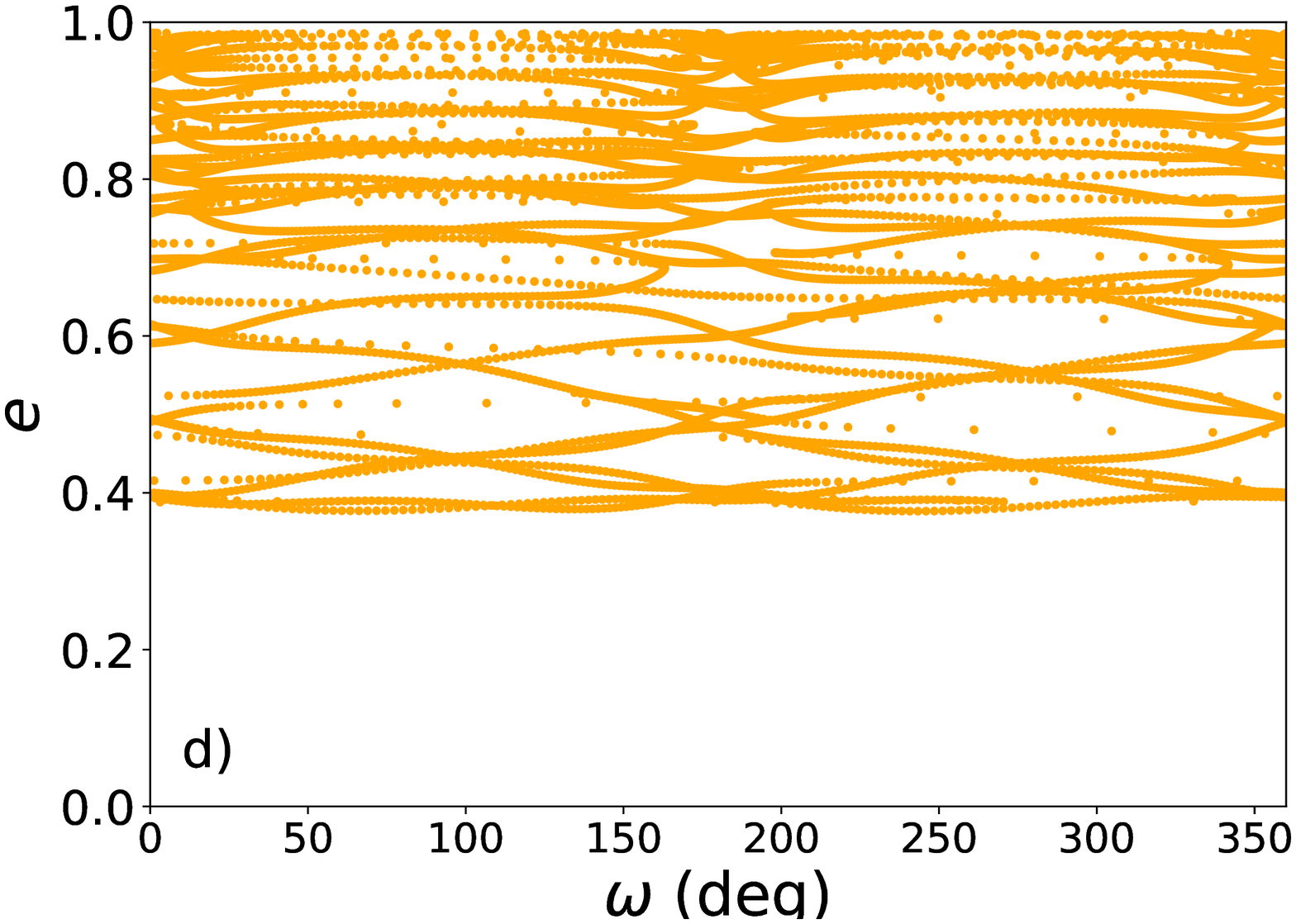}\\
	\end{tabular}
    \caption{Dynamic evolution of a system with four planets in an initial planetary configuration of increasing mass. In the dynamic instability phase, planets 1 and 2 are ejected at 0.063 and 0.005 Myr, respectively. In panel a) we show the evolution of the semimajor axis and pericentric distances for the surviving planets, being one of them the HJ candidate. The horizontal dashed line is located at 0.05 au. In panel b) the periodic evolution of its inclinations (with respect to the invariant plane).  For the HJ candidate, the evolution of its inclination includes values higher than $30^{\circ}$.  Panels c) and d) show the phase diagrams $e$ vs $\Delta \varpi$ and $e$ vs $\omega$ of the HJ candidate after 0.063 Myr, respectively.}
    \label{fig:E1_1}
\end{figure*}

\begin{figure*}
	\textbf{E1 mechanism - case 2}\par\medskip
	\begin{tabular}{cc}
	\includegraphics[width=6.35cm,height=4.5cm]{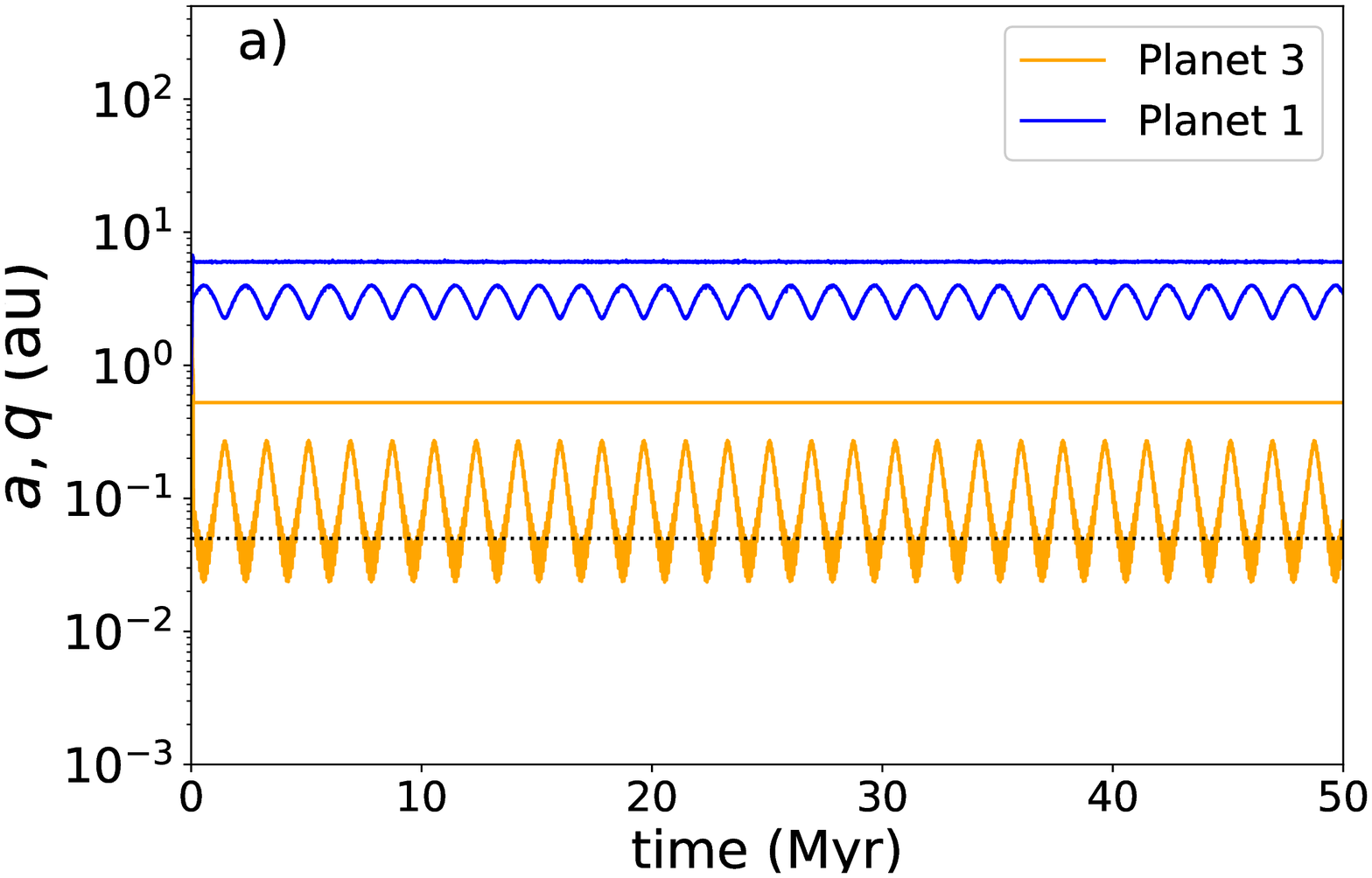} &
	\includegraphics[width=6.35cm,height=4.5cm]{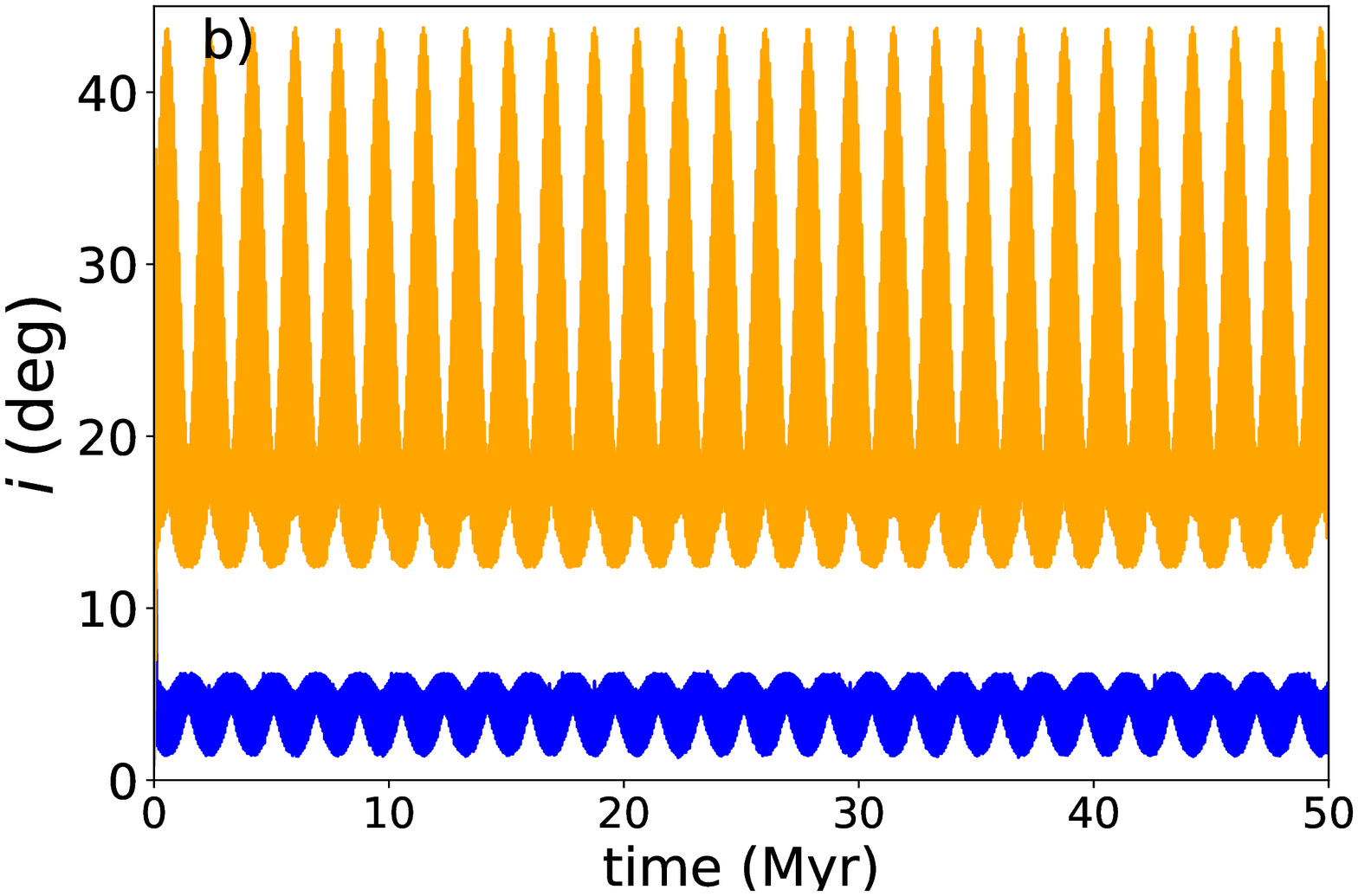}\\
	\includegraphics[width=6.35cm]{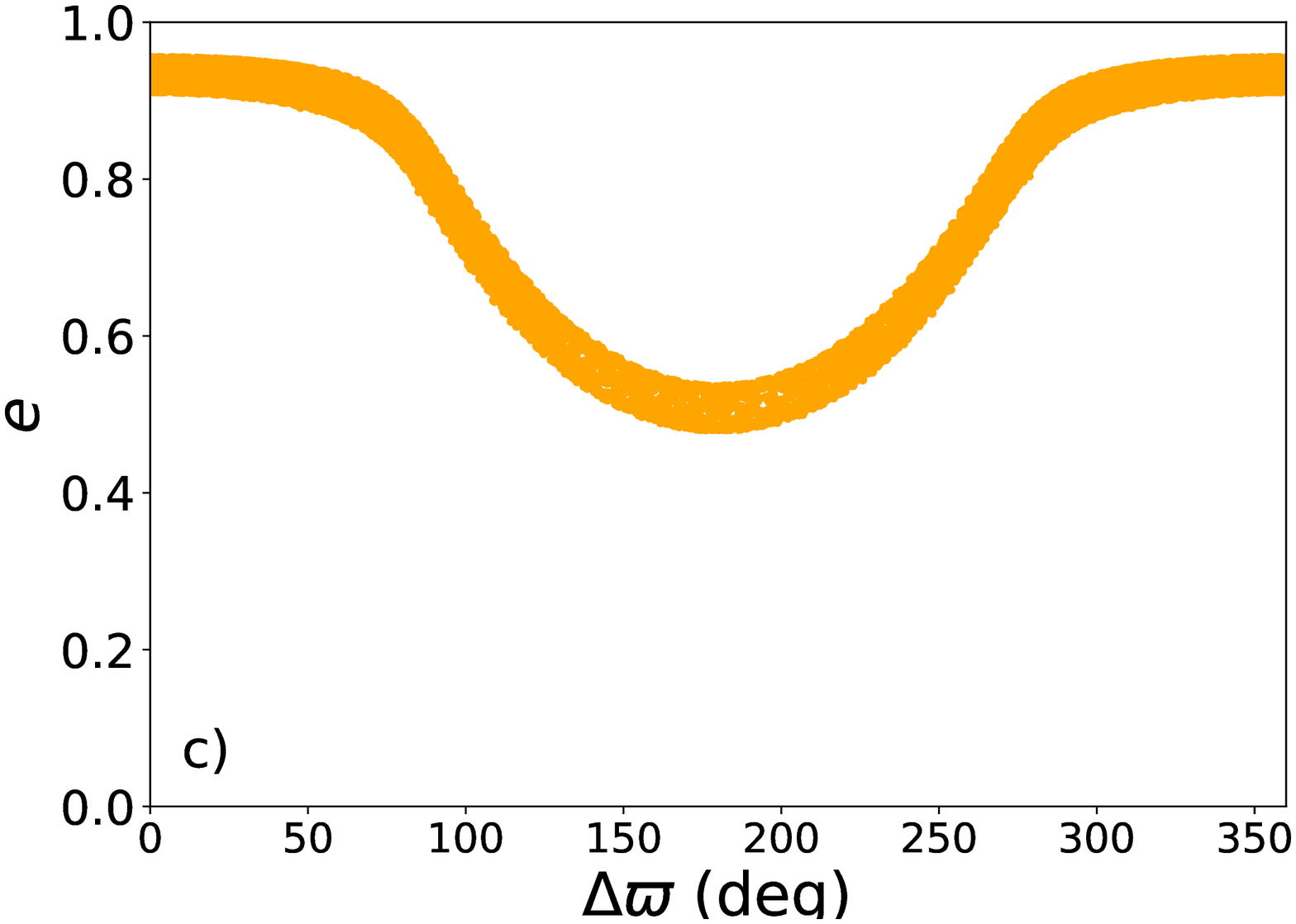} &
	\includegraphics[width=6.35cm]{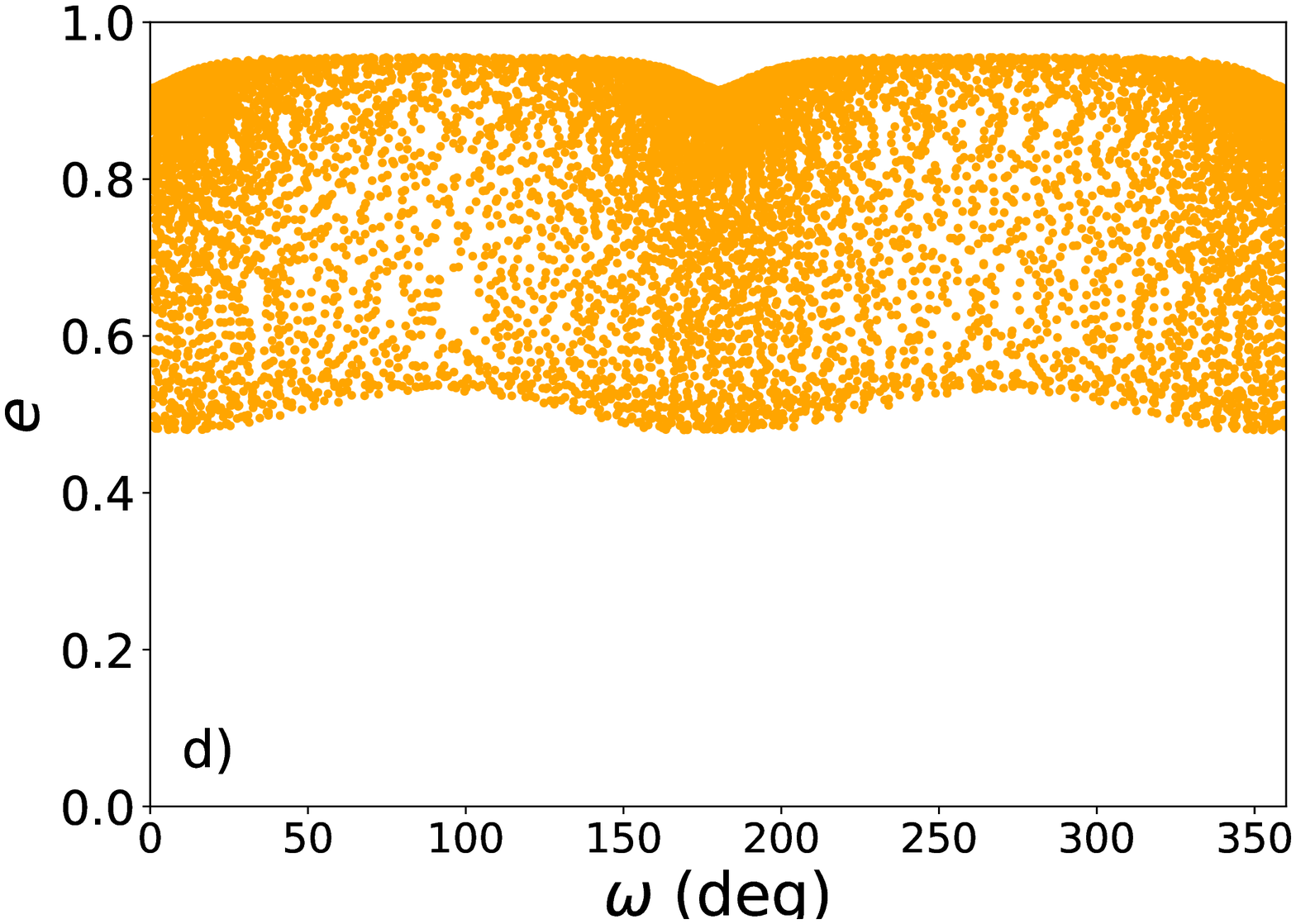}\\
	\end{tabular}
    \caption{Dynamic evolution of a system with three planets in an initial planetary configuration of equal mass. In the dynamic instability phase, planet 2 is ejected at 0.097 Myr. In panel a) we show the evolution of the semimajor axis and pericentric distances for the surviving planets, being one of them the HJ candidate. The horizontal dashed line is located at 0.05 au. In panel b) the periodic evolution of its inclinations (with respect to the invariant plane). For the HJ candidate, the evolution of its inclination includes values higher than $30^{\circ}$. Panels c) and d) show the phase diagrams $e$ vs $\Delta \varpi$ and $e$ vs $\omega$ of the HJ candidate after 0.097 Myr, respectively.}
    \label{fig:E1_2}
\end{figure*}

\begin{figure*}
	\textbf{E2 mechanism}\par\medskip
	\begin{tabular}{cc}
	\includegraphics[width=6.35cm,height=4.5cm]{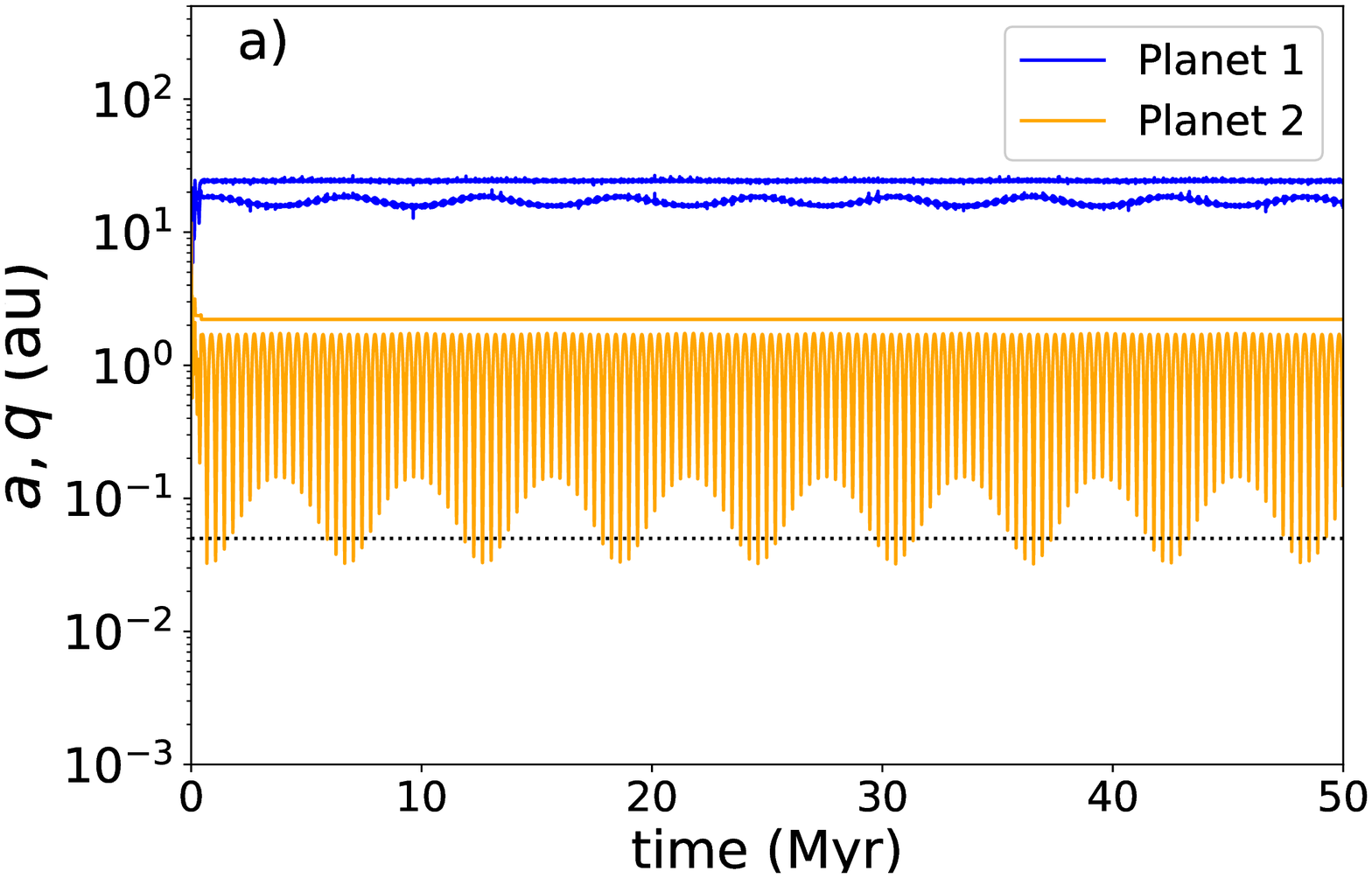} &
	\includegraphics[width=6.35cm,height=4.5cm]{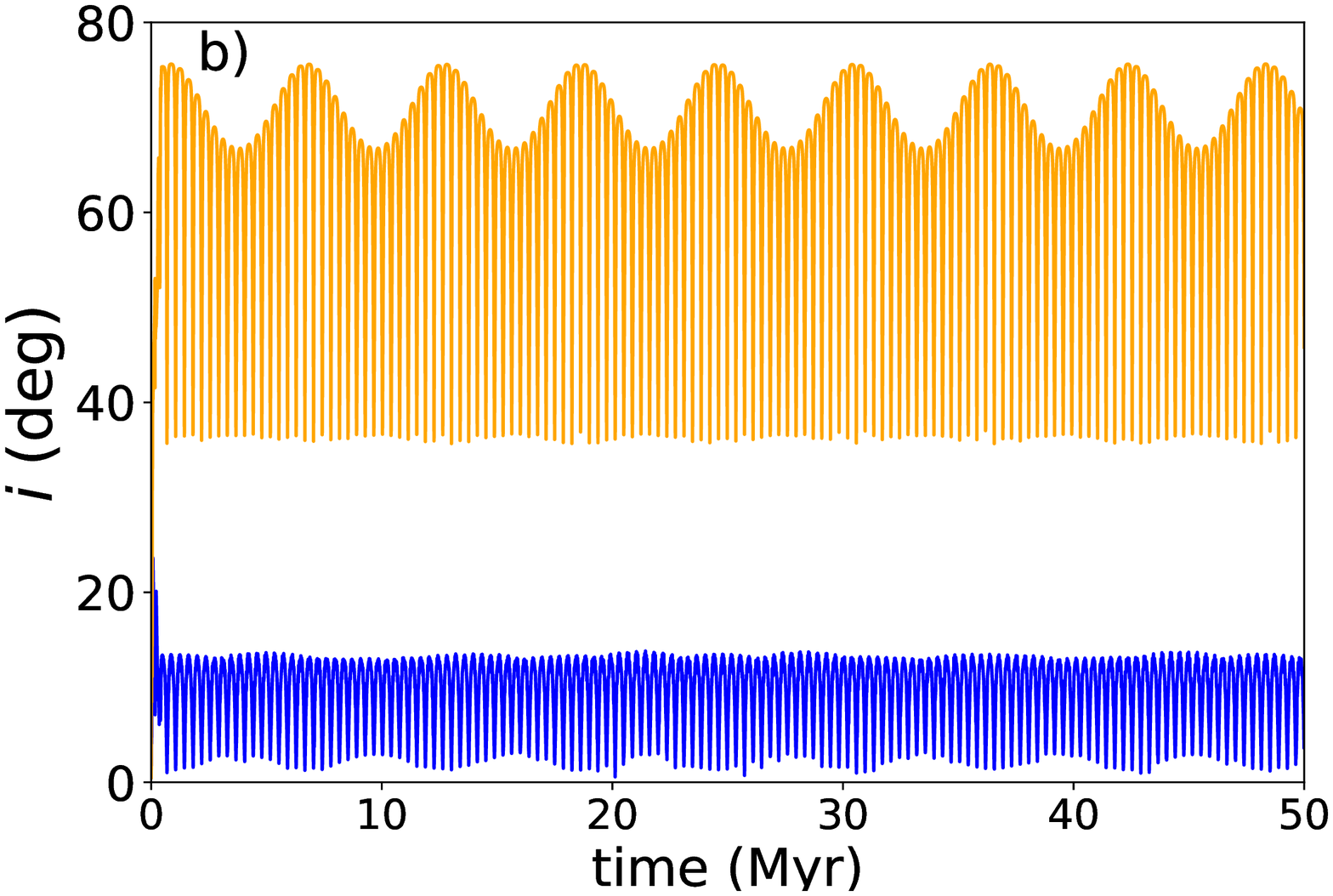}\\
	\includegraphics[width=6.35cm]{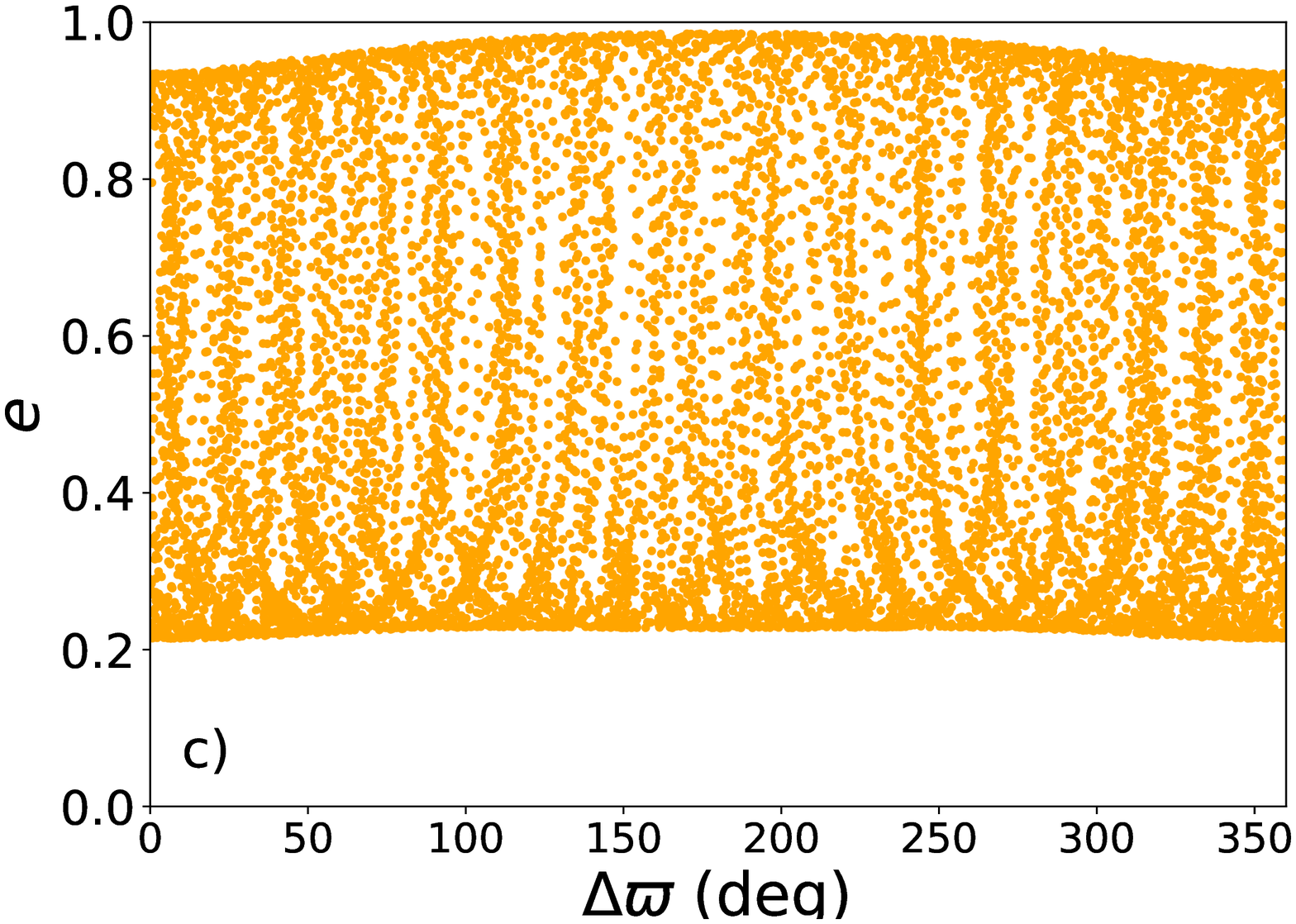} &
	\includegraphics[width=6.35cm]{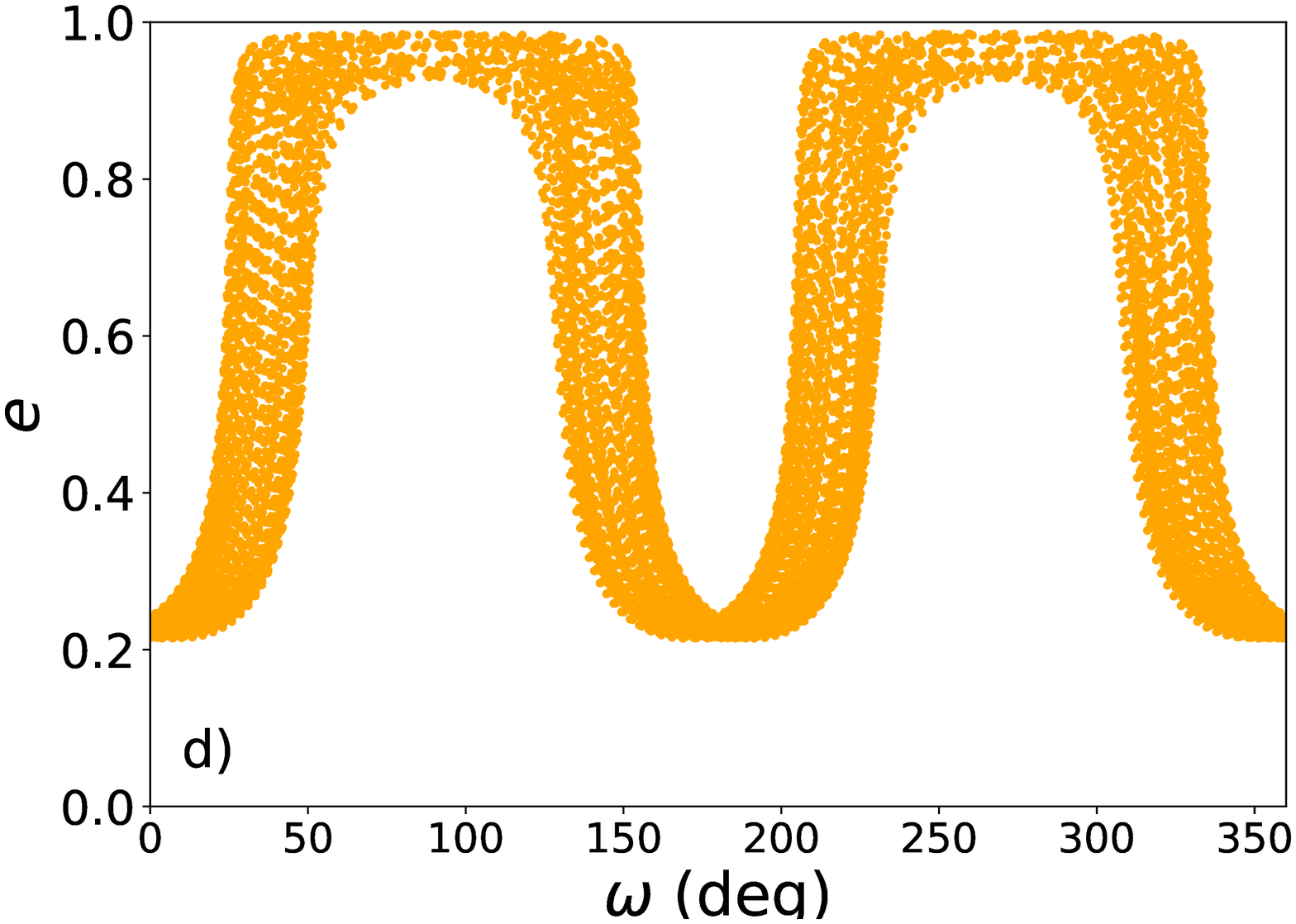}\\
	\end{tabular}
    \caption{Dynamic evolution of a system with five planets in an initial planetary configuration of decreasing mass. In the dynamic instability phase, planets 3, 4 and 5 are ejected at 0.182, 0.082 and 0.440 Myr, respectively. In panel a) we show the evolution of the semimajor axis and pericentric distances for the surviving planets, being one of them the HJ candidate. The horizontal dashed line is located at 0.05 au. In panel b) the periodic evolution of its inclinations (with respect to the invariant plane). Panels c) and d) show the phase diagrams $e$ vs $\Delta \varpi$ and $e$ vs $\omega$ of the HJ candidate after 0.440 Myr, respectively.}
    \label{fig:E2_1}
\end{figure*}

We show in Fig.~\ref{fig:scattering} to~\ref{fig:E2_1} examples of HJ candidates produced by each of the six high-eccentricity mechanisms described in Section~\ref{sec:identify_mechanism}.

\bsp	
\label{lastpage}
\end{document}